\documentclass[epj]{svjour}
\usepackage{amsmath,amssymb,epsfig,bm,mathrsfs,feynmp,slashed,color,graphicx,mathtools}
\usepackage[numbers,sort&compress]{natbib}
\usepackage{filecontents}
\usepackage{tikz}
\usepackage[utf8]{inputenc}
\newcommand{\be}{\begin{equation}}
\newcommand{\ee}{\end{equation}}
\newcommand{\bea}{\begin{eqnarray}}
\newcommand{\eea}{\end{eqnarray}}
                                                               
\newcommand{\aap}{Astron.\& Astrophys.}

\newcommand{\ep}{\epsilon}

\newcommand{\Real}{\Re{\rm e}}
\newcommand{\Img}{\Im{\rm m}}
\newcommand{\vecq}{{\bm q}}

\newcommand{\vecA}{{\bm A}}
\newcommand{\vecnabla}{{\bm \nabla}}
\newcommand{\vecv}{{\bm v}}
\newcommand{\vecp}{{\bm p}}
\newcommand{\vecP}{{\bm P}}
\newcommand{\vecK}{{\bm K}}

\newcommand{\veck}{{\bm k}}
\newcommand{\vecQ}{{\bm Q}}

\newcommand{\vecr}{\bm r}

\newcommand{\vecsigma}{{\bm \sigma}}

\newcommand{\vectau}{{\bm \tau}}

\newcommand{\ie}{{\it i.e.}}
\newcommand{\eg}{{\it e.g.}}

\newcommand{\Fd}{F^{\dagger}}

\newcommand {\pp} {\parallel}
\newcommand {\rY} {{\rm Y}}

\definecolor{red}{rgb}{0.8,0,0}
\definecolor{orange}{rgb}{0.8,0.2,0.0}
\definecolor{blue}{rgb}{0.3,0.0,0.8}
\definecolor{violet}{rgb}{0.4,0,0.4}
\definecolor{green}{rgb}{0,0.5,0.0}

\usepackage[normalem]{ulem}  

\def\araa{ARA\&A}%
\def\apj{ApJ~}%
\def\apjl{ApJ Lett.~}%
\def\apss{Ap\&SS}%
\def\aap{A\&A~ }%
\def\jcap{J. Cosmo. Astropart. Phys.~}%
\def\mnras{MNRAS~}%
\def\pra{Phys.~Rev.~A}%
\def\prb{Phys.~Rev.~B}%
\def\prc{Phys.~Rev.~C~}%
\def\prd{Phys.~Rev.~D~}%
\def\prl{Phys.~Rev.~Lett.}%
\def\nat{Nature~}%
\def\nphysa{Nucl.~Phys.~A}%
\def\physrep{Phys.~Rep.~}%
\def\physrev{Phys.~Rev.~}%
\def\rmp{Rev.~Mod.~Phys.}%
\def\anphys{Ann.~Phys.~(NY)}%
\def\ijmp{Int.~J.~Mod.~Phys.~}%
\def\epja{Eur.~Phys.~J.~A}%
\def\jltp{J.~Low~Temp.~Phys.}%
\def\ptp{Prog.~Theor.~Phys.}%
\def\ptps{Prog.~Theor.~Phys.~Suppl.}%
\def\ppnp{Prog.~Part.~Nucl.~Phys.~}%
\usepackage[bookmarks=false,colorlinks=true,linkcolor=black,citecolor=blue,urlcolor=blue]{hyperref}
\journalname{Eur. Phys. J. A}
\begin{document}
\title{Superfluidity in nuclear systems and neutron stars }
\author{Armen Sedrakian\inst{1} \and John W. Clark\inst{2,3}
}                     
%
%
\institute{Frankfurt Institute for Advanced Studies, Ruth-Moufang 
  str.\,\,1, D-60438 Frankfurt am Main, Germany \and Department of Physics and McDonnell Center for the Space
Sciences, Washington University, St.~Louis, MO 63130, USA\and
Centro de Investiga\c{c}\~{a}o em Matem\'{a}tica e Aplica\c{c}\~{o}es,
University of Madeira, 9020-105 Funchal, Madeira, Portugal}
\date{Received: 23 June 2019 / Revised: 8 August 2019 /
Published online: 30 September 2019\\
\textcopyright \, 
Societ\`a Italiana di Fisica / Springer-Verlag GmbH Germany, part of Springer Nature, 2019 
\\
 Communicated by N. Alamanos}
%
\abstract{
Nuclear matter and finite nuclei exhibit the property of
  superfluidity by forming Cooper pairs. We review the microscopic
  theories and methods that are being employed to understand the basic
  properties of superfluid nuclear systems, with emphasis on the
  spatially extended matter encountered in neutron stars, supernova
  envelopes, and nuclear collisions. Our survey of quantum many-body
  methods includes techniques that employ Green functions, correlated
  basis functions, and Monte Carlo sampling of quantum states.  With
  respect to empirical realizations of nucleonic and hadronic
  superfluids, this review is focused on progress that has been made
  toward quantitative understanding of their properties at the level
  of microscopic theories of pairing, with emphasis on the
  condensates that exist under conditions prevailing in neutron-star
  interiors. These include singlet $S$-wave pairing of neutrons in the
  inner crust, and, in the quantum fluid interior, singlet-$S$ proton
  pairing and triplet coupled $P$-$F$-wave neutron pairing.
  Additionally, calculations of weak-interaction rates in neutron-star
  superfluids within the Green function formalism are examined in
  detail. We close with a discussion of quantum vortex states in
  nuclear systems and their dynamics in neutron-star superfluid
  interiors.  
} 
\maketitle
\setcounter{tocdepth}{3}
\tableofcontents

\newpage
\section{Introduction}
\label{sec:intro}

Pairing phenomena play an important role in experimental and
observational manifestations of neutron stars and finite nuclei. Their
theoretical understanding is rooted in the microscopic theory of
superconductivity advanced by Bardeen, Cooper, and Schrieffer
(BCS)~\cite{1957PhRv..108.1175B}.  However, strong correlations, which
are generic to nuclear systems, and the complex dynamics of finite
systems such as nuclei, require developments beyond this theory.  The
study of nuclear systems is built on our understanding of the
underlying nuclear forces and the quantum many-body theory of
fermionic systems -- both aspects having undergone immense advances
during the past several decades.  In parallel with these improvements
in theoretical and computational techniques, the scope of the problems
considered has broadened over the years and now includes such
traditionally condensed-matter issues as the crossover from BCS
pairing to a Bose-Einstein condensate (BEC), inhomogeneous phases with
broken spatial symmetries, pair-breaking in strong magnetic fields and
resistive flow of quantum vorticity.

Fundamental insights into nuclear pairing were put forward shortly
after the advent of the BCS
theory~\cite{1958PhRv..110..936B,1959NucPh..13..655M,1958SPhD....3..279B}.
The overwhelming success of the BCS theory in explaining the
properties of metallic superconductors provided experimental support
for the Cooper pairing mechanism~\cite{1956PhRv..104.1189C}, by which
two species of fermions interacting via an attractive interaction form
bound states with zero total momentum at sufficiently low temperature.
Since the long-range part of the nuclear interaction is attractive, it
is natural to conclude that nucleons will form Cooper pairs in nuclei
and neutron-star matter, as these systems possess an ensemble of
quantum-degenerate states of nucleons bound by either the nuclear
force (nuclei) or gravity (neutron stars)~\footnote{Pairing mechanisms
  that arise from {\it repulsive} fermion-fermion interactions have been
  proposed in condensed matter
  systems~\cite{Al-Hassanieh2009,kagan2014modern,Kagan2PhyU2015}.}.
In due course, essential aspects of modern quantum many-body theory
were introduced, such as the Fermi-liquid theory of nuclear systems
~\cite{1962NucPh..30..239M,LarkinMigdal1963}, quantum vorticity in
superfluid neutron matter~\cite{GK1965}, and type-II superconductivity
of the proton component of neutron-star
matter~\cite{1969Natur.224..673B}.

A new impetus to the theory of fermionic pairing was provided by the
discovery of pulsars in 1967~\cite{1968Natur.217..709H} and their
identification with neutron stars~\cite{1968Natur.218..731G}. In
particular, observation of long time scales for the recovery of
regular pulse frequencies following pulsar ``glitches'' provided the 
first evidence of possible superfluidity of neutron star
interiors~\cite{1969Natur.224..872B}.  Initial many-body calculations
of pairing already predicted the correct magnitude of the gap in
neutron and proton fluids of about 1~MeV, although the nuclear
interactions available at the time were not very realistic. The
initial theoretical treatments indicated that neutron pairing in the
inner crust of a neutron star would occur in the $^1S_0$
state~\cite{Kennedy1968,ClarkYang1970,YangClark1971,Clark_err1971}
and in the $^3P_2$ state~\cite{1970PhRvL..24..775H,
1971PThPh..46..114T,Takatsuka1972} at higher densities present in the stellar
core.  Because of the low abundance of protons relative to neutrons in
$\beta-$stable neutron-star matter, protons were predicted to pair in
the $^1S_0$ state over some range of densities within the
core~\cite{Chao1972proton,Takatsuka1973}. 

The uncertainty in the values of the pairing gaps predicted for
various models was substantially reduced with the advent of potentials
that are realistic in the sense that they provide high-precision fits
to the energy dependence of the experimental scattering phase shifts
(with $\chi^2\simeq 1$). Within the BCS models of pairing there is, in
fact, a direct relation between scattering phase shifts for nucleonic
scattering and the magnitudes of the pairing
gaps~\cite{KhodelKhodelClark1996,1998PhRvC..57.1174E}. Nevertheless, 
our quantitative microscopic understanding of the way in which gap values
are affected by correlations among nucleons produced both by their
interactions and Pauli exclusion, is still 
incomplete.  We shall examine this situation at considerable depth
in Secs.~\ref{sec:pairing_patterns}, \ref{sec:higher_partial_waves}, 
and \ref{subsec:H_pairing}.

What can be learned about nuclear superfluidity and pairing from
observations of neutron stars? In fact, the observed rotational
anomalies in pulsar periods and X-ray measurements of their surface
temperatures provide us with significant evidence of superfluidity of
their interiors.  The pulsed emission of pulsars (with periods of
seconds or less) is locked to the rotation period of the star.
Pulsars are nearly perfect clocks, with periods increasing gradually
over time due to the secular loss of rotational energy.  However, some
pulsars undergo abrupt increases (glitches) in their rotation and
spin-down rates that are followed by slow relaxation toward their
pre-glitch values, on a time scale of order weeks to years. These
recoveries, when they occur, are not perfect in general, \ie, some
permanent residual shifts of either sign may remain.\footnote{Glitches
have been observed since 1969 in about 180 different pulsars with
the number of such events exceeding 500. The most prolific glitching
pulsar is the Vela pulsar, with typical changes in the spin
$\Delta\nu/\nu\simeq 10^{-6}$ and spin-derivative
$\Delta\dot\nu/\dot\nu\simeq 10^{-2}$ \cite{Cordes1988}.  Smaller
glitches with $\Delta\nu/\nu\simeq 10^{-8}$ were observed in the
Crab pulsar. For a contemporary review of glitch observations, see
\cite{Manchester2018}.}  Such behavior is attributed to a
component within the star that is only weakly coupled to the rigidly
rotating normal-matter component responsible for the emission of
pulsed radiation~\cite{1969Natur.224..673B}.  A natural candidate for
such a phase is the neutron superfluid either in the core (triplet
$P$-$F$-wave) or in the crust (singlet $S$-wave). Furthermore, young
neutron stars cool by neutrino emission from their dense interior, and
the cooling histories of neutron stars appear to be consistent with
the existing data only if the neutrino emission rates incorporate the
superfluidity of their
interiors~\cite{Pehick1992RvMP,Page2013,1996NuPhA.605..531S,YakovlevPethick2004,Sedrakian2007PrPNP,Yakovlev2001,SchmittShternin2017}.

The study of neutron-star matter evokes the astonishing universality
of quantum many-body phenomena, most intensely expressed in fermionic
pairing and superfluidity and superconductivity.  This generic
phenomenon extends across vast scales of temperature or energy: from
the atomic level, below $\sim {\rm mK}$ or $10^{-1} \mu{\rm eV}$ in
the case of liquid $^3$He, to the electronic regime exhibited
originally at $\sim 10$~K or $\sim 1$~meV (but now realized at
critical temperatures an order of magnitude higher), and to the
nucleonic or hadronic scale at $\sim 10^{10}$ K or
$\sim 1$~MeV.  Indeed, at the extremes, cold atomic gases admit
critical temperatures of order nanokelvin, while color
superconductivity at the quark-gluon level is anticipated at
temperatures of order $10^{11}$~K.

Interdisciplinary connections abound when one considers the progress
made toward quantitative microscopic description of the forms of
matter existing in the interior of a neutron star, including the
determination of the equation of state in its distinct regions.
Description of the crust of the star borrows methods from solid-state
physics as well as nuclear physics.  The material in the outer crust
resembles that in the interior of a white dwarf, with (neutron-rich)
nuclei forming a crystal lattice embedded in a Fermi sea of
relativistic electrons.  With neutron pairing in play, the inner crust
may be viewed as a nuclear analog of a terrestrial BCS superconductor,
with neutrons instead of electrons permeating lattices, some likened
to various pastas, formed from (some exotic) neutron-rich nuclei. The
band structure, which arises in this case not only for electrons but
also for free (unbound) neutrons, needs to be taken into
consideration.  The quantum fluid interior of the star is expected to
contain both neutron and proton superfluids.  A terrestrial
counterpart of the neutron component of this regime is provided by
liquid $^3$He, in both normal and superfluid phases. These two
many-body systems, nucleonic and atomic, share the feature of triplet
rather than singlet pairing, such that there is advantageous synergy
in the {\it ab intio} microscopic analysis and calculation of their
superfluid phases.  If the neutron star contains a quark core, quantum
chromodynamics also enters the picture as a crucial theoretical
ingredient.

The last decades have seen impressive advances in
  both experimental and theoretical research on pairing in the novel
  fermionic systems realized in ultracold fermionic gases, which
  exhibit many features in common with nucleonic superfluids.  Such
  atomic systems allow for remarkable control within the relevant
  parameter space, notably by tuning of the strength of the pairing
  force via a Feshbach
  resonance~\cite{BlochDalibardZwerger2008,Leggett2012,Giorgini2008};
  for general discussions see~\cite{PitaevskiiBEC,pethick_smith_2008}.
  Importantly, these systems can provide a test-bed for the repertoire
  of theoretical approaches being used to describe nucleonic pairing
  at the microscopic level.  Existing parallels have been explored in
  the context of several phenomena, especially the transition from a
  BCS-paired state to a BEC. Another parallel involves the quantum
  vortex states in ultracold atomic gases that can be explored {\it in
    situ} by imaging techniques, thereby providing an analog of
  neutron vorticity in rotating neutron stars.  Yet another parallel
  between nuclear systems and ultracold fermionic gases involves the
  unitary limit, which can be strictly realized experimentally in the
  latter systems.  At very low densities, the $S$-wave component of a
  short-range two-body interaction dominates, and its effective range
  $r_e$ becomes negligible compared to the average interparticle
  separation $r_s$ and hence the inverse Fermi wavelength $k_F^{-1}$.
  The unitary limit is reached when the density becomes so low that
  the scattering length $a$ of the interaction satisfies $k_Fa \gg 1$.
  In this unitary limit, the physics of the interacting Fermi gas
  becomes universal, with all quantities depending on a single scale,
  which may be taken as the Fermi energy.  All measurable
  thermodynamic quantities are then determined by a single quantity
  known as the Bertsch parameter, given by the ratio of the energy
  density of the unitary Fermi gas to its Fermi energy.  Because
  interacting neutrons have an anomalously large scattering length, a
  dilute neutron gas may be regarded as close to the unitary limit.
  We will address this limit in Sec.~\ref{sec:S_wave_overview}.

Radioactive-ion-beam facilities have opened an exciting new arena for
nuclear physics -- the study of exotic nuclei close to the proton and
neutron drip lines.  They enable acquisition of vital information on
the nature of the pairing in neutron/proton-rich stable and unstable
nuclei, which is of great importance in nuclear astrophysics, especially 
for an understanding of neutron-star crusts~\cite{ChamelHaensel2008}.  
Hartree-Fock-Bogolyubov (HFB) theories have evolved into a standard tool 
that incorporates pairing in the description of medium-to-heavy 
nuclei~\cite{Bulgac2002,Bender2003,GorielyChamel2013,Dobaczewski2013,Pei2014,Goriely2016,GorielyChamel2016,Bennaceur2017}.
However, modern HFB codes still employ simplistic pairing interactions
that are matched phenomenologically to more rigorous computations in
infinite nuclear matter based on realistic nuclear interactions.
While some consideration will be given to the role of pairing in
exotic nuclei and the neutron-star crust in Sec.~\ref{sec:astro}, this
will not be a topic of emphasis in the present review.

In this review, we concentrate on recent developments in the quantum
many-body problem associated with nuclear pairing and on the roles
played by pairing in macroscopic manifestations of neutron stars.
With respect to the phenomenology of neutron stars, this review will
explore the roles of pairing in their neutrino and axion emission, as
well as quantum vorticity and superfluid dynamics.  The first set of
topic relates to neutrino physics and to particle physics beyond the
standard model; the second, to phenomena that are also displayed in
terrestrial quantum fluids at liquid-$^4$He temperatures and below.
Naturally, discussion of these topics will be supported by our
concentration on microscopic many-body methods developed for
computation of the superfluid properties of neutron-star matter.

There exist a number of previous reviews that cover different stages
of development of pairing theory in the nuclear context, with emphasis
placed on varied aspects of pairing
phenomena~\cite{LombardoSchulze2001LNP,2003RvMP...75..607D,Page2013,1993PThPS.112...27T}.
The reader will benefit from consulting them for an alternative or
supplementary exposition of selected topics.

Natural units $\hbar = c = k_B = 1$ will be used throughout, unless
otherwise indicated.

\section{Basic BCS theory for nuclear systems}
\label{sec:basics}

\subsection{Pairing Hamiltonian and the gap equation}
\label{sec:contact_gap}

We start with a brief description of the simplest model of
superconductivity, based on Bogolyubov's method of canonical
stransformations~\cite{1958NCim....7..794B}.  This method has served as
a prototype for the treatment of pairing in finite
nuclei~\cite{Ring80}.  Consider a system of fermions with macroscopic
number $N$ described by the pairing Hamiltonian $\hat H$, defined by
\bea\label{eq:H_mf}
\hat H-\mu \hat N &=& \sum_{\vecp,\sigma}\ep_p
\hat a^{\dagger}_{\vecp,\sigma}\hat a_{\vecp,\sigma}\nonumber\\
&&\hspace{-2.0cm}  +\frac{1}{V}\sum_{\vecp_1+\vecp_2=\vecp_3+\vecp_4} \!\!\!
v(\vecp_3,\vecp_4;\vecp_1,\vecp_2)
\hat a^{\dagger}_{\vecp_3,\uparrow}\hat  a^{\dagger}_{\vecp_4,\downarrow} 
\hat  a_{\vecp_1,\downarrow} \hat a_{\vecp_2,\uparrow}, 
\eea 
where $\hat a^{\dagger}_{\vecp,\sigma}$ and $\hat a_{\vecp,\sigma}$
are respectively the creation and annihilation operators for particles
having spin $\sigma = \uparrow\downarrow$ and momentum $\vecp$, and
$\mu$ is the chemical potential.  The first term on the right includes
the kinetic energy, with $\ep_p = p^2/2m-\mu$ and $m$ the particle
(effective) mass, while the second term represents the potential
energy, with $v(\vecp_3,\vecp_4;\vecp_1,\vecp_2)$ denoting the
attractive pairing interaction and $V$ the volume.  If one considers
only pairing with zero total momentum (see
Sec.~\ref{sec:Unconv_BCSBEC} where this restriction is lifted), then
the sum in Eq.~\eqref{eq:H_mf} should be constrained to momenta
fulfilling the condition $\vecp_1+\vecp_2=0$.

The method of canonical transformations introduces two new creation and 
annihilation operators $\hat\alpha_{\vecp,\sigma}^{\dagger}$ and $\hat\alpha_{\vecp,\sigma}$ through
\be\label{eq:qp_operators} \hat a_{\vecp,\uparrow}=
u_{\vecp} \hat\alpha_{\vecp,\uparrow}
+v_{\vecp} \hat\alpha^{\dagger}_ {-\vecp,\downarrow}\quad
{\rm and} \quad \hat a_{\vecp,\downarrow}=
u_{\vecp} \hat\alpha_{\vecp,\downarrow}-v_{\vecp}
\hat\alpha^{\dagger}_{-\vecp,\uparrow}.  
\ee
These new operators obey the fermionic commutation relations
\bea
\{\hat\alpha_{\vecp,\sigma},\hat\alpha^{\dagger}_{\vecp',\sigma'}\}&=&
\delta_{\vecp\vecp'}\delta_{\sigma\sigma'},\\
\{\hat\alpha_{\vecp,\sigma},\hat\alpha_{\vecp',\sigma'}\}&=&
\{\hat\alpha^{\dagger}_{\vecp,\sigma},\hat\alpha^{\dagger}_{\vecp',\sigma'}\}=0,
\eea
provided the Bogolyubov amplitudes $u_{\vecp}$ and $v_{\vecp}$ (which
can be chosen real in the absence of flow, or for $S$-wave pairing)
satisfy the normalization condition $u_{\vecp}^2+v_{\vecp}^2 =
1$. This implies that the thermodynamic potential of the system is a
functional of only one
amplitude, conventionally $v_{\vecp}$.  

At a given temperature $T$, this amplitude may be determined 
by minimization of the expectation value of the free energy
(see \cite{Abrikosov:Fundamentals}, $\S 16.4$)
\be\label{eq:free_energy}
E-\mu N - TS= \langle \hat H-\mu \hat N -T \hat S\rangle
\ee
where $\langle \cdots \rangle$ denotes a statistical average over the
operator enclosed in brackets, $\hat N$ is the particle-number and
$\hat S$ is the entropy operator. The quasiparticle occupation numbers
are defined by
$
\langle \hat\alpha^{\dagger}_{\vecp,\downarrow}\hat\alpha_{\vecp,\downarrow}
\rangle = n_{\vecp,\downarrow}$  and 
$\langle \hat\alpha^{\dagger}_{\vecp,\uparrow}\hat\alpha_{\vecp,\uparrow}\rangle
 = n_{\vecp,\uparrow}.
$
Minimization, which requires ${\delta (E-\mu N)}/{\delta} v_{\vecp} =
0$, leads to the {\it gap
  equation.}\footnote{
Note that the minimization at constant
  quasiparticle occupation numbers automatically requires that the
  entropy of the system, given by
 $$
  S = - \sum_{\vecp,\sigma}
 \left[
  n_{\vecp,\sigma} \log n_{\vecp,\sigma}
 + (1-n_{\vecp,\sigma} )\log (1- n_{\vecp,\sigma})
 \right],
 $$
 is held constant~\cite{LifshitzPitaevskiiStat2}, \S 39.  It is also
 worthwhile to note that the terms involving operations of the type
 $\alpha^{\dagger}\alpha^{\dagger}$ and $\alpha\alpha$ that emerge in
 the interaction part of the Hamiltonian when evaluating
 Eq.~\eqref{eq:free_energy} in terms of Bogolyubov operators vanish.
 Such terms would account for fluctuations in the system, but are
 beyond the scope of the present mean-field treatment.}  For the case
of an $S$-wave pairing interaction $v_0(p,p')$, the gap equation takes
the form (with $p$ the modulus of $\vecp$)
\be \label{eq:gap_swave}
\Delta_{p} = -
\frac{1}{V}\sum_{\vecp'} v_0(p,p')
u_{p'}v_{p'} (1-n_{p',\downarrow} - n_{p',\uparrow}),
\ee
with
\be\label{eq:u_and_v} 
u_p^2 = \frac{1}{2} \left(1+\frac{\ep_p}{E_p}\right),\qquad
v_p^2 = \frac{1}{2} \left(1-\frac{\ep_p}{E_p}\right).
\ee
The quasiparticle energy is given by
\be
\label{eq:qpe}
 E_p = \sqrt{\ep_p^2+\Delta_p^2}, 
\ee
\ie, the spectrum of the system features an energy gap~\footnote{Note that the variation 
${\delta (E-\mu N)}/{\delta n_{\vecp,\uparrow}}$, with $u_p$ and
$v_p$ held constant, yields the quantity $E_p$, confirming its
interpretation.} $\Delta_p$. Consequently, fermionic excitations can be
created in the system if a Cooper pair breaks, for which an energy of
at least $2\Delta_p$ must be supplied to the system.
We now observe that the spectrum $E_p$ reaches a minimum at the
Fermi momentum $p_F$, such that the minimal value of $E_p/p$, given
by $\Delta_{p_F}/p_F$, is positive definite.  Accordingly, the
Landau criterion for superfluidity is fulfilled: it is impossible to
create excitations for velocities less than $\Delta_{p_F}/p_F$.
(For an extended discussion of this criterion see 
\cite{LifshitzPitaevskiiStat2}, $\S~23$.)  This behavior ensures 
an important property of conventional superconductors -- the absence 
of resistance to an electrical current, or the absence of dissipative 
fluid flow in neutral fermionic fluids. 

We should note that the existence of a gap is sufficient {\it but not
  necessary} for occurence of superfluidity or superconductivity in
attractive, one-component, homogeneous fermionic systems at weak
coupling.  In other words, fermionic systems possessing the spectrum
$E_p$ are superconducting (superfluid), but not every fermionic
superconductor (superfluid) needs to have such a spectrum.  For
example, superconductivity in some materials could be {\it gapless} in
the sense that the gap vanishes at least in some segments of the Fermi
surface, see \cite{Abrikosov:Fundamentals}, $\S 21.2$ and
Sec.~\ref{sec:Unconv_BCSBEC} below.

Before proceeding, we may recall that as an alternative to the
Bogoliubov canonical transformation method, one may start with 
a variational {\it Ansatz} for the superfluid ground state of the
form
\bea
\label{eq:BCSket0}
|\Phi_{\rm BCS} \rangle =
\prod_\vecp \left[ u_\vecp +  v_\vecp a_{\vecp,\uparrow}^\dagger
 a_{-\vecp,\downarrow}^\dagger \right] \vert 0 \rangle ,
\eea
where $|0 \rangle$ denotes the vacuum state.  This may be recognized 
as the original BCS trial ground state \cite{1957PhRv..108.1175B}, 
expressed in Bogoliubov amplitudes.  Constrained functional minimization of 
$\langle \Phi_{\rm BCS}|{\hat H}-\mu{\hat N}|\Phi_{\rm BCS} \rangle$
leads to a gap equation identical to that arrived at above, under 
identical assumptions for the interaction $v$, generalization 
to finite termperature being straightforward.  By construction,
both formulations are mean-field approximations, in the respect 
that a given Cooper pair is considered to move in the mean field of the 
corresponding normal system -- in this sense, one is treating
one Cooper pair at a time.

The arguments above apply strictly to pure, isotropic fermionic
systems without impurities or a periodic lattice. Such complications
are prominent in terrestrial solid-state superconductors and have been
predicted to exist in neutron-star crusts (see
Sec.~\ref{sec:pairing_patterns}).  In simulations by cold-atom
systems, they may be created, \eg, by optical lasers.  The theory of
superconductivity in the presence of impurities has been discussed,
for example, in \cite{Abrikosov:QFT}, where it is shown that
gapless superconductivity can arise. If fermions are embedded in a
periodic lattice, the material is characterized by energy bands, which
are separated by band gaps where fermionic states are forbidden.  The
position of the Fermi surface with respect to the valence and
conduction bands then determines the electrical, optical, and other
properties of the material. As is well known, BCS superconductivity
does not arise in semiconductors or insulators where the chemical potential
is located within a band gap. By contrast, in conductors the chemical
potential  is located outside the band gap, and the Cooper mechanism
takes effect at temperatures below its critical value.

Uncharged fermionic superfluids have low-lying bosonic excitations,
the Anderson-Bogolyubov modes, which we will address in more detail in
Sec.~\ref{sec:collective_modes}. These modes have a
linear-in-wave-vector spectrum with velocity given by
$c_{s}\simeq v_F/\sqrt{3}$, where $v_F$ is the Fermi velocity.  They
have a critical velocity equal to the mode velocity, which is greater
than $\Delta_{p_F}/p_F$.  Consequently, they do not negate the
argument given above that it is impossible to create excitations for
velocities less than $\Delta_{p_F}/p_F$.  (Here, as above, we assume
the weak-coupling BCS regime where the gap is much smaller than the
chemical potential.)  From the phenomenological standpoint, the
Anderson-Bogolyubov modes play a role analogous to that of phonons in
liquid $^4$He. These modes constitute the normal component of the
Landau-Tisza two-fluid model of liquid $^4$He, which coexists with the
superfluid component, \ie, the Bose condensate of $^4$He atoms.

Although it gives fundamental insights into the nature of pairing and
superfluidity in many-fermion systems, the simple pairing model
\eqref{eq:H_mf}-\eqref{eq:qpe} developed above is not suited for
quantitative microscopic description of these phenomena in the nuclear
systems that are the subject of this review, for reasons that will
become apparent.  It is nevertheless of interest to apply this model
to the case of spin-1/2 fermions interacting through a contact
interaction characterized by a free-space scattering length $a_0$,
specific examples being cold atomic gases and neutron matter in the
dilute gas limit $\vert a_0\vert p_F \ll 1$. For neutron matter, the
value of the scattering length, $ a_0 \simeq -19$~fm, implies
$p_F\ll 0.054$ fm$^{-1}$, which translates to a number density
$n\ll 10^{-5}n_0 $, where $n_0=0.16$ fm$^{-3}$ is nuclear saturation
density. Therefore, the range of applicability of this model in the
case of neutron matter is
limited to the asymptotically dilute regime.  To proceed, we first
recall that at finite temperature $T$, the equilibrium occupation 
numbers for fermion quasiparticles take the  Fermi-Dirac form 
$f(p) = (e^{E_p/T}+1)^{-1}$. The gap equation can then be written as
\bea \label{eq:gap_contact}
1 = t_{\rm sc} \nu(p_F) \int_0^{\infty}
\frac{d\ep_p}{2} \left(\frac{1-2f(E_p)}{\sqrt{\ep_p^2+\Delta^2}} -
\frac{1}{\ep_p}\right), 
\eea
where $t_{\rm sc}= 4\pi \vert a_{0} \vert/m$ is the magnitude 
of the two-body scattering matrix ($t$-matrix), and $\nu(p_F)= 
mp_F/\pi^2$ is the density of quasiparticle states summed over 
spins. In the zero-temperature limit, \ie, when $f(E)\to 0$, 
Eq.~\eqref{eq:gap_contact} can be solved for the 
gap, to obtain~\cite{Gor'kovMB1961} 
\be \label{eq:gap_contact_zeroT}
\Delta_0 = \tilde \epsilon  \exp\left(-2/\lambda_{c}\right),
\ee
where $\lambda_{c} = 4p_F|a_0|/\pi$ is the dimensionless contact 
pairing interaction and the prefactor $\tilde \epsilon = (8\ep_F/e^2) 
\beta_{\rm GM}$ is proportional to the Fermi energy $\ep_F$ and a 
factor $\beta_{\rm GM} = (4e)^{-1/3}$ that takes into account 
the in-medium modification of the interaction due to polarization.  

The result \eqref{eq:gap_contact_zeroT} is reminiscent of the BCS
weak-coupling formula for the gap in the phonon-mediated electronic
pairing model. It reveals a property of BCS pairing that is awkward
from the computational standpoint: the exponential sensitivity of the
energy gap to variations of the pairing interaction. Studies of
pairing in neutron matter within Gor'kov-Melik-Barkhudarov theory
\cite{Gor'kovMB1961} and its extensions, especially to finite-range
corrections, have been carried out
in~\cite{SchulzePolls2001,FanKrotscheck2017}. For extensions to
multicomponent systems and Fermi-Bose mixtures of cold gases, see
\cite{Heiselberg2000}.

In the asymptotic regime $T \to T_c$, where $T_c$ is the critical
temperature for destruction of pairing, the gap equation can be
linearized by setting $\Delta = 0$ in the denominator of
Eq.~\eqref{eq:gap_contact}. Straightforward integration leads to
$ T_c = ({\tilde\epsilon\gamma}/{\pi}) \exp\left(-\pi/2p_F\vert
  a_0\vert\right) ={\gamma}/{\pi}\Delta_0, $
where $\gamma \equiv e^C$ and $C \simeq 0.577$ is the Euler constant.
Keeping the next-to-leading order term in the $T \to T_c$ regime gives
$\Delta(T) =2\pi \sqrt{2/{7\zeta(3)}}\left[T_c(T_c-T)\right]^{1/2}$.
This implies that the critical exponent of the order parameter is
$1/2$, which is a well-known universal feature of the mean-field
theories, and that the gap closes with infinite slope. Note also that for
asymptotically low temperatures $T\to 0$ the temperature dependence
the gap is given by
$ \Delta(T) - \Delta_0 = - \sqrt{2\pi\Delta_0 T}\,\,\exp(-\Delta_0/T).
$

\subsection{Nucleon-nucleon pairing in different partial waves}
\label{sec:various_pw}

The complexity of the problem of pairing in nuclear systems stems
largely from the complexity of nuclear interactions.  In practice, the
assumed interactions divide roughly into those employed in density
functional studies and those designed for microscopic
computations. With the pairing interaction given directly by the bare
nucleon-nucleon (NN) potential, computation of the gap and other
superfluid properties of infinite nuclear systems (e.g.\ neutron
stars) has become routine {\it at the mean-field BCS level} in the
energy range where the interactions are well constrained by the
elastic nucleon-nucleon scattering data, \ie, for laboratory energies
$E_{\rm lab.}< 350$ MeV. For a one-component, isotropic and
homogeneous system interacting with an $S$-wave interaction
$v_0(p,p')$, where $p$ and $p'$ are the magnitudes of the relative
incoming and outgoing momenta of the particles, the pairing gap
$\Delta_{p}$ in the quasiparticle spectrum is given by the
mean-field BCS {\it gap equation}~(see
\eg~\cite{Abrikosov:Fundamentals,LifshitzPitaevskii})
\be \label{eq:gap_swave}
\Delta_{p} = -
\frac{1}{V}\sum_{\vecp'} v_0(p,p')
\frac{\Delta_p'}{2E_p'} (1-2f_{p'}),
\ee
where $V$ is the volume, the quasiparticle energy is given by
$ E_p = \sqrt{\varepsilon_p^2+\Delta_p^2}$, with $\varepsilon_p$ being
its counterpart in the unpaired state, and the equilibrium occupation
numbers for fermions at temperature $T$ are given by the Fermi-Dirac
form $f_{p} = (e^{E_p/T}+1)^{-1}$.

Such mean-field BCS calculations performed with NN-interaction models
that fit the scattering data with high precision, including the
Argonne $V_{18}$ \cite{1995PhRvC..51...38W}, Paris
\cite{1980PhRvC..21..861L}, Nijmegen \cite{1994PhRvC..49.2950S} and
Bonn \cite{2001PhRvC..63b4001M} potentials, converge to nearly
identical results for the pairing gaps in partial waves with $L\le 3$.
Note that for $E_{\rm lab.} >350$~MeV the elastic scattering
phase-shifts predicted by these NN-interactions deviate from each
other which results in deviations in the predictions for the
$^3P_2$--$^3F_2$-wave pairing gaps~\cite{BaldoElgaroy1998_Pwave}.  The
low-energy sector of the nuclear force is accurately described by
potentials that are based on chiral perturbation theory, in which the
interactions are modeled in terms of pion and nucleon fields and are
organized in powers of the ratio of a typical momentum scale of the
nuclear problem to a cutoff $\Lambda_{\rm QCD} \sim 1~{\rm GeV}/c$
provided by the chiral symmetry breaking scale
\cite{2011PhR...503....1M}.  At sufficiently high order (third or
fourth in the chiral expansion), the nuclear potentials constructed
using chiral effective field theory may have precision comparable to
that achieved with the high-precision NN phenomenological potential
models mentioned above.

At this point, it must be made clear what is considered mean-field BCS
theory in the context of actual, strongly interacting many-fermion
systems at meaningful densities, especially nuclear matter.  The
actual NN interaction exhibits very strong momentum
dependence. Specifically, NN interaction models designed to fit the NN
scattering data and deuteron properties contain a strong short-range
repulsion in competition with an outer attractive well, plus tensor
and spin-orbit components, along with crucial dependence on total spin
and isospin $S$ and ${\sf T}$. Consequently, the simple exponential behavior
characteristic of the pairing gap obtained from
Eq.~\eqref{eq:gap_swave} in the weak-coupling BCS theory (i) at
asymptotically low densities, (ii) for a contact interaction, and
(iii) in phonon-mediated electronic pairing, can be misleading when
conducting realistic microscopic studies of nuclear systems
\cite{KhodelKhodelClark1996}.  Even so, the strong sensitivity of
predictions of the pairing gap to inputs for the pairing interaction
and the density of states persists. The microscopic approaches to
pairing outlined in Sec.~\ref{sec:Methods} have the collective goal of
transcending the limitations of mean-field BCS theory in terms of
parquet-consistent \cite{Jackson1982,Jackson1985} irreducible
interactions and corresponding self-energies.

\begin{figure}[t] 
\begin{center}
\includegraphics[width=8.4cm]{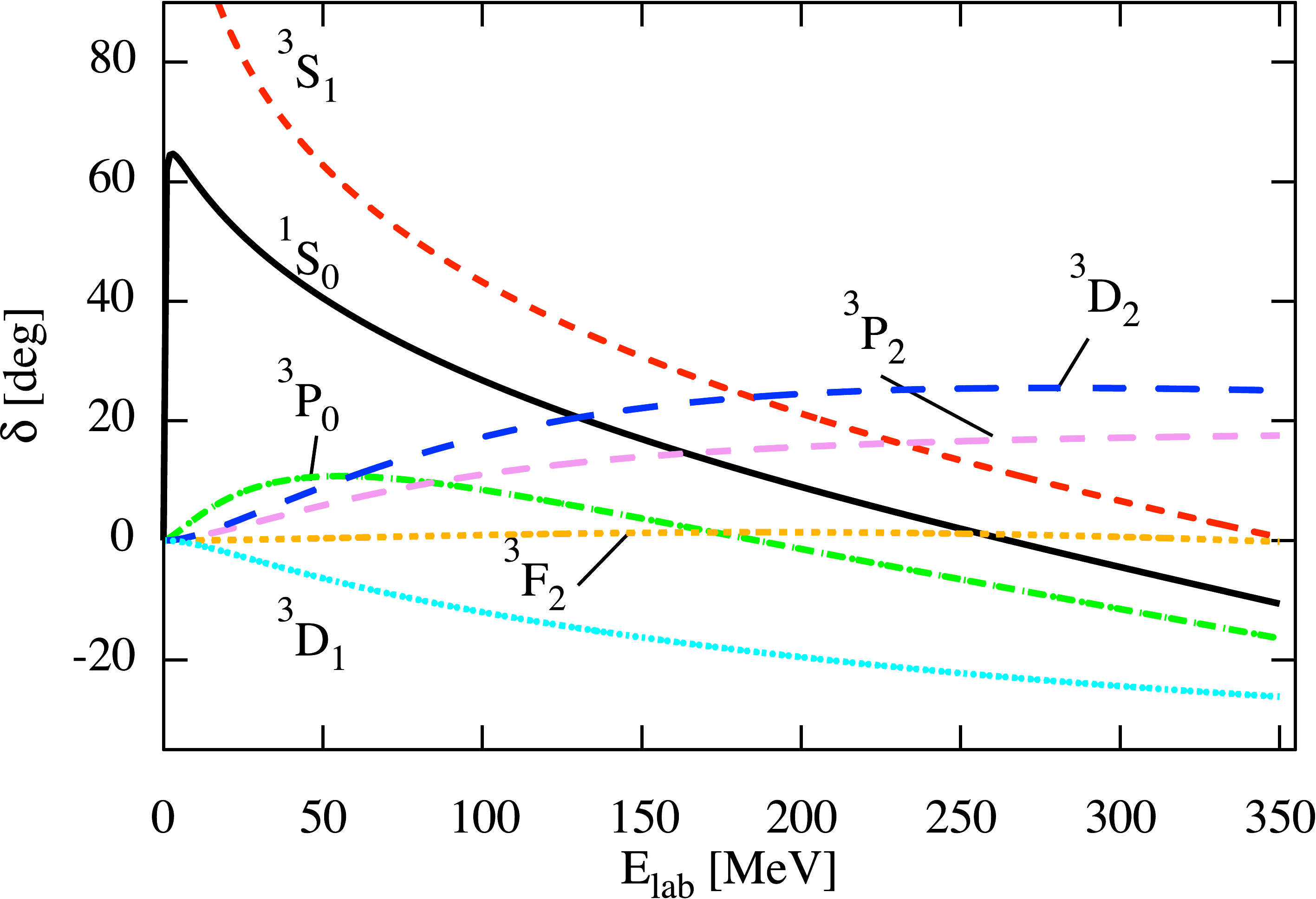}
\caption{Dependence of nucleon-nucleon scattering phase shifts on the
  laboratory energy of a two-nucleon system for the channels
  relevant to the pairing problem.}
\label{fig:phaseshifts} 
\end{center}
\end{figure}

The pairing patterns in nuclear matter and neutron-star matter can be
understood qualitatively on the basis of partial-wave analysis of 
NN scattering data.  Phase shifts derived from this
analysis for different partial-wave channels $^{2S+1}L_J$ of  
the two-nucleon scattering problem are identified using standard 
spectroscopic notation.  The relative orbital angular momentum quantum 
number $L = 0,1,2,\ldots$ is mapped successively to $S$, $P$, $D$, $F$, 
$G$, ..., while the total spin quantum number $S=0,1$ maps to singlet
and triplet spin states.  The allowed values of the total angular
moment quantum number, $J = 0,1,2 \ldots$, follow from the 
quantum-mechanical vector sum of the relative orbital and total
spin angular momentum operators.  

The experimental scattering phases in the range of laboratory energies 
$0< E_{\rm lab}\le 350$~MeV are shown in Fig.~\ref{fig:phaseshifts} 
for partial waves that are relevant to pairing in nuclear and 
neutron matter. As discussed in the next subsection, isospin ${\sf T}=1$ pairing
dominates in neutron-rich matter, whereas in symmetrical nuclear matter 
${\sf T}=0$ pairing competes with ${\sf T}=1$ pairing. Before assessing the 
roles of various pairing channels, we focus on the consequences of the
Pauli principle for the scattering of two nucleons, whose total wave 
function has spin and isospin components besides its spatial component.

To satisfy the Pauli principle, the total wave function, including
isospin, must be antisymmetrical under interchange of the two
nucleons.  The antisymmetry of the two-nucleon wave function implies
that the sum $L + S + {\sf T}$ must be odd.  At low energies, $L=0$ states
dominate.  Necessarily symmetrical in spatial dependence under
exchange, the allowed possibilities are the $^1S_0$ partial wave and
the $^3S_1$--$^3D_1$ coupled partial wave.  (Coupling in the latter
case reflects the presence of a tensor component in the nuclear
force, required to explain the quadrupole moment of the deuteron.)
With $L=0$, consider now the cases of neutron-neutron and
proton-proton scattering, trivially implying ${\sf T}=1$ and hence an
isospin-symmetric wave function.  To satisfy the Pauli principle, the
spin component must then be asymmetric under exchange, thus excluding
occupancy of the $S=1$ state of spin (in which case the sum $L+S+{\sf T}$
would be even) and thereby ruling out the triplet $^3S_1$--$^3D_1$ coupled
partial wave.

Consequently, the dominant attractive $^3S_1$--$^3D_1$ partial wave 
channel (see Fig.~\ref{fig:phaseshifts}) cannot lead to pairing in 
neutron-dominated matter, where neutrons and protons have Fermi surfaces 
of vastly different radii.  On the other hand, at the opposite extreme of 
symmetrical nuclear matter, these Fermi surfaces coincide, and one may 
expect strong ${\sf T}=0$ pairing to occur in this channel.  Moreover, since 
the deuteron is bound in this partial wave with energy $E_d = -2.2$ MeV, 
one may also expect a transition to a Bose-Einstein condensate of 
deuterons at asymptotically low density
\cite{BaldoLombardoSchuck1995,SedrakianClark2006PRC,SteinSedrakian2014}. 
(Note that higher-order clustering in low-density nuclear matter is 
expected; therefore a pure condensate of deuterons is an idealization.) 

Only ${\sf T}=1$ Cooper pairs can form 
\cite{LombardoSchulze2001LNP,2003RvMP...75..607D}  
at the large isospin asymmetries typically found in neutron stars,
where the neutron number density is around $95\%$ of the total
baryonic density below and at saturation density and gradually
decreases to $\sim 70\%$ at higher densities.\footnote{The nuclear
physics aspects of the composition of neutron star interiors is
discussed, for example, in \cite{ShapiroTeukolsky1983}, 
\cite{Glendenning_book}, and \cite{weber_book}.} At relatively 
low densities, ${\sf T}=1$ pairing is driven by the attraction in the 
$^1S_0$ partial-wave channel.  It is seen in Fig.~\ref{fig:phaseshifts} 
that the attractive $^3P_0$ channel remains sub-dominant to the 
$S$-wave channel in the low-energy regime below $E_{\rm lab} = 70$ MeV, 
where this $P$-wave competitor is overtaken by the $^3P_2$--$^3F_2$ 
coupled partial wave as the most attractive $L=1$ channel.  However, 
it is only at around $E_{\rm lab}=170$ MeV that the $^3P_2$--$^3F_2$ 
partial wave starts to dominate the ${\sf T}=1$ scattering, as the $^1S_0$-wave 
interaction loses its attractive component and eventually becomes 
repulsive (having negative phase shifts) for $E_{\rm lab} >250$ MeV.

Thus, the dominant ${\sf T}=1$ channel above $E_{\rm lab}=200$ MeV 
is the coupled $^3P_2$--$^3F_2$ partial-wave channel, for which the 
spatial wave function is antisymmetric, whereas the total spin $S=1$ 
and isospin ${\sf T}=1$ imply symmetrical components of the wave function 
in their respective spaces.  Accordingly, pairing in the triplet 
spin-1 channel is allowed by the Pauli principle for two neutrons or 
two protons. In contrast, if the nuclear system has equal populations
of neutrons and protons, $S=1$ and ${\sf T}=0$ pairs may be formed in the 
$^3D_2$ channel, which applies exclusively to neutron-proton scattering, 
being forbidden for like-isospin particles by the Pauli principle.
Note that the $^1P_1$ and $^3P_1$ partial waves, not shown in
Fig.~\ref{fig:phaseshifts}, are repulsive within the relevant energy
range and are therefore inconsequential for the pairing problem.

Up to this point, we have referred to specific features of the nuclear
interaction exhibited in two-nucleon scattering over ranges of
laboratory energy.  How does one translate this behavior into density
ranges in neutron stars?  This can be done semi-quantitatively by
observing that the center-of-mass energy of two scattering fermions,
given by $E_{\rm lab}/2$, should be roughly twice the Fermi energy of
the nuclear medium.  With applications to neutron stars in mind, we
may focus on the high-density, low-temperature regime of highly
degenerate nucleonic matter.  Neutron Fermi energies are roughly
$\epsilon_{Fn}\simeq 60$~MeV in neutron-star matter at the nuclear 
saturation density, $n_0=0.16$ fm$^{-3}$.  From this, we can already 
predict the result, borne out in microscopic many-body calculations, 
that neutron pairing in the $^1S_0$ partial wave will expire at depths 
slightly above the crust-core interface, where the density is about 
half $n_0$.  The low proton fraction in the neutron-star core,$^2$
$x_p\simeq 5$-$10\%$, and the correspondingly low proton Fermi energies, 
imply that proton pairing occurs in the $^1S_0$ state up to quite high 
densities. It is also conceivable that at neutron-star densities in 
excess of a few times the nuclear saturation density, pairing can occur 
in higher even-$L$ partial waves such as the $^1D_2$ channel (not shown 
in Fig.~\ref{fig:phaseshifts}). On the other hand, isospin-symmetric 
nuclear matter with $n_p = n_n$, where $n_n$ and $n_p$ are the number 
densities of neutrons and protons, may support pairing in the attractive 
$^3D_2$ partial wave, with a wave function which is symmetrical in space, 
antisymmetrical in isospace (${\sf T}=0$) and symmetrical in spin space ($S=1$).  
Some models of dense matter might support pairing in the $^3D_2$ partial 
wave~\cite{AlmSedrakian1996}.  Indeed, the abundance of protons can be 
equal (or even exceed) that of neutrons if $K^-$ condensation takes
place~\cite{GlendenningSchaffner1999,weber_book}. 

Should a neutron star feature a pion-condensed core, the ground
state of matter in that regime could be a superposition of neutron-proton 
quasiparticles filling a single Fermi sphere.  Such matter is 
conventionally described by a single type of ``nucleonic'' quasiparticle 
\cite{Campbell1975,Sawyer1977,BaymFlowers1974,BaymAnnRev1975}.

\subsection{Effects of isospin asymmetry and neutron stars}
\label{sec:isospin_asym}

Much of the research on nuclear pairing is concerned with neutron
stars, so it is important to review the state of matter in such
objects. The interiors of neutron stars are approximately in
equilibrium with respect to the weak interactions during their
lifetimes. Small deviations from such equilibrium may be important
in some problems, such as the bulk viscosity of matter, but for the
most part we will assume strict $\beta$-equilibrium.  The resulting
disparity between the neutron and proton numbers (breaking the $SU(2)$
symmetry in matter) has profound influence on the pairing patterns in
neutron stars.

\begin{figure}[tb]
\begin{center}
\includegraphics[width=8.5cm,height=6cm]{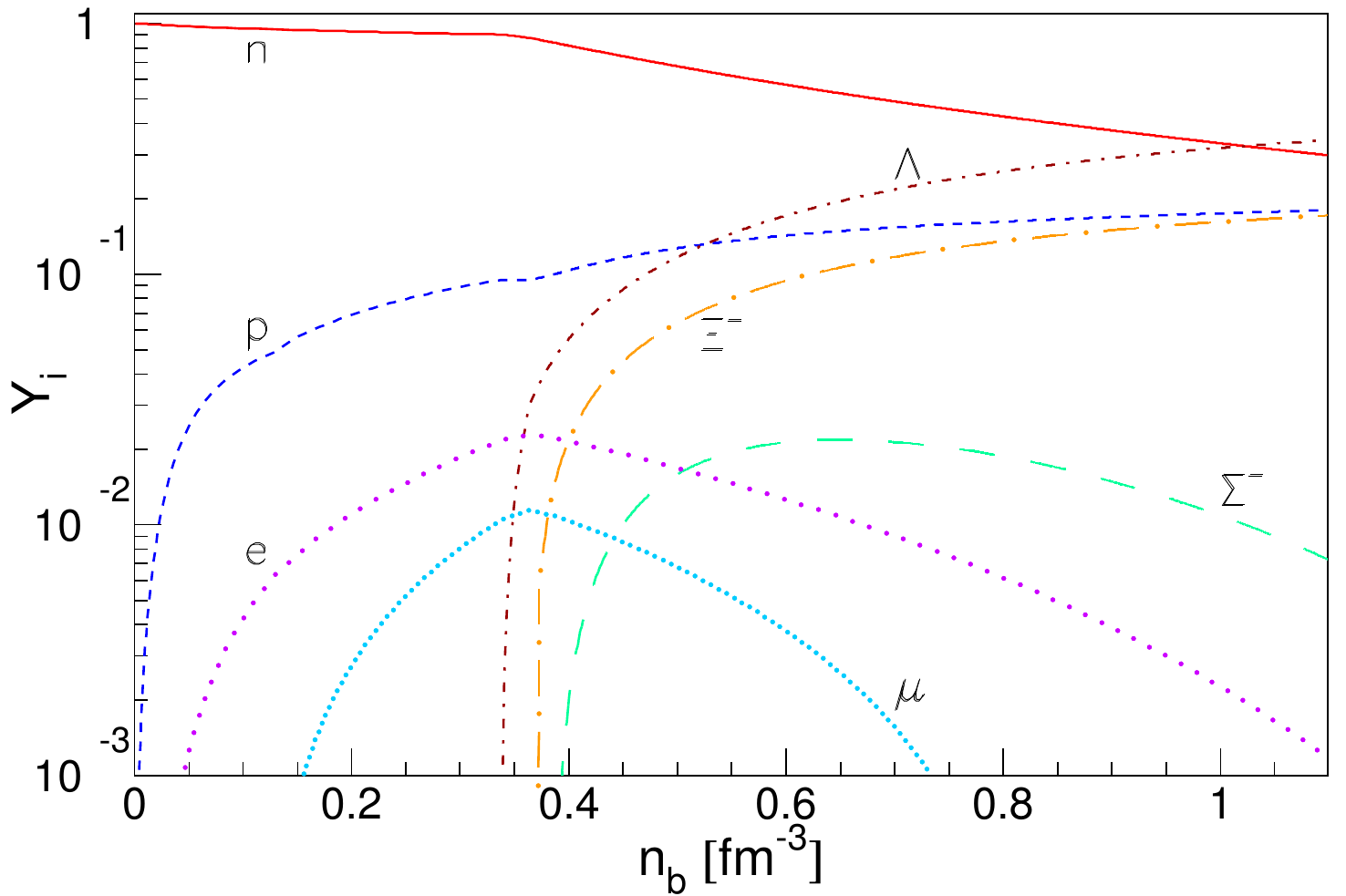}
\caption{ Dependence of the baryon and lepton fractions on the total
  baryonic density $n_b$ for the DDME2 covariant density functional.
}\label{fig:amba}
\end{center}
\end{figure}

In Fig.~\ref{fig:amba} we illustrate the abundances of various species
in a mixture of baryons and leptons in the interior of a neutron star
in the case of density-dependent covariant functional
theory~\cite{RadutaSedrakian2017}.\footnote{The basics of covariant
density functional theory for nuclear systems are discussed, for
example, in \cite{Glendenning_book,weber_book,SerotWalecka1997,Meng2016}.}
As already discussed above qualitatively, the neutrons and protons
forming the dominant component of matter at low densities are subject
to a large disparity in their densities, and hence in their chemical
potentials. Therefore, pairing with quantum numbers $S=1$ and ${\sf T}=0$,
specifically in the partial-wave channels $^3S_1$--$^3D_1$ and
$^3D_2$, is strongly suppressed. Thus, the two channels that provide
the largest attraction in symmetrical nuclear matter are ineffective
in neutron-star matter (see Fig.~\ref{fig:phaseshifts}).  In other
words, within the BCS approximation, pairing in neutron stars is
dominated by the $^1S_0$ and $^3P_2$--$^3F_2$ partial waves in the
${\sf T}=1$ channel at low and high densities, respectively. The disparity
in the neutron and proton densities also implies that the transition
from $S$-wave to $P$--$F$-wave pairing takes place at quite different
densities for the neutron and proton components.  For neutrons this
transition occurs at $n \simeq n_0$, whereas for protons the required
density is not reached in neutron star interiors in the majority (but
not all) models.  Notwithstanding the arguments above, it has been argued 
that in the low-density and low-isospin-asymmetry nuclear matter that 
may be created in low to intermediate heavy-ion collisions, supernova, 
and proto-neutron-star matter, $^3S_1$--$^3D_1$ pairing may persist if 
the pairing interaction does not differ strongly from that in free 
space~\cite{BaldoLombardoSchuck1995,SedrakianClark2006PRC,SteinSedrakian2014}. 
There is no direct evidence for such pairing in these
  systems, in particular, heavy-ion collisions, where the measured
  deuteron distributions are well described by simple statistical
  models. 
In analogy with nucleonic pairing, a hyperonic component of neutron-star 
interiors will develop BCS condensates when the mutual interaction of 
hyperons is attractive, as will be discussed in Sec.~\ref{subsec:H_pairing}.

\subsection{Finite nuclei}
\label{sec:finite_nuclei}

Although this review is concerned primarily with pairing in infinite
nuclear systems, it will be helpful to recapitulate the basic facts
about pairing in finite nuclei.  Validation of pairing theory in
direct terrestrial experiments on accessible nuclides (characterized
by neutron number $N$, proton number $Z$, and mass number $A=N+Z$)
provides a valuable source of constraints and methods potentially
relevant to the study of infinite nuclear matter.  For in-depth
expositions of the pairing in finite nuclei see
\cite{BrogliaZelevinsky,Ring80,2003RvMP...75..607D}.

At the most basic level, pairing correlations in finite nuclei 
express themselves in the odd-$A$-even-$A$ staggering of the measured 
binding energies of nuclei. The neutron ``pairing gaps'' in the 
cases of odd and even neutron numbers are commonly defined as 
\bea
\label{eq:delta_odd}
\Delta^{\rm odd}_{Z,N} = \frac{1}{2}\left( {\cal E}_{Z,N+1}+{\cal E}_{Z,N-1}\right) -
{\cal E}_{Z,N}, \, (\textrm{odd}\, N),\\
\label{eq:delta_even}
\Delta^{\rm even}_{Z,N} = -\frac{1}{2}\left({\cal E}_{Z,N+1}+{\cal E}_{Z,N-1}\right) +
{\cal E}_{Z,N},\, (\textrm{even}\, N),
\eea
where ${\cal E}_{Z,N}$ is the binding energy of a nucleus with proton
number $Z$ and neutron number $N$. The pairing gaps for changes of
proton number are defined by the same Eqs.~\eqref{eq:delta_odd} and
\eqref{eq:delta_even}, with the roles of $N$ and $Z$ interchanged.
Evaluation of these differences in the case of neutron-number
increments shows that the odd-$N$ nuclides are less bound than their
even-$N$ neighbors by about 1 MeV on average.  Gap values for each
fixed $N$ can fluctuate by a factor two.  Enhancement of the pairing
effect on binding is observed for nuclei having neutron magic numbers
$N=28,~50,~82,$ and 126.  Proton pairing shows the same energetic
systematics, with somewhat smaller values of the odd proton gap than
the odd neutron gap, presumably due to the Coulomb repulsion between
protons.  The pairing gaps decrease with the mass number of nuclei, a
behavior described phenomenologically by fits to gaps. A simple form
of such fit, applicable to both neutrons and protons, suggests
$\Delta \simeq 12 A^{-1/2}$ MeV. However, this functional form tends
to overestimate the magnitude of the gap in region of nuclei with
$N\le 30$ in case of neutron gaps and $Z\le 30$ in case of proton
gaps. A fit that accounts for even/odd differences \cite{Bertsch2013}
reads, in MeV units,
\bea &&\Delta^{n,\rm
  even/odd}_{Z,N}
= 12 A^{-1/2}+ a_{\rm even/odd},\quad \textrm{(neutrons)} \nonumber\\
&&\Delta^{p,\rm even/odd}_{Z,N} = (0.96 \pm 0.28)/(1.64 \pm 0.46),
\quad\textrm{(protons)} \nonumber 
\eea 
with $a_{\rm even} = \pm 0.28$ MeV and $a_{\rm odd} = \pm 0.25$ MeV.
This fit suggests that in a first approximation the proton pairing
gap, is $A$-independent.  An alternative fitting formula that provides
scaling intermediate between $\Delta\propto A^{-1/2}$ and
$\Delta\neq \Delta(A)$ is
$\Delta^{n,p} = \alpha_{\Delta} + \beta_{\Delta} A^{-1/3}$, with the
best fit values $\alpha_\Delta=0.3$ and
$\beta_{\Delta}=3.1$~\cite{HILAIRE200261}.  This dependence can be
justified by a straightforward expansion of the weak-coupling formula
for the gap, i.e., $\Delta \propto \exp(-1/G \nu)$, with respect to
the small parameter $\chi_{\Delta}$, where $G \propto A^{-1}$ denotes
the pairing matrix element and
$\nu \propto A(1 + \chi_{\Delta} A^{-1/3})$ is the level density at
the Fermi energy.

Differences in the excitation spectra of even-$N$ and odd-$N$ nuclei
provide another source of evidence for pairing correlations in nuclei.
For $N$ even, the excited states are separated from the ground state
by a gap that can be interpreted as the energy needed to break a pair
of neutrons, whereas for $N$ odd, the lowest of the discrete (but
dense) energy levels are found well within the range of 1 MeV
characteristic of gaps in nuclei. Additionally, it should be noted
that the excited states of nuclei may have collective nature that is
reminiscent of the phonon modes present in macroscopic
superfluids. Because of the finite nature of nuclei, these modes are
not necessarily bulk modes, \ie, they could be associated with the
lowest-order quadrupolar shape oscillations of the nucleus with
angular momentum and parity quantum numbers $J^{\pi} = 2^+$.

Theoretical studies of pairing properties of nuclei in the range of
intermediate and large mass number are generally performed within the
framework of density functional theory (DFT) either in
non-relativistic~\cite{Bulgac2002,Bender2003,GorielyChamel2013,Dobaczewski2013,Pei2014,Goriely2016,GorielyChamel2016,Bennaceur2017}
or relativistic formulations
~\cite{LongMeng2016,Li2015,LongRing2010,1991ZPhyA.339...23K}.  Such
approaches may be based purely on Hartree-Fock (HF) functionals,
nuclear properties (energy states and associated densities and
currents) being computed in the absence of pairing, with pairing
included in a final step within a simplified BCS approach.
Alternatively, nuclear pairing studies may utilize
Hartree-Fock-Bogolyubov (HFB) functionals, performing computations
that iterate the normal and anomalous states of the system in a manner
that allows for feedback of pairing correlations in the resulting mean
fields, guaranteeing a self-consistent solution.  The pairing
interactions are typically modeled as contact interactions. The two
parameters of this theory, namely the dimensionless pairing
interaction (or coupling) and the energetic range over which the
pairing is effective, are adjusted to the phenomenology of the nuclei
being considered. Note that the energy range of the pairing
  interaction becomes finite after  regularization of the integration
  in the gap equation, which otherwise is divergent for contact
  interactions~\cite{BulgacYu2002}.  Alternatively, these parameters
can be chosen to reproduce the results of pairing calculations in
infinite symmetric nuclear and neutron
matter~\cite{Garrido1999,Margueron2008,GorielyChamel2016,GorielyChamel2013}.

As we shall discuss in the later sections, there are two effects that
influence the results obtained with simple two-body contact
interactions. First, there could be substantial corrections to the
pairing interaction coming from polarization
effects.  Secondly, three-nucleon interactions are non-negligible in
nuclear systems, as they have been found to be important in
high-precision fits to the properties of light nuclei and to some
extent for the saturation of nuclear matter. (Apart from the generic
three-body forces originating at the level of quark substructure,
there are also ``effective'' three-body forces generated in diverse
theoretical treatments of two-body interactions that feature strong
short-range repulsive components.) In addition, the energetic scale
over which the contact-interaction is non-zero is expected to depend
on the occupancy of states in the vicinity of the Fermi surface.

As explained in Sec.~\ref{sec:various_pw}, at sub-saturation densities
the dominant attractive NN interaction is in the $^3S_1$--$^3D_1$
channel, \ie, the channel supporting a $np$ bound state in free space
-- the deuteron.  However, the foregoing discussion of pairing in
nuclei has involved only isospin-triplet ($nn$ or $pp$), spin-singlet
pairing. Noteworthy in this connection is the empirical fact that the
binding energies of nuclei on the $N=Z$ line are larger than those of
their neighbors by an amount known as the Wigner
energy~\cite{Satula1997}.  This could be interpreted as evidence for
pairing in the $^3S_1$--$^3D_1$ channel, which is otherwise suppressed
for $N\neq Z$ nuclei by the mismatch in the neutron and proton
energy-level occupancies.  Obviously, the pairing interaction may be
modified by the ambient medium differently in different isospin-spin
channels, one consequence being a less attractive force in the
$^3S_1$--$^3D_1$ than in the $^1S_0$ channel. Moreover, the spin-orbit
field of the nucleus may affect the spin coupling of nucleonic Cooper
pairs differentially, suppressing the $^3S_1$--$^3D_1$ neutron-proton
pairing more than $S$-wave pairing of like-isospin pairs. Neutron-proton 
pairing is expected also from HFB computations for
large nuclei~\cite{FriedmanBertsch2007}.

Besides its influence on static properties of nuclei, pairing and the
accompanying superfluidity are known to affect the dynamics of nuclei,
including rotation, shape oscillations, and fission. In contrast to
neutron stars (addressed intensively in Sec.~\ref{sec:astro}), where
the effects of nuclear superfluidity extend over macroscopic scales,
the characteristic scale of Cooper pairs, \ie, the coherence length,
is of the order of the size of the nucleus or somewhat larger.
Accordingly, one would expect the breakdown of superfluidity in nuclei
to have little or no effect on their global dynamics. Surprisingly,
self-consistent cranking HFB models, which reproduce the $2^+$
excitations of nuclei with good accuracy, require moments of inertia
which are half the rigid-body value~\cite{Delaroche2010}.

In its study, sub-barrier fission offers another tool to assess the
degree to which various nuclei are superfluid~\cite{Schunck2016,Bulgac2016}. 
Specifically, superfluidity enhances the probability of fission, as it
produces a larger overlap between different nearly degenerate
configurations.  Since quantum-mecha\-ni\-cal tunneling probability
depends exponentially on the energy difference between configurations,
one would expect a high sensitivity of the empirical results for the
fluid parameters of a given nucleus.  In particular, theoretical
interpretation of the fission of $^{234}$U and $^{240}$Pu requires
inclusion of an enhancement from superfluidity to account for the 
observed decay lifetimes~\cite{Bertsch2013,Sadhukhan2016,Bulgac2016}.

\subsection{Interface between nuclear systems and cold atomic gases}
\label{sec:interface}

The realization of BCS pairing in ultracold atoms in
2004-2006~\cite{JinRegal,Zwierlein2006} was a major development that
has considerably enlarged and diversified the scope of fermionic
pairing as exemplified in strongly correlated quantum many-body
systems. Indeed, prior to this discovery, the domain of application of
fermion pairing had been limited to specific examples considered to
arise in nature, specifically in nuclei, neutron-star matter,
color-superconducting quark matter, liquid $^3$He, and electrons in
solids. Quantum gases of fermionic atoms offer the freedom to
transcend nature by tuning the interaction between atoms via the
Feshbach resonance
mechanism~\cite{BlochDalibardZwerger2008,Leggett2012,Giorgini2008},
notably to the strongly interacting regime $p_F \vert a\vert \gg 1$,
where $p_F$ is the Fermi momentum and $a$ the scattering length of the
interaction.  In this regime the gas particles can no longer be
described as a weakly interacting gas.  Remarkably, the maximally
strong-coupling regime -- the unitary limit corresponding to
$p_F\vert a\vert \to \infty$ -- has become accessible for ultracold
fermions because three-body collisions are strongly suppressed in
these systems {\it precisely because} of the Pauli principle.  (The
opposite situation applies for the case of bosonic atoms, where the
lifetime of such cold-atom systems tends to zero due to three-body
collisions.)  This new possibility is of special significance for
nuclear physics, because pure neutron matter, having an anomalously
large scattering length $a_n \simeq - 19$ fm, is close to the unitary
regime at very low density.

Furthermore, by adjusting the magnitude of the magnetic field to tune
Feshbach resonances~\cite{ChinGrimm2010RvM}, it has become feasible to
drive a trapped cold atomic gas experimentally from the weakly
interacting BCS regime, where the gas consists of loosely bound Cooper
pairs, to the strongly interacting BEC regime of tightly bound
dimers. Thus, the theoretical ideas put forward several decades ago in
support of a hypothetical BCS-BEC
transition~\cite{Nozieres1985,Leggett2012,BlochDalibardZwerger2008,Strinati2018}
have been validated in experimental
realizations~\cite{REGAL2007,Zwierlein2006}.

The experimental prospects opened by techniques developed to
manipulate cold atoms also include the possibility of creating a
trapped atomic gas, for example composed of $^6$Li atoms, that has
unequal populations of two different hyperfine states -- thereby
simulating an interacting Fermi gas with unequal numbers of spin-up
and spin-down particles.  Such systems are expected to exhibit a rich
variety of unconventional pairing phases, such as the FFLO
phase~\cite{LO1964,FuldeFerrell1964} predicted in 1964, which features
Cooper pairs with non-zero center-of-mass momentum. Importantly, the
combination of these two features  -- the BCS-BEC
crossover and population imbalance -- will allow one to explore
regimes of strongly interacting paired fermionic matter that have
never been accessible in other systems, yet are of high interest for
the phenomenological understanding of pairing in asymmetric nuclear
matter and spin-polarized neutron matter (see
Sec.~\ref{sec:phase_diagram} and \ref{sec:spin_pol_NM}).

By placing fermions in an optical lattice of suitable design, one is
now able to simulate the effect of a periodic potential on the
properties of strongly correlated fermions subject to tunable
interactions~\cite{Bloch04,Ketterle06,Schreiber842,Zwerger2003}.  So
far, experimental studies of quantum many-body systems along such
lines has concentrated mainly on properties of Hubbard
models~\cite{Chiu2017} and the Mott
transition~\cite{Greiner2002,Esslinger2010}.  With dense-matter
astrophysics in mind, one potential application of this new ability is
a cold-atom laboratory model of the matter in the crust of a neutron
star (see Sec.~\ref{sec:astro}). Insight could be gained into the
interplay of the periodic potential and pairing in a strongly
interacting gas under freely adjustable conditions, including lattice
spacing, strength of interaction, various shapes of lattice potentials
that may induce non-spherical ``nuclei'' (pasta phases),
etc.\footnote{ Non-spherical nuclear pasta was initially studied in
  \cite{Ravenhall1983,Hashimoto1984,Lorenz1993,Pethick1995}. Recent
  advances in studies of these phases in neutron-star crusts are
  discussed in \cite{Schuetrumpf2015,Schneider2016,Fattoyev2017}.}

Another area of overlap between the nuclear superfluids in neutron
stars and those created in cold-atom traps involves the presence of
quantum vortices.  Experimental realization of quantum fermionic
vortices in trapped gases and their evolution through the BCS-BEC
crossover was initially instrumental in proving the very existence of
superfluidity in a Fermi gas of $^6$Li~\cite{Zwierlein2006}. However,
the range of phenomena that can be probed experimentally is vast.  For
example, it embraces studies of: (i) core quasiparticle excitations in
different interaction regimes and with respect to imbalance, (ii)
mutual friction in superfluid-normal mixtures of gases, (iii)
higher-spin vortices, and (iv) mixtures of fermionic superfluids and
Fermi-Bose fluids. In fact, vortices were realized recently in
mixtures of Fermi-Bose fluids~\cite{Yao2016}. In anticipation of the
aforementioned experimental studies, theoretical work has been carried
out on vortex-core quasiparticle excitations in different interaction
regimes and with respect to
imbalance~\cite{Iskin2008,Takahashi2007,Yu2003,Warringa2011,Warringa2012,BulgacYu2003}. Macroscopic
dynamics of rotating superfluids featuring vortex lattices has been
investigated in great detail both theoretically and
experimentally~\cite{2009RvMP...81..647F}. The corresponding studies
in ultracold bosonic gases have focused on vortex-lattice oscillations
(Tkachenko modes), quadrupolar modes of oscillations, rapid-rotation
induced Landau quantization of states, etc.; for reviews
see~\cite{Cooper2008,Tempere2017}. These experimental studies find
analogs in the physics of neutron stars, as will be explained in
Section~\ref{sec:astro}.

\section{Methods for strongly correlated systems}
\label{sec:Methods}
\subsection{Green Functions approach and Gor'kov formalism
} \label{sec:GF}
In this section we outline and discuss the Green functions
method~\footnote{Introductions to the method of Green functions 
can be found, for example,  in the texts \cite{Abrikosov:QFT},
\cite{1971qtmp.book.....F}, and \cite{Mahan}.} 
for
the treatment of superfluid systems.  The method was originally
introduced by Gor'kov and by Nambu~\cite{Nambu61,Gor'kov58}. Their
formulation is based on thermodynamic Green functions (GF) defined in
the imaginary-time formalism.  The starting point of this formalism is
the set of coupled Dyson-Schwinger equations for the normal and
anomalous GF which contain the self-energies of the system.
The self-energies allow for diagrammatic representation which provides
a systematic way to account for the correlations in the system in
terms of resummations of diagrams in the relevant dynamical
channels. A variant of the zero-temperature GF theory of pairing
appropriate for nuclear systems was developed in
\cite{LarkinMigdal1963}, on the basis of the Landau Fermi-liquid
theory for normal systems. Already in this early work a number of
important aspects of the fermionic pairing problem were introduced,
including wave-function renormalization and summations in the
particle-hole and particle-particle channels, with results expressed
in terms the phenomenological parameters of the Landau Fermi-liquid
theory.  This approach was further adapted to finite Fermi systems
(nuclei), and a number of nuclear observables were evaluated using the
Landau parameters for nuclear systems~\cite{migdal1967theory}.

In the following decades the GF method was largely abandoned in the
context of nuclear pairing.  It was revived in the early 1990s by a
number of research groups, specifically in the context of
$^3S_1$--$^3D_1$ pairing in isospin symmetric and asymmetric
systems~\cite{BaldoBombaciLombardo1992,BaldoLombardoSchuck1995,SedrakianAlmLombardo1997},
as well as for $^1S_0$ and $^3P_2$--$^3F_2$
pairing~\cite{Elgaroy1996_Swave,Elgaroy1996_Pwave,Elgaroy1996_betamatter,BaldoElgaroy1998_Pwave}.
These studies were already based on realistic (\ie, phase-shift
equivalent) NN interactions and included single-particle spectra
renormalized within Brueckner-type theories of nuclear matter.
Somewhat earlier, the real-time  GF treatment of
nuclear pairing was introduced in \cite{SuYangKup1987}, but the
interactions were treated at the level of the Skyrme effective contact
forces commonly used for computations on finite nuclei.  The
particle-particle and particle-hole resummations in the GF theory are
related to the microscopic determination of the Landau Fermi-liquid
parameters (see Sec.~\ref{sec:polarization} for details). This task
was taken up within GF theory at about the same time 
~\cite{AinsworthWambach1989,WambachAinsworthPines1993,SchulzeCugnonLejune1996}.

The class of theories of {\it unpaired} matter formulated in terms of
GF allows one to deduce only the critical temperature of the superfluid
phase transition, as signaled by poles that emerge in the
medium-modified scattering matrix of two
nucleons~\cite{Dickhoff1988,Schmidt1990,Alm1993,SedrakianRopkeAlm1995,AlmSedrakian1996,SteinSchenll1995}.
We relegate to Section~\ref{sec:Tmatrix} the discussion of theories in
which pairing is inferred indirectly from instability of the normal
state.

An important feature of the GF formulation is that it admits a description 
beyond the concept of quasiparticles inherent to the Landau Fermi-liquid 
theory by accounting for the finite width of particle states.  This may 
strongly affect pairing when it is addressed at the level of 
self-energies~\cite{Bozek2002,Bozek2003,BozekCzerski2002, Bozek2000,MuetherDickhoff2005,Sedrakian2003}.  We relegate the
discussion of these theories to subsection~\ref{sec:SCGF}.  Excellent
reviews of GF methods applied and results obtained up to the turn of
the century have been provided
in~\cite{LombardoSchulze2001LNP,2003RvMP...75..607D}.

The following two decades have seen wide application of GF theory to
superfluid nuclear systems.  One approach is to accurately incorporate
many-body corrections while enforcing consistency between various
ingredients, especially vertex corrections and renormalization of
single-particle energies, as has been done for $S$-wave
channels~\cite{CaoLombardo2006,ShenLombardoSchuck2005,SedrakianLombardo2000}.
Another line of development has employed soft effective interactions
to account for the resummations in the particle-hole channel in the
framework of Landau Fermi-liquid theory, specifically for $S$- and
$P$-wave channels~\cite{Schwenk2003,SchwenkFriman2004}. The effects of
phonons and retardation of the interaction on pairing have also been
explored based on effective
interactions~\cite{Sedrakian2003,Barranco2005,Pankratov2015}. More
recently, the following aspects of the problem of nucleonic pairing
have been brought into focus: (i) Incorporation of effects on the
pairing interaction and self-energies produced by three-body (3N) forces,
either of fundamental origin or generated by the many-body method used
to treat strong
correlations~\cite{DongLombardZuo2013,Papakonstantinou2017,Drischler2017},
(ii) calculation of pairing gaps based on a variety of soft, chiral NN
interactions~\cite{Finelli2015,Maurizio2014,SrinivasRamanan2016,Drischler2017},
which in part explore the influence of the cutoff of these interactions, 
and (iii) studies of the effects on pairing of short-range 
correlations~\cite{Ding2016,RiosPollsDickhoff2017}, as accounted for 
in terms of spectral functions (to be considered in Sec.~\ref{sec:SCGF}). 
The general trends that emerge from these studies will be discussed 
at a later stage (see Sec.~\ref{sec:S_wave_overview}).

We turn now a discussion of a Green-functions formulation of pairing
theory that is applicable to superfluid Fermi systems at finite
temperature for finite-range two-body
interactions~\cite{SedrakianClarkReview2006}.  This formulation, is an
extension to finite temperatures of the pioneering work of
Ref.~\cite{LarkinMigdal1963}.

\subsubsection{Green functions formalism}
\label{sec:GF_formalism}

The Gor'kov GF describing the superfluid state formally obey the
Dyson-Schwinger equations
\begin{eqnarray}\label{eq:dyson_6a}
 G_{\alpha\beta}(P) &=&  G^N_{\alpha\gamma}(P)
\left[\delta_{\gamma\beta}+\Delta_{\gamma\delta}(P)
\Fd_{\delta\beta}(P)\right],\\
\label{eq:dyson_6b}
\Fd_{\alpha\beta}(P) &=&  G^N_{\alpha\gamma}(-P) 
\Delta^{\dagger}_{\gamma\delta}(P)  G_{\delta\beta}(P),
\end{eqnarray}
where $P = (\omega, \vecp$) is the four-momentum, the Greek indices
$\alpha,\beta\dots$ label spin and isospin states, and the GF in the
normal state and given by
$G^N_{\alpha\beta}(P) = \delta_{\alpha\beta}[\omega
-\varepsilon(\vecp)]^{-1}$,
in effect, defines the single-particle energy
$\varepsilon(\vecp) = \epsilon_p + \Sigma(\vecp)$, where $\epsilon_p$
is the free single-particle spectrum. Note that the self-energy
$\Sigma(\vecp)$ is diagonal in spin and isospin spaces, given
spin-isospin conserving forces.
Eqs.~\eqref{eq:dyson_6a}-\eqref{eq:dyson_6b} have the solutions
\bea \label{eq:propagators_21}
G_{\alpha\beta}(\omega, \vecp) &=& \delta_{\alpha\beta} 
\frac{\omega-E_A(\vecp)+E_S(\vecp)}
{\left[\omega-E_A(\vecp)\right]^2-E_S(\vecp)^2-\Delta^2(\vecp)},\\  
 \label{eq:propagators_22}
\Fd_{\alpha\beta}(\omega, \vecp) &=& 
\frac{\Delta^{\dagger}_{\alpha\beta}(\vecp)}
{\left[\omega-E_A(\vecp)\right]^2-E_S(\vecp)^2-\Delta^2(\vecp)},
\eea
where 
$E_{S/A} = \left[\varepsilon(\vecp)\pm\varepsilon(-\vecp)\right]/2$
denotes the symmetric ($S$) and antisymmetric ($A$) parts of the
single-particle spectrum $\varepsilon(\vecp)$ in the normal state,
and the gap $\Delta(\vecp)$ satisfies
$\Delta(\vecp)\Delta^{\dagger}(\vecp) \equiv -\Delta^2(\vecp)$.
The Green functions $G_{\alpha\beta}$ and $\Fd_{\alpha\beta}$ in 
Eqs.~\eqref{eq:propagators_21} and \eqref{eq:propagators_22} 
share the same poles at
\be \label{eq:poles}
\omega_{\pm} = E_A(\vecp)\pm \sqrt{E_S(\vecp)^2 + \Delta^2(\vecp) },
\ee
thereby determining the excitation spectrum. If the normal self-energy
is invariant under reflections in space (\ie\ even under
$\vecp \to -\vecp)$ {\it and} time-reversal invariant (\ie\ even under
$\omega \to -\omega$), then component $E_{A}$ is zero.  Accordingly,
there is a non-zero energy cost $\sim 2\Delta$ for creating a fermionic
excitation from the ground state of the system. If by some physical
mechanism it occurs that $E_A \neq 0$, the superconductivity may be
{\it gapless} \cite{Abrikosov:Fundamentals} (for a recent discussion of
gapless superconductivity in the nuclear context see
\cite{SteinSedrakian2014,SteinHuang2012,Clark2016JP}.)

Superconductivity is
inherently a Fermi-surface phenomenon, so one natural approximation
entails an expansion of the self-energy $\Sigma(\omega,\vecp)$ of
the normal state around its on-shell value, assuming that the
off-mass-shell contribution is small. Since the imaginary part of this
self-energy vanishes quadratically on the mass shell, the expansion is
carried out for the real part by writing
\bea\label{eq:sigma_exp}
{\Real}\Sigma(\omega,\vecp) = {\Real}\Sigma(\varepsilon_p) +
\frac{\partial {\Real} \Sigma(\omega, \vecp)}{\partial\omega}
\Big\vert_{\omega = \varepsilon_p} (\omega-\varepsilon_p),
\eea
where $\varepsilon_p = \epsilon_p + {\Real}\,\Sigma (\varepsilon_p)$
is the on-mass-shell single-particle spectrum in the normal state.

Within this approximation, the self-energies contain only on-shell
self-energies $\Sigma(\ep_p, \vecp)$ and are multiplied by a
wave-function renormalization, \ie,
$ G_{\alpha\beta} \to {\cal Z} (\vecp)G_{\alpha\beta}$ and
$ F_{\alpha\beta} \to {\cal Z} (\vecp)F_{\alpha\beta}$, where
\bea\label{eq:renormalization}
{\cal Z}(\vecp)^{-1} 
\equiv 1-\frac{\partial {\Real}\Sigma(\omega,\vecp)}{\partial\omega}
\Big\vert_{\omega = \varepsilon_p}.
\eea
A similar expansion may be implemented for the anomalous self-energy,
\ie, the gap function $\Delta(\omega,\vecp)$.  It should be
noted, however, that for time-local pairing interactions (essentially
all bare or soft effective NN interactions) the gap function is
energy-independent. Non-local interactions are naturally generated
from local ones, if they are constructed via summations of series, as
in models of medium polarization (see Sec.~\ref{sec:polarization}.)

The existence of a Fermi surface
also implies an approximation of the momentum dependence
of the self-energy, although this approximation can be trivially
avoid\-ed. Expanding the normal self-energy at the Fermi surface one
finds
\bea \label{eq:fermi_li}
&&\varepsilon(p) = v_F(p-p_F) -
\mu^*,\nonumber\\
&& \frac{m}{m^*} = 1+\frac{m}{p}
\frac{\partial{\Real}\Sigma(\omega,\vert\vecp\vert)}{\partial p}\Bigg\vert_{p=p_F},
\eea
where $\mu^*\equiv -\epsilon(p_F)+\mu - {\Real} \Sigma(\ep_F,p_F)$,
$v_F$ is the Fermi velocity, and $m^*$ is an effective mass.  Here we
assumed that the system is homogeneous and isotropic. Therefore,
the self-energy depends only on the magnitude of the momentum, \ie,
the dispersion can be characterized by a single effective mass. In
more general situations, an effective mass tensor should be used.  The
spectrum \eqref{eq:fermi_li} now has the proper form for a Fermi
liquid, although there are no significant computational gains from
this effective-mass approximation.

\subsubsection{Mean-field BCS theory}
\label{BCS}

The next essential step is to establish the prescription for computing
the self-energies.  BCS theory is a mean-field theory for the
anomalous self-energy, which in its most general form can be written
as
\be
\label{eq:delta_mf}
\Delta(P) = -2\int\!\frac{d^4P'}{(2\pi)^4} \Gamma(P,P')
\,{\Img}F^{\dagger}(P')f(\omega'),
\ee
where $\Gamma(P,P')$ is a four-point interaction vertex function to be
determined from the nucleon-nucleon interaction, $P=(\omega, \vecp)$
is the four-momentum and $f(\omega) = [1+{\rm exp}(\beta\omega)]^{-1}$
is the Fermi distribution at inverse temperature $\beta$. 

Consider next time-local (but space non-local) interactions, in which
case the replacement $\Gamma (P,P')\to V(\vecp,\vecp')$ can be made
and, moreover, $V(\vecp,\vecp')$ can be expanded in partial waves.
Performing wave-function renormalization of the GF, integrating over
the energy variable considering a single uncoupled channel, and
recalling that $p=\vert \vecp\vert$, we arrive at the integral
equation
\bea\label{eq:gap_partial} 
\Delta(p)&=& Z(p)\int \frac{dp'
  \,p'^2}{(2\pi)^2} V(p,p')\nonumber\\
&&Z(p')
\frac{\Delta(p')}{\omega_+(p')}
\left\{f[\omega_+(p')]-f[\omega_-(p')]\right\}, 
\eea
$\omega_{\pm} (p)$ being given by
Eq.~\eqref{eq:poles} with $E_A = 0.$ In a number of cases, \eg in
low-density nuclear systems, it is necessary to solve for the density
\bea \label{eq:density} \rho &=& -2\sum_{\alpha}\int\!\! \frac{d^4 P}{(2\pi)^4}
{\Img}G(P) f(\omega)\nonumber\\ &=& \frac{1}{2}\sum_{\alpha}\int\!\! \frac{d^3
  p}{(2\pi)^3}Z(\vecp) \sum_{i = +, -} \left(1 +
  \frac{\varepsilon_p}{\omega_i}\right)f(\omega_i)
\eea
to obtain the chemical potential, which is modified by the effects of 
pairing on the single-particle energies. (Here $\alpha$ denotes
a sum over all spin/isospin states.)  This ``back-reaction'' of the 
density on the chemical potential is small in the weak-coupling regime, 
but becomes important with strong coupling.  For an input pairing 
interaction $V(p,p')$ and the spectrum $\varepsilon_p$ in the unpaired 
state, Eqs.~\eqref{eq:gap_partial} and \eqref{eq:density} fully determine 
the gap and the chemical potential, from which all the thermodynamic 
functions of the system can be computed.

In the foregoing development, we implicitly assumed that the normal
self-energy $\Sigma(P)$, and hence the normal-state spectrum
$\varepsilon(p)$, do not depend on the properties of the paired state,
e.g., the gap $\Delta(P)$.  The replacement of $G(P)$ by $G^N(P)$ when
computing the normal-state spectrum is an approximation, known as the
{\it decoupling approximation}, which is only valid when the pairing
is a small perturbation on the normal ground-state. This approximation
should work well for nuclear systems at high densities (implying weak 
coupling), but might not be adequate at lower densities where the 
strong-coupling corrections are significant. 

Qualitatively, the renormalization of the single-particle spectrum in
momentum space (accounted for, in particular, through the effective mass 
ratio $m^*/m$ for nucleons) acts to reduce the density of states, 
therefore the magnitude of the gap, by factors up to two or three, 
depending on density. Additional reduction comes from the wave-function 
renormalization ${\cal Z} (\vecp) \le 1$.

\subsubsection{Polarization effects}
\label{sec:polarization}

The interaction between nucleons is modified in the nuclear medium. 
Therefore the replacement $\Gamma(P,P')$ by the free-space 
interaction, which describes correctly only the asymptotic states 
of the nucleons, is an approximation that needs further elaboration.  
The leading class of modifications of the pairing interaction in the medium 
arises from ``polarization effects'' or ``screening.''  Let us
examine this type of modification.

We start with a simple but instructive approach based on ideas from
the Landau theory of Fermi liquids.  Consider the integral
equation~\cite{Nozierez,LarkinMigdal1963,BaymPethick}
\bea \label{eq:ph_ladders}
\Gamma (\vecp,\vecp',Q) 
&=& U(\vecp,\vecp',\vecq)
- i\int\!\frac{d^4P''}{(2\pi)^4}\,U(\vecp,\vecp'',\vecq)
\nonumber\\
&&\hspace{-1.5cm}
G^N(P''+Q/2)G^N(P''-Q/2)\Gamma(\vecp'',\vecp',Q), 
\eea
which sums the particle-hole diagrams to all orders, with
$Q = (\omega,\vecq)$ being the four-momentum transfer.  The driving term
$U(\vecp,\vecp',\vecq)$ must be devoid of blocks that contain
particle-particle ladders, to avoid double summation in the gap
equation. In general this driving interaction depends on spin and
isospin and can be decomposed as
\be \label{eq:U_block} U_{\vecq} = f_{\vecq} + g_{\vecq}
(\vecsigma \cdot \vecsigma') + \left[f'_{\vecq} + g'_{\vecq}
  (\vecsigma \cdot \vecsigma') \right](\vectau\cdot \vectau')\,, 
\ee
where $\vecsigma$ and $\vectau$ are the vector observables represented
by Pauli matrices in the spin and isospin spaces.
Equation~\eqref{eq:U_block} is written assuming the block $U_{\vecq}$
depends only on the three-momentum transfer.  This is a good approximation
for highly degenerate Fermi systems, where the remaining momentum
arguments of $U(\vecp,\vecp',\vecq)$ are restricted to the
Fermi surface and the angle formed by them can be expressed in terms
of the magitude of the momentum transfer (as seen below). For illustrative
purposes, the tensor component of the interaction and the spin-orbit
terms are ignored in Eq.~\eqref{eq:U_block}. The solution of
\eqref{eq:ph_ladders} is given by
\bea
\nu(p_F)\Gamma_{Q} &=& \frac{F_{\vecq}}{1+L(Q)F_{\vecq}} +
\frac{G_{\vecq}}{1+L(Q)G_{\vecq}}
(\vecsigma \cdot \vecsigma')\nonumber\\
&&\hspace{-1cm} \left[\frac{F'_{\vecq}}{1+L(Q)F'_{\vecq}} +
  \frac{G'_{\vecq}}{1+L(Q)G'_{\vecq}} (\vecsigma \cdot
  \vecsigma') \right](\vectau\cdot \vectau'), \nonumber\\ 
\eea 
where 
$F_{\vecq} = \nu(p_F)f_{\vecq}$, 
$G_{\vecq} = \nu(p_F)g_{\vecq}$,
$F'_{\vecq} = \nu(p_F)f'_{\vecq}$, 
and
$G'_{\vecq} = \nu(p_F)g'_{\vecq}$ are the dimensionless
particle-hole interactions (Landau parameters), $\nu(p_F)$ is the
density of states, and
\be 
L(Q) = \nu(p_F)^{-1}\int\!\frac{d^4P''}{(2\pi)^4}
\,G^N(P''+Q/2)G^N(p''-Q/2) 
\ee
is the polarization tensor, given in the present case by the Lindhard
function~\cite{FetterWalecka1971,BaymPethick}.  The momentum transfer is related
to the scattering angle $\theta$ and Fermi momentum $p_F$ according to
$q = 2p_F \sin\theta/2$, assuming the particle momenta are restricted
to the Fermi surface. The parameters $F$, $F'$, $G$, and $G'$ can be
expanded in Legendre polynomials with respect to the scattering angle
$\theta$, writing
\be \label{eq:expansion}
\left(\begin{array}{c}F(q)\\
        G(q)\end{array}
    \right) =\sum_l \left(\begin{array}{c}F_l\\
                            G_l\end{array}\right) P_l(\cos \theta)\,,
 \ee 
 and similarly for $F'(q)$ and $G'(q)$.  The Landau parameters $F_l$,
 $G_l$, $F_l'$, and $G_l'$ depend on the Fermi momentum. In neutron
 matter one has $\vectau \cdot \vectau' = 1$, and the number of
 independent Landau parameters for each $\vecq$ or $l$ can be reduced
 to two by defining $F^{n} = F + F'$ and $G^{n} = G + G'$.  Keeping
 the dominant lowest-order polynomials in the expansion
 (\ref{eq:expansion}), the interaction in a singlet pairing state
 (total spin of the pair $S = 0$ and $\vecsigma\cdot\vecsigma' = -3$)
 becomes
\bea \label{eq:pairing_int_Landau} \nu(p_F)\Gamma_Q &=&
 F_0^{n}\left[1- \frac{L(Q) F_0^{n}}{1+L(Q)F_0^{n}}\right]\nonumber\\
 &-&3G_0^{n}\left[1-\frac{L(Q)
     G_0^{n}}{1+L(Q)G_0^{n}}\right]\,.  
\eea 
In general, the polarization tensor $L(Q)$ is complex-valued.
However, it is real in the limit of zero energy transfer (at fixed 
momentum) and at zero temperature, being given (with 
$q \equiv \vert \vecq\vert$) by
\be\label{eq:Lindhard} L(q) = -1 + \frac{p_F}{q}\left(1 -
   \frac{q^2}{4p_F^2}\right) {\rm
   ln}\Bigg|\frac{2p_F-q}{2p_F+q} \Bigg|.
\ee 
The pairing interaction \eqref{eq:pairing_int_Landau} consists of two
pieces, namely the direct part $F_0^n - 3G_0^n$ generated by the 
terms $1$ inside the square brackets and the remaining induced 
part arising from density and spin-density fluctuations, respectively 
the terms $\propto (F_0^n)^2$ and $\propto (G_0^n)^2$.

Given the Landau parameters, the effect of polarization can be assessed
by defining a pairing interaction averaged over momentum transfers
and evaluated at zero energy transfer, \ie
\be\label{eq:gamma_induced}
 \Gamma(q,q') = \frac{1}{2qq'}\int_{\vert q-q'\vert}^{q+q'} dp\, p
\Gamma (p) .  
\ee 
Using the formalism outlined above, the impact of such fluctuations 
on pairing in neutron matter below nuclear saturation density $n_0$ 
has been established by \cite{ClarkKallman1976}, who showed that 
the density fluctuations enhance the attraction between neutrons, 
whereas the spin-density fluctuations suppress it. 
Using the available values of Landau parameters in neutron matter, 
they concluded that the suppression of pairing via spin-density 
fluctuations is the dominant effect.

We turn now to studies that employ more refined approximations for the
induced part of the
interaction~\cite{AinsworthWambach1989,WambachAinsworthPines1993,SchulzeCugnonLejune1996,Schwenk2003}.
First, while the structure of Eqs.~\eqref{eq:pairing_int_Landau} and
\eqref{eq:gamma_induced} remains the same, the replacement
\bea
F_0^n -3 G_0^n  \to V_s - 3V_a \equiv \Gamma_{\rm dir} (p) 
\eea
is made, with $V_s$ and $V_a$ set equal to the spin-symmetrical and 
anti-symmetrical parts of the bare (phase-shift equivalent) nuclear 
potential or its low-momentum reduction.  Then the induced interaction 
is determined from
\bea
\nu(p_F)\Gamma_{\rm ind} (p) = 
\frac{F(p)^2L(p)}{1+L(p)F(p)}-\frac{3G(p)^2L(p)}{1+L(p)G(p)},
\eea
where we have dropped the subscript $n$ on the particle-hole interactions
$F(q)$ and $G(q)$, which now depend on the {\it magnitude of the momentum
transfer $q$}.

The method used to compute the induced interaction was developed in
the 1970s and accounts for the mostly repulsive effect of screening on
the direct
interaction~\cite{BabuBrown1973,Backman1973,BackmanBrown1985,SchulzeCugnonLejune1996},
which by itself contains sufficient attraction to guarantee pairing.
In these approaches the driving term in the series summing up the
induced interaction is computed from the Brueckner-Bethe-Goldstone
theory of nuclear matter~\cite{Bethe1971} and is represented by the
${\cal G}$-matrix. For example, the spin-symmetric interaction $F(q)$
is determined through the coupled integral equations
 \bea F = {\cal G} -
{\cal A } G_{ph}F, \quad {\cal A} = F + F G_{ph} {\cal A} , 
\eea 
(written for simplicity in operator form), where ${\cal A}$ represents
the particle-hole scattering amplitude and $G_{ph}$ is the two-body
particle-hole GF;
see~\cite{AinsworthWambach1989,WambachAinsworthPines1993} for
details. The spin-antisymmetric channel is treated in complete
analogy.

Numerical computations of the $^1S_0$ gap in neutron matter that
include the induced interaction at various levels of sophistication
indicate that its dominant repulsive character produces a strong
reduction of the gap.  The resulting maximum of the gap is around
$1-2$~MeV; however, the density at which the maximum is attained
varies
substantially~\cite{AinsworthWambach1989,WambachAinsworthPines1993,SchulzeCugnonLejune1996,Schwenk2003,CaoLombardo2006}.

\subsubsection{Boson-exchange theories}
\label{sec:boson_exchange}

In reality, the pairing interaction is retarded in time, not only
because the mesons, as mediators of the nuclear force, propagate at
finite speed, but also because any induced interaction which
embodies resummation of a certain class of diagrams is frequency
dependent. Such induced pairing interactions can also be framed within
a theory of effective phonon exchange between nucleons, as is commonly
done in the theories of pairing in finite nuclei. Therefore it is of
interest to consider boson-exchange theories in general and leave
the nature of bosons arbitrary for the time being. 

Generic theories of pairing based on a boson-exchange model
originated in the work of \cite{Eliashberg1960} on
electron-phonon superconductivity in metals. The Dyson-Schwin\-ger
Eqs.~\eqref{eq:dyson_6a} and \eqref{eq:dyson_6b} remain intact in this
model.  However it is now convenient to split the retarded self-energy
into components even ($S$) and odd ($A$) in $\omega$, \ie\ $\Sigma(P) 
= \Sigma_S(P) + \Sigma_A(P)$, and define the wave-function 
re\-nor\-ma\-liza\-tion 
$\mathsf {Z}(P)= 1-\omega^{-1}\Sigma_A(P)$.  The single-particle energy is 
then renormalized as $E_S= \epsilon_p+\Sigma_S(E_S,\vecp )$.  Accordingly, 
the propagators now take the forms
\begin{figure}[t]
 \includegraphics[width=0.6\hsize,angle=90]{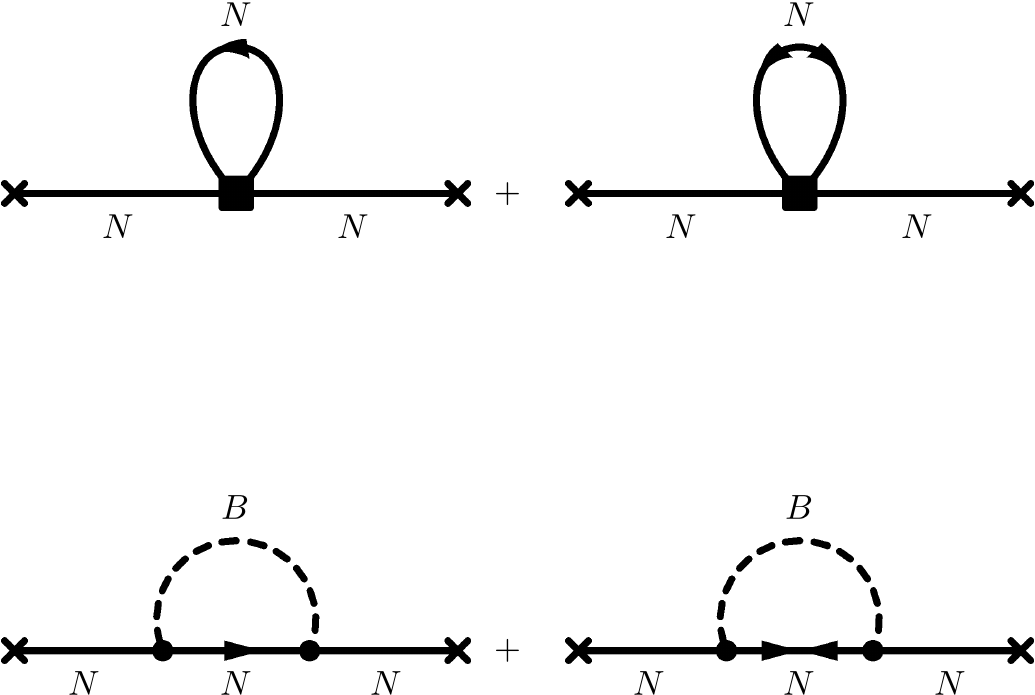}
 \caption{ Examples of Hartree (top) and Fock (lower panel)
   self-energies for normal (left) and anomalous (right) sectors.  The
   solid lines correspond to nucleons ($N$) and the dashed lines to
   bosonic mode ($B$). The lines with single and double arrows belong
   respectively to normal and anomalous propagators. The square vertex
   stands for the time-local part of the pairing interaction; crossed
   propagators do not belong to self-energies and are shown for
   clarity.\label{fig:boson_exchange} }
\end{figure}
\bea\label{PROP1}
G(P) &=& \frac{\omega \mathsf Z(P)+E_S(\vecp)}
{(\omega+i\eta)^2 \mathsf Z(P)^2-E_S(\vecp)^2-\Delta(P)^2}\,,\\
 F(P) &=&- \frac{\Delta(P)}
{(\omega+i\eta)^2 \mathsf Z(P)^2-E_S(\vecp)^{2}-\Delta(P)^2}\,,
\eea
where $\Delta\Delta^{\dagger}\equiv -\Delta^2$.  Next we need to
specify the pairing interaction. The time-local part of the interaction
appears in the Hartree self-energy (Fig.~\ref{fig:boson_exchange},
upper diagrams). The retarded boson-exchange interaction contributes
to the Fock self-energy (Fig.~\ref{fig:boson_exchange}, lower
diagrams).

We do not discuss the Hartree self-energies, as they can be 
readily calculated from any given nuclear interaction that is local 
in time (e.g., a phase-shift equivalent nuclear potential).  Using 
the fact that neutron matter is a highly degenerate Fermi system,
the normal and anomalous Fock self-energies can be expressed
in the following form~\cite{Sedrakian2003}:
\bea\label{eq:fock_normal} 
\Sigma(p_F,\omega) &=& -\int_0^{\infty}d
\omega' K_{\rm int}(\omega')
\Bigl\{g(\omega')\Big[G(\omega+\omega') \nonumber\\
&&\hspace{-1.5cm}+G(\omega-\omega')\Big]
+\int_{-\infty}^{\infty}\frac{d\ep}{\pi} {\Img~}[ G(\ep) ]J_E(\omega
, \omega' , \ep)
\Bigr\}, \\
\label{eq:fock_abnormal}
\Delta (p_F,\omega) &=& \int_0^{\infty}d \omega' K_{\rm int}(\omega')
\Bigl\{g(\omega')\Big[F(\omega+\omega') \nonumber\\
&&\hspace{-1.5cm}+F(\omega-\omega')\Big]
+\int_{-\infty}^{\infty}
\frac{d\ep}{\pi} {\Img~}[ F(\ep) ]J_E(\omega , \omega' , \ep)
\Bigr\},
\eea
where $\Sigma  (\omega,p_F)$ and $\Delta  (\omega, p_F)$ are 
respectively the normal and anomalous retarded self-energies,
while
\be\label{eq:kernel}
J_E(\ep, \omega , \omega') =
\frac{f(\ep)}{\ep-\omega-\omega'-i\eta}
+\frac{1-f(\ep)}{\ep-\omega+\omega'-i\eta} ,
\ee
where $g(\omega)$ and $f(\omega)$ and the Bose and Fermi distribution
functions.  Additionally, we have introduced a momentum-averaged
(real) interaction kernel defined by
\be\label{eq:kernel2} K_{\rm int}(\omega) =
\frac{m^*}{p_F}\int_0^{2p_F}\!\!\!\!  \frac{dq~ q}{(2\pi)^3}
\int_0^{2\pi}\!\!  d\phi~B(\vecq,\omega)~{\rm Tr}~
\{\Gamma^B_{0}(\vecq)\Gamma^B(\vecq)\} , \ee
in which $\Gamma^B_{0}$ and $\Gamma^B$ are the bare and full boson-fermion
vertices and $B(\omega, \vecq)$ is the spectral function of the
bosons.  Eqs.~\eqref{eq:fock_normal} and \eqref{eq:fock_abnormal}
provide a set of nonlinear coupled integral equations for the
complex pairing amplitude and the normal self-energy (or, equivalently
the wave-function renormalization). 
\begin{figure}[tb]
\includegraphics[width=0.98\hsize]{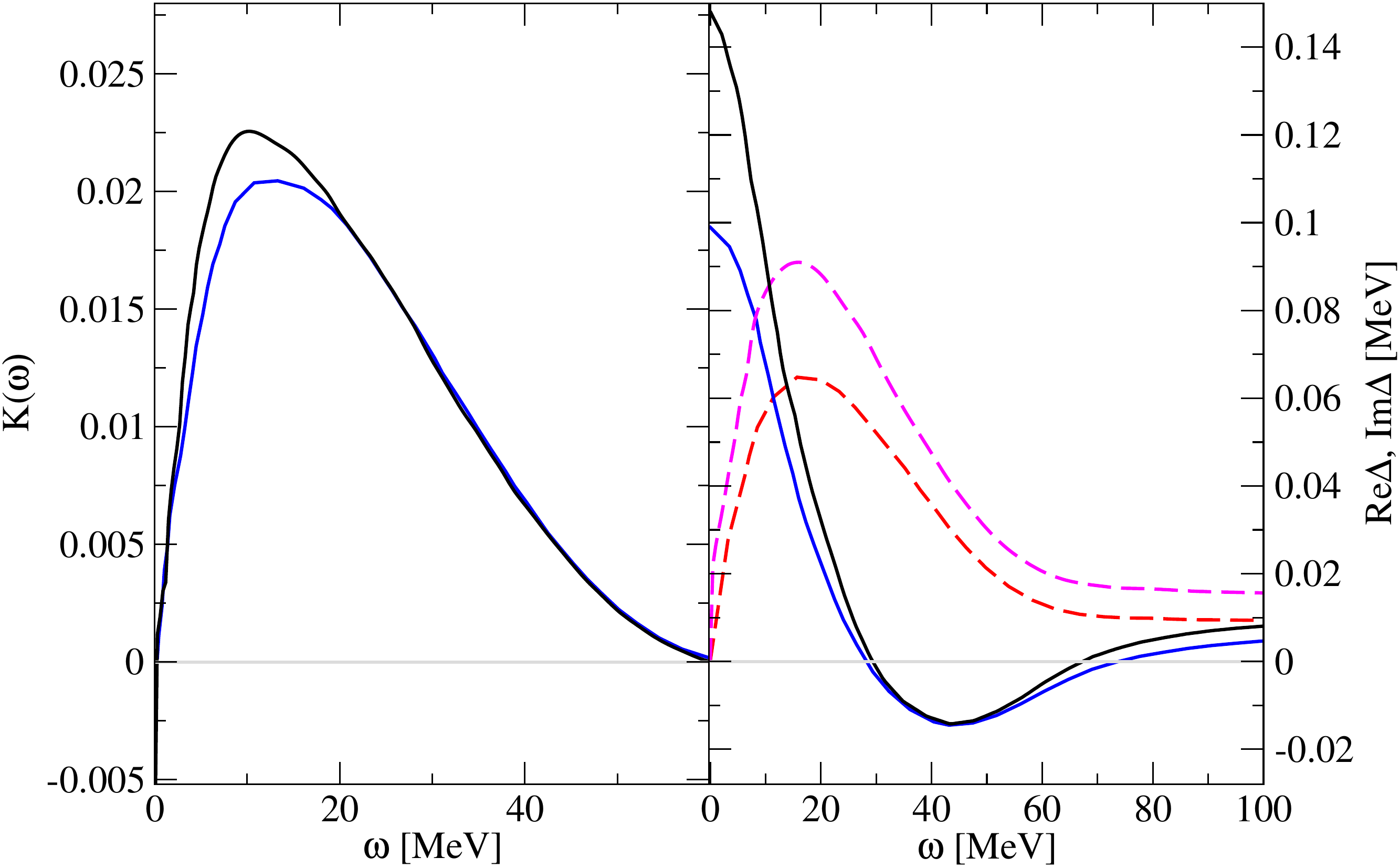}
\caption{Left panel: Frequency dependence of the effective interaction
  kernel $K(\omega)$, Eq.~\eqref{eq:kernel2}, for two different
  strengths of pairing interaction. Right panel: real (solid lines)
  and imaginary (dashed lines) components of the pairing gap in
  neutron matter for Fermi-momentum $p_F = 0.4$ fm$^{-1}$ for pairing
  interactions given in the left panel. In the on-shell limit
  $\omega\to 0$ one finds $\Img\Delta \to 0$ and
  $\Real\Delta \to \Delta_{0}$, where $\Delta_{0}$ is the on-shell value
  of the gap.
}
\label{fig:eliashberg}
\end{figure}
To illustrate some numerical solutions, consider a model in which
neutrons interact via soft-pion
exchange~\cite{Sedrakian2003,Pankratov2015}.  Given a spectral
function for the bosons, the kernel \eqref{eq:kernel2} is constructed
as input to Eqs.~\eqref{eq:fock_normal} and \eqref{eq:fock_abnormal}.
The input kernel for this specific model is shown in the left panel of
Fig.~\ref{fig:eliashberg}, while its right panel shows the
zero-temperature solutions of Eqs.~\eqref{eq:fock_normal} and
\eqref{eq:fock_abnormal}.  The imaginary component of the gap tends to
zero on the mass shell ($\omega = 0$); its real part gives the
on-shell value of the gap.  For non-zero energies these functions have
complex structure that reflects the features of the input kernel
$K_{\rm int}$.  Knowledge of the frequency dependence of the pairing gap
in nuclear and neutron matter could be important for the analysis of
frequency-dependent observables, especially for the description of
their dynamical response to various perturbations.

\subsection{T- and G-matrix approaches, Thouless criterion}
\label{sec:Tmatrix}

The onset of pairing correlations, and in particular the critical
temperature of the superfluid phase transition, can be determined from
properties of the normal (unpaired) state, notably from the {\it
scattering matrix}, defined here as an extension of the free-space
$\cal T$-matrix to a medium of strongly correlated fermions. As considered
in more detail below, generalization to the medium can be implemented
at different levels.  An important class of $\cal T$-matrix theories is
obtained when the propagation of particles and holes in intermediate
states is included symmetrically~\cite{Galitskii58}.  An alternative
extension, introduced historically in the context of nuclear matter
calculations, is based on the $K$-matrix (or ``reaction matrix'') -- 
the ${\cal G}$-matrix, in current notation -- where only
particle-particle propagation is taken into
account~\cite{BruecknerGammel1958}.  The relation between
superconductivity and singularities of the $\cal T$- and
${\cal G}$-matrices was recognized quite early in the development of
quantum many-body theory and considered in detail in
\cite{Emery1959,Emery1960}.  Singular behavior of the $\cal T$-matrix 
is directly related to the pairing properties of the system, in that
it can signal the onset of the superfluid phase.  In fact, the critical
temperature $T_c$ for the onset of pairing in attractive fermionic
systems, including nuclear systems, can be extracted as the temperature 
at which the $\cal T$-matrix of the normal state diverges as
$T_c$ is approached from above (\ie, from a higher temperature state).
This condition for the determination of the onset of superconductivity
is known as the Thouless criterion~\cite{Thouless1960}.

In vacuum, both these choices for the scattering matrix reduce
trivially to the $\cal T$-matrix of two nucleons interacting in free
space, which is fitted to the experimental elastic NN phase shifts for
laboratory energies below 350~MeV.  In the case of the ${\cal G}$-matrix,
the singularities are not directly related to the coherently paired
state, and it is still meaningful to perform calculations at
$T \le T_c$ without introducing a pairing gap in the fermion energy
spectrum~\cite{Bethe1971}.

With the advent of phase-shift equivalent, high-precision NN
potential models, $\cal T$-matrix theory was revived and employed to 
predict the critical temperature of the phase transition to the 
superfluid state in nuclear matter in the attractive interaction
channels~\cite{Schmidt1990,Alm1990ZPhys,Alm1993,SedrakianRopkeAlm1995,AlmSedrakian1996,SteinSchenll1995,Roepke1998,Rubtsova2017}. 
It is interesting that evidence of a di-neutron bound state has been
revealed in ${\cal G}$-matrix calculations that exhibit poles of this
quantity lying below the Fermi
energy~\cite{Isaule2016,Arellano2016}. The conditions for such
singular behavior are analogous to those for $\cal T$-matrix poles, the
difference being in the treatment of the intermediate states, as
we discuss now in some detail.  

The integral equation determining the $\cal T$-matrix can be written in
momentum space as
\bea \label{eq:tmat_1} {\mathcal  T}(\vecp,\vecp';K) &=& {V}(\vecp,\vecp') +
\int\!\!\frac{d\vecp''}{(2\pi)^3}
~V(\vecp,\vecp'')  \nonumber\\
&\times& G_2(\vecp'';K) \mathcal T(\vecp'',\vecp';K), 
\eea
where $V(\vecp,\vecp'')$ is the two-particle interaction and 
the two-particle GF  is given by 
\bea\label{eq:G2}
G_2(\vecp;K) &=&
\int\!\!\frac{d^4K'}{(2\pi)^4}\int\!\!\frac{d\omega}{(2\pi)}  
\Big[G^>(P_+)G^>(P_-) \nonumber\\
&-&  G^<(P_+)G^<(P_-)
 \Big] \frac{(2\pi)^3\delta(\vecK-\vecK')}{\Omega-\Omega'+ i\eta},
\eea
having introduced the four-vectors $P_{\pm} = {K}/{2}\pm P$ and
$P = (\vecp, \omega)$, with $K = (\vecK,\Omega)$ denoting the
center-of-mass four-momentum.  Equation~\eqref{eq:tmat_1} has the
familiar form of the Bethe-Salpeter integral equation appearing in
scattering theory. The GF $G^{>,<}(P)$ are the off-diagonal
GF in the non-equilibrium Keldysh-Schwinger 
formalism~\cite{Danielewicz1984,Botermans1990}.  In equilibrium
they can be written identically as 
\bea\label{eq:KB_ansatz}
-iG^{<}(P) &=& a(P)f(\omega), \\ 
iG^{>}(P) &=& a(P)\left[1-f(\omega)\right],
\eea
where $a(P)$ is the spectral function of fermions and $f(\omega)$ is the
equilibrium Fermi distribution function.  The spectral function of
quasiparticles (in the unpaired state) is given by
\be\label{eq:a_quasiparticles}
a(P) = 2\pi {\cal Z}(\vecp)\delta(\omega-\ep(\vecp)),
\quad
\ep(p) = v_F(p-p_F) - \mu^*.
\ee
Here the wave-function renormalization ${\cal Z}(\vecp)$ is defined in terms
of the normal-state self-energy by Eq.~\eqref{eq:renormalization}, while the
effective mass and chemical potential are as defined in
Eq.~\eqref{eq:fermi_li}. With these approximations, Eq.~\eqref{eq:G2}
reduces to
\bea\label{eq:G_quasiparticles}
G_2(\vecp;P) 
= \mathcal Z(\vecp_+)\mathcal Z(\vecp_-) 
\frac{\mathcal Q (\vecp_+,\vecp_-)}{\Omega-\ep(\vecp_+)-\ep(\vecp_-)+ i\eta},
\eea
where
\bea
\mathcal Q (\vecp_+,\vecp_-)  = \left[1 - f(\vecp_+)][1 - f(\vecp_-)\right]
- f(\vecp_+) f(\vecp_-)
\eea
is the Pauli-blocking function, which accounts for the phase-space
occupation in the intermediate scattering states of the
$\cal T$-matrix. The first and second terms of the latter expression 
refer to particle-particle and hole-hole propagations, respectively.   

In Brueckner-Bethe-Goldstone theory, a diagrammatic expansion 
of the normal ground-state energy is carried out in the number of hole 
lines, and hole-hole propagation terms are neglected,
\ie, one considers a ${\cal G}$-matrix equation
\bea \label{eq:gmat_1} {\cal G}(\vecp,\vecp';P) 
&=& {V}(\vecp,\vecp') +
\int\!\!\frac{d\vecp''}{(2\pi)^3}
~V(\vecp,\vecp'')   \nonumber\\
&\times&\frac{\tilde {\mathcal Q}(\vecp_+,\vecp_-)}{\Omega-\ep(\vecp_+)-\ep(\vecp_-)+ i\eta} {\cal G}(\vecp'',\vecp';P), \nonumber\\
\eea
with $\tilde {\mathcal Q}(\vecp_+,\vecp_-)  = \left[1 - f(\vecp_+)][1 -
  f(\vecp_-)\right]$.

Returning to the $\cal T$-matrix equation \eqref{eq:tmat_1}, we consider
the poles that this equation might develop as the temperature is
reduced from a temperature $T > T_c$.  This can be illustrated
analytically by assuming a rank-one separable interaction
$V(\vecp , \vecp') = \lambda_0 v(\vecp)v(\vecp')$.  The solution of
Eq.~\eqref{eq:tmat_1} is then given by
\bea\label{eq:tmat_separable}
\mathcal T(\vecp, \vecp', P) &=& \frac{V(\vecp, \vecp')}{1-J(P)},\\
J(P) &=& \lambda_0 \int\!\!\frac{d\vecp}{(2\pi)^3} v^2(\vecp)
G_2(\vecp, P).  
\eea 
The $\cal T$-matrix depends parametrically on the chemical potential and
temperature of the matter through the two-particle propagator
$G_2(\vecp, P)$.  At the critical temperature of a phase transition to
the superfluid state, the $\cal T$-matrix develops a pole for the
energy-momentum arguments $\Omega_c = 2\mu^*$ and $\vert \vecP_c\vert= 0$,
 which is equivalent to the conditions
\bea\label{eq:J_re_im}
{\Real} ~J(\vert \vecP_c\vert) = 1 \quad {\rm and}  \quad {\Img}
~J(\vert \vecP_c\vert) = 0.
\eea

One may conclude that for the given interaction, the critical temperature
of the superfluid phase transition can be determined as the
temperature $T_c$ at which the $\cal T$-matrix is
divergent~\cite{Thouless1960}.  Note that $\cal T$-matrix poles may also
appear for $\vecP\neq 0$, indicating an onset of the superfluid
phase in which pairs carry non-zero total momentum.
\begin{figure}[tb] 
 \includegraphics[width=1.\hsize]{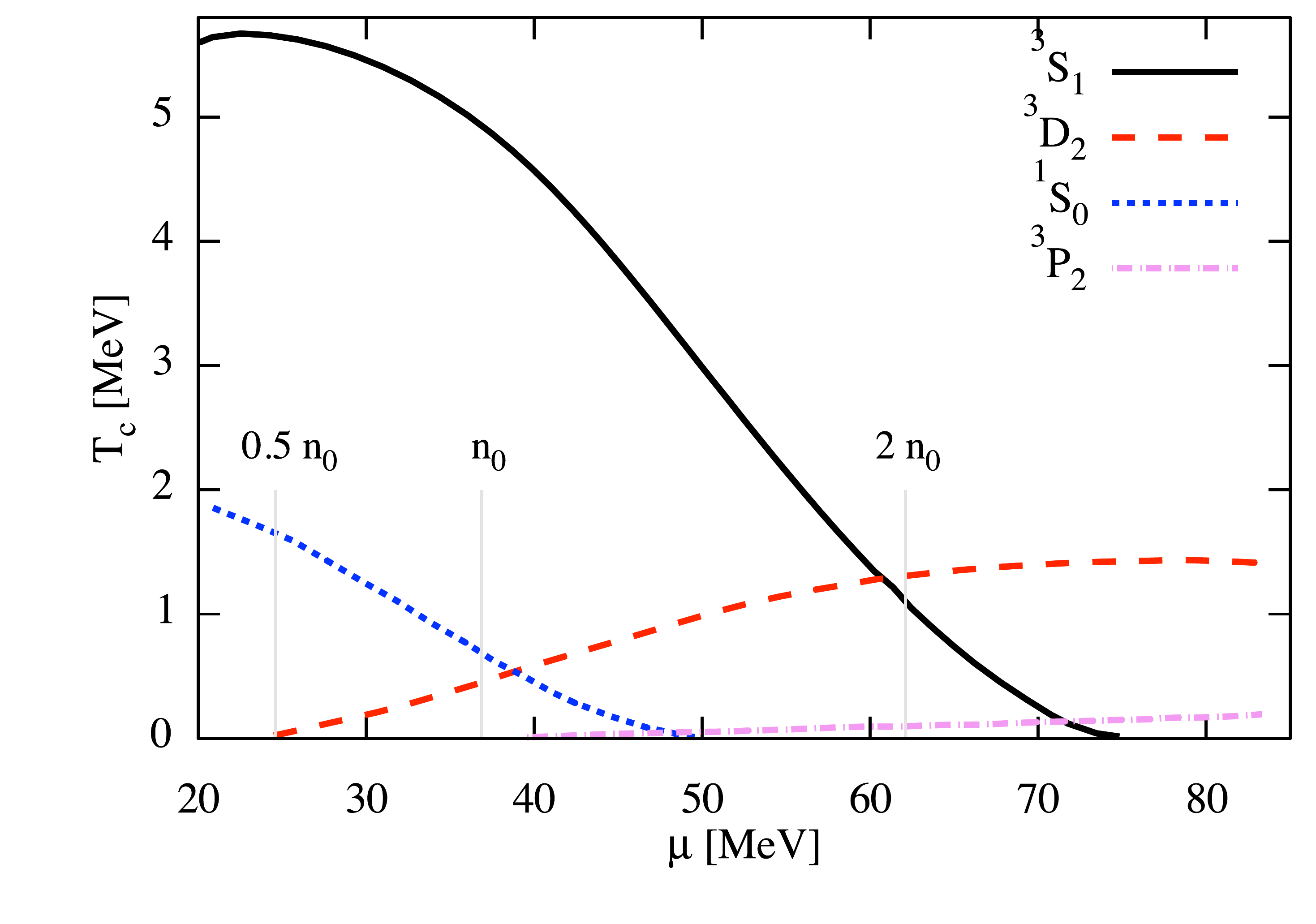}
 \caption{Dependence of the critical temperature $T_c$ of superfluid
   phase transitions on the chemical potential in symmetrical nuclear
   matter in attractive NN channels, as determined from the $\cal T$-matrix
   instability.  Vertical lines indicate
   densities in units of the nuclear saturation density $n_0$. }
\label{fig:Tmatrix_tc}
\end{figure}

Figure~\ref{fig:Tmatrix_tc} shows the critical temperatures of the
dominant channels of pairing in nuclear matter as a function of
chemical potential obtained from the singularity of the
  $\cal T$-matrix~\cite{SedrakianRopkeAlm1995}. Among isospin-singlet
(${\sf T}=0$) states, the highest critical temperatures are obtained in the
$^3S_1$--$^3D_1$ and $^3D_2$ partial-wave channels at low and high
densities respectively, for isospin-symmetric nuclear matter. In
neutron-rich matter, these channels are suppressed by the strong
isospin asymmetry, such that the ${\sf T}=1$ channels $^1S_0$ and
$^3P_2$--$^3F_2$ become dominant, at low and high densities
respectively.

In case of the ${\cal G}$-matrix, the absence of hole-hole propagation
in the intermediate states breaks particle-hole symmetry.
Consequently, the instability of the $\cal T$-matrix that signals the onset
of the superfluid state is suppressed and the ${\cal G}$-matrix can be
computed at temperatures below $T_c$, down to $T=0$.
Brueckner-Bethe-Goldstone theory utilizes the ${\cal G}$-matrix as an
effective interaction in generating the perturbative hole-line
expansion.  That is, the diagrams in the expansion for the energy are
ordered according to the number of hole lines present, each hole line
implying a convergence factor given roughly by the ratio of the volume, 
per particle, excluded by the repulsive component of the NN interaction 
to the mean volume per particle, known as the wound parameter.
On one hand, this has the apparent virtue of wiping away the
instability associated with pairing; on the other, the resulting
theory is non-conserving in that it entails self-energies and
scattering amplitudes that are asymmetric with respect to interchange
of particles and holes.  In fact, any collision integral constructed from scattering
amplitudes (or non-equilibrium self-energies in the language of the
Keldysh-Schwinger formalism) must vanish in the equilibrium limit. 
This condition fails to be met if particle-hole symmetry is broken.  
One consequence of such broken symmetry in theories of nuclear matter 
based on the ${\cal G}$-matrix, where only particle-particle propagation 
is taken into account~\cite{BruecknerGammel1958}, is violation of 
the Hugenholz-van Hove theorem \cite{HugenholtzvanHove1958}, which 
requires coincidence of the chemical potential and the Fermi energy 
in the presence of arbitrarily strong interactions.  Even so, 
as long as the hole-line expansion is valid, such violation ought 
to be small.

\subsection{Self-consistent Green functions theory}
\label{sec:SCGF}

\begin{figure}[t] 
\begin{center}
 \includegraphics[width=0.7\hsize]{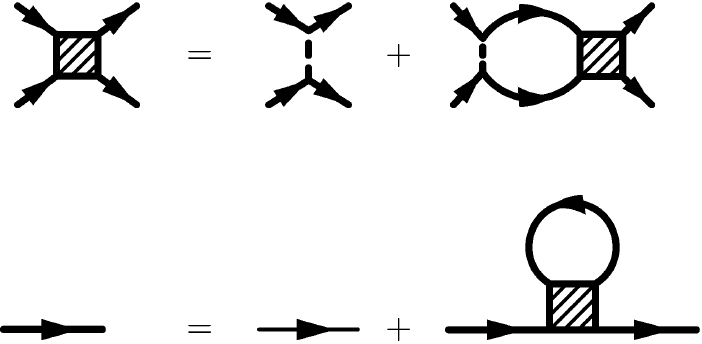}
\end{center}
 \caption{Diagrammatic representation of the coupled equations for the
   $\cal T$-matrix (top relation) and Green function of fermions (bottom
   relation) in the self-consistent Green function (SCGF) theory. The
   vertical dashed line stands for the bare interaction, while the
   thick and thin lines correspond respectively to the full and bare
   propagators.}
\label{fig:SCGF}
\end{figure}

The foundations of self-consistent Green functions (SCGF) theory 
were established long ago (see especially \cite{kadanoff1962quantum}).  
It can be applied to nuclear matter at finite temperatures above 
the critical temperature for 
pairing~\cite{Botermans1990,Danielewicz1984,Dickhoff_book}. 
SCGF theory is a microscopic approach to properties of the normal 
(unpaired) state in which the interactions between nucleons are 
accounted for via the two-body $\cal T$-matrix constructed from 
the bare NN interaction. The single-particle spectrum is obtained 
from the self-energy computed in the $\cal T$-matrix approximation.  
Equations~\eqref{eq:tmat_1}-\eqref{eq:KB_ansatz} determining
the $\cal T$-matrix remain intact, but the spectral function 
is now completely general, \ie
\bea \label{eq:spectral_func}
a(p) &=&i\left[ G^{R}(p) - G^{A}(p)\right] =  i\left[ G^{>}(p) -
  G^{<}(p)\right] \nonumber\\
&=& 
-\frac{2\Img \Sigma (p)}{[\omega - \epsilon(p) - \Real\Sigma(p)]^2 
+ [\Img \Sigma(p)]^2},
\eea
where $G^{R/A}(p)$ are the retarded and advanced GF and $\Sigma(p)$ is
the self-energy. Consequently, in SCGF theory the two-particle GF is
given by
\bea\label{eq:propagator2}
G_2(\vecp;P) = \int\!\!\frac{d\Omega' d\omega}{(2\pi)^2} a(p_+)a(p_-) 
\frac{Q(p_+,p_-)}{\Omega-\Omega'+ i\eta},
\eea
with $Q(p_+,p_-) = 1 - f(p_+) - f(p_-)$. The self-energy in the
$\cal T$-matrix approximation is expressed as
\bea \label{eq:sigmaT}
\Sigma(p) &=& \int\frac{d^4p'}{(2\pi)^4} \Biggl[
T(\vecq,\vecq;p+p') a(p') f(\omega')
\nonumber\\
&&\hspace{-1.5cm} +2g(\omega+\omega')\Img
T(\vecq,\vecq;p+p')
\int \frac{d\bar\omega}{2\pi}
\frac{a(\vecp',\bar\omega)}{\omega'-\bar\omega}
\Biggr],
\eea
where $\vecq \equiv (\vecp-\vecp')/{2}$ and $g(\omega)$ is the Bose
distribution function.  Equations~\eqref{eq:tmat_1},
\eqref{eq:propagator2}, and \eqref{eq:sigmaT} form a closed system of
coupled integral equations requiring as input the interaction between
the nucleons, see Fig.~\ref{fig:SCGF}. These equations can be solved
numerically by iteration for phase-shift equivalent two-body
potentials~\cite{BozekCzerski2002,Frick2003,Frick2004,Rios2006,Rios2009}
and two-body plus three-body potentials~\cite{Carbone2013}.

To obtain a closed set of equations for investigation of pairing, 
it is necessary to specify an approximation to the anomalous self-energy. 
In the mean-field (BCS) approximation, one has 
\bea \label{GAP}
\Delta^{\dagger}(p) &=&  
i\int \frac{d^4p'}{(2\pi)^4} ~V(\vecp,\vecp') \Fd (p') 
\nonumber\\
&=&  i\int \frac{d^4p'}{(2\pi)^4} 
V(\vecp,\vecp') G^N(-p')\Delta^{\dagger}(p') G(p') ,
\nonumber\\
\eea
where, in the second step, the anomalous GF has been
replaced by an equivalent expression in terms of the GFs
$G^N$ and $G$.
Thus, in the on-shell limit we have
\bea \label{eq:gap_gg}
\Delta^{\dagger}(\vecp) 
&=&  i\int \frac{d^3p'}{(2\pi)^3}
V(\vecp,\vecp') \Delta^{\dagger}(\vecp') G^S_2(\vecp'),
\eea
the two-particle GF  in the superfluid state being given by 
\bea\label{eq:GS}
G^S_2(\vecp)  = \int \frac{d\omega}{2\pi}G^N(-\omega,\vecp) 
G(\omega, \vecp).
\eea
The BCS gap equation \eqref{eq:gap_partial} is recovered if the
two-particle GF $G^S_2(\vecp)$ is evaluated in the quasiparticle
approximation. Going beyond the quasiparticle approximation within the
SCGF theory entails replacement of the two-particle GF in
the superfluid state $G^S_2(\vecp)$ by its counterpart in the normal
state~\eqref{eq:propagator2}.  The main advantage of the SCGF approach
is that the gap equation is solved while keeping the off-mass-shell
information contained in the full spectral function of the normal
state.

However, we have seen that the $\cal T$-matrix from which the spectral
function is computed in SCGF theory is divergent below $T_c$. 
Accordingly, the spectral function apparently needs to be computed at
temperatures {\it above} $T_c$. This problem is dealt with by
extrapolating the imaginary part of the normal-state self-energies to
temperatures $T\le T_c$ using the fact that $\Img\Sigma(\omega)$ must
vanish on the Fermi surface at $T=0$. The real part of the self-energy
is then computed from the Kramers-Kronig dispersion relation and,
in this way, the complete spectral function is constructed 
below $T_c$ \cite{Frick2003,Frick2004,Rios2006,Rios2009}.

Numerical calculations demonstrate that upon going beyond the
approximation that employs on-shell quasiparticles with a renormalized
spectrum by adopting the GF given by Eq.~\eqref{eq:propagator2}, the
pairing gap is suppressed by about ten percent in the isospin-singlet
$^3S_1$--$^3D_1$ state, as well as in the isospin-triplet $^1S_0$ and
$^3P_2$--$^3F_2$ channels. These results can be attributed to the shift
of some spectral weight from the quasiparticle peak toward other
energies, upon implementing full spectral
functions~\cite{Ding2016,RiosPollsDickhoff2017}. The corresponding 
numerical results are discussed below in
Sec.~\ref{sec:S_wave_overview} for the $S$-wave case and 
in Sec.~\ref{sec:higher_partial_waves} for the $P$--$F$-wave channels. 

\subsection{Correlated Basis Functions Theory}
\label{sec:Var_CBF}

The method of correlated basis functions (CBF) provides a powerful tool 
for studying strongly correlated fermionic or bosonic quantum 
systems~\cite{ClarkWesthaus1966,Feenberg1969,Clark1979,Clark1979PPNP,KrotscheckClark1980,Krotscheck1981,Krotscheck2002book,Fantoni1981,FantoniRosati1975,FantoniRosati1978}.  
It was applied to nucleonic pairing at the early stages of theoretical 
development of this generic quantum many-body theory~\cite{Chao1972proton,YangClark1971,ClarkYang1970,Clark_err1971}. 
Since that time, the CBF method has undergone extensive further developments,
with applications in diverse physical contexts, in particular to nuclear 
systems~\cite{ChenClark1986,ChenClark1993,Fabrocini2008,Fabrocini2005,
FanKrotscheck2017, PavlouClark2017,BenharDeRosi2017} and the low-density
fermionic gas~\cite{Fan2015}.

An important feature of CBF theory is that it implements a strategy  
for building essential normal-state correlations into the description
of a strongly interacting Fermi system through the 
action of a correlation operator $F$.  Pairing correlations are then
superimposed on the correlated normal ground state, in full analogy to
the original approach of BCS theory.  In particular, CBF theory
is designed for inclusion of the strong short-range  correlations
produced by repulsive cores in nuclear systems, and the effects
of induced long-range interactions can be treated on the same
footing.  The discussion below will focus on pairing in nuclear
matter at zero temperature.

Consider a complete set of correlated normal states defined 
for each particle number $N$ 
\be
\label{eq:NormCBF}
\vert \Psi_{m}^{(N)} \rangle = \frac{F_N \vert \Phi_{m}^{(N)}
\rangle}{\langle \Phi_{m}^{(N)} \vert F_N^\dagger F_N \vert 
\Phi_{m}^{(N)} \rangle^{1/2}},
\ee
where the $\vert \Phi_{m}^{(N)}\rangle$ represent eigenstates of the
noninteracting Fermi gas, $F_N$ is a correlation operator and
$m^{(N)} = \{m_1\dots m_N\}$ specifies the set of plane-wave orbitals
entering $|\Phi_m^{(N)} \rangle$. The states $\vert \Psi_{m}^{(N)} \rangle$ 
are normalized to unity, but generally not orthogonal.  The correlation 
operator $F_N$ is commonly taken to be of Jastrow-Feenberg form, 
depending only on radial distances between pairs of particles, 
\be
\label{JFeen1}
F_N({\bf r}_1,\ldots,{\bf r}_N) = 
\exp\left[U_N({\bf r}_1,\ldots,{\bf r}_N)/2\right], 
\ee
with
\bea
\label{eq:JFeen2}
U_N &=& \sum_{i<j}  u_2(r_{ij}) + \sum_{i<j<k} u_3(r_{ij},r_{jk},r_{ki}) +
\cdots \nonumber \\
&& + \sum_{i_1<\ldots<i_N}u_N(r_{i_1i_2},\ldots,r_{i_{N-1}i_N)}. 
\eea 
This series is usually truncated at the two-body or three-body level.
The familiar Jastrow two-body correlation function is
$f(r_{ij}) = \exp[u_2(r_{ij})/2]$, with limiting behavior
$\lim_{r\to 0} f(r)\to 0$ and $\lim_{r\to \infty} f(r) \to 1$.
Dependence on spin and isospin, \ie, state dependence, may also be
incorporated, as in
 \be
\label{eq:opexpans}
F_N = {\cal S} \{\Pi_{i<j}f(ij)\},\qquad f(ij) = 
\sum_\alpha f_\alpha(r_{ij})O_\alpha(ij),
\ee
where $\cal S$ is the symmetrization operator and
the index $\alpha$ runs over the set of two-body operators 
$O_\alpha(ij)$ entering the NN interaction adopted (or a subset 
of them), these being formed with appropriate symmetries from spin, 
iso\-spin, tensor, and spin-orbit operators.

The next step is to construct a correlated superfluid ground state
residing in Fock space, which allows for consistent derivation of a
gap equation in the presence of both pairing correlations that
introduce off-diagonal long-range order and conventional correlations
(of short or long range) that preserve $U(1)$ symmetry.  One begins with
the $|{\rm BCS \rangle}$ ground state 
\be
\label{eq:BCSket}
\vert {\rm BCS} \rangle =
\prod_\veck \left[ u_\veck +  v_\veck a_{\veck \uparrow}^\dagger
 a_{-\veck \downarrow}^\dagger \right] \vert 0 \rangle 
\ee
expressed in terms of Bogolyubov amplitudes
\be\label{eq:u_and_v} 
u_{\veck}^2 = \frac{1}{2}
\left(1+\frac{\ep_{\veck}}{E_{\veck}}\right),
\qquad 
v_{\veck}^2 = \frac{1}{2} \left(1-\frac{\ep_{\veck}}{E_{\veck}}\right), 
\ee
where $\ep_{\veck}$ and $E_{\veck}$ are respectively the single-particle 
spectra in the normal and superconducting states.   A robust choice
for the correlated superfluid trial ground state has proven to be
\be
\label{eq:CBCSdef}
\vert {\rm CBCS}\rangle =  \sum_{{m},N} \vert {\Psi_{m}^{(N)}} \rangle 
\langle \Phi_{m}^{(N)} \vert {\rm BCS} \rangle, 
\ee
formed as a superposition of the correlated normal states defined
by Eq.~\eqref{eq:NormCBF}.  This trial ground state superposes the 
correlated basis states $\vert \Psi_m^{(N)}\rangle$ with the same
amplitudes that the model normal states $\vert \Phi_m^{(N)} \rangle$ 
have in the corresponding expansion of the original BCS state 
vector.\footnote{An alternative CBF formalism for the description
of fermionic pairing \cite{Fantoni1981,Fabrocini2008} replaces the normalized 
CBF basis state $| \Psi_{m}^{(N)} \rangle$ in Eq.~\eqref{eq:CBCSdef} by 
$F_N|\Phi_m^{(N)}\rangle$.  As applied, this approach and the one outlined 
here have complementary strengths and
 weaknesses~\cite{Clark2013,Fan2015,FanKrotscheck2017}.} 

Given the {\it Ansatz} \eqref{eq:CBCSdef} for the correlated superfluid 
ground state and a Hamiltonian operator $\hat H$ in Fock space containing 
a two-body interaction $v(ij)$, the thermodynamic potential 
of the pair-correlated system can be evaluated with the 
result \cite{Krotscheck1981,Fan2015}
\bea
\label{eq:HmuN}
\langle  {\hat H} - \mu {\hat N}  \rangle
&=& H_{00}-\mu N+ 2\!\!\sum_{\veck,\,|\veck|>k_F}v_\veck^2\ep_\veck\nonumber\\
&&\hspace{-1cm}- 2\!\!\!\!\sum_{\veck,\,|\veck|<k_F}  u_\veck^2 \ep_\veck 
+  \sum_{\veck,\veck'}  V_{\veck\veck'} u_\veck v_\veck u_{\veck'} v_{\veck'} ,
\eea
where $\hat N$ is the number operator with expectation value $N$
and $H_{00}= \langle \Psi_{0}| \hat H | \Psi_{0}\rangle$ is the
expectation value of the Hamiltonian in the normal $N$-particle ground 
state as described by $| \Psi_{0}\rangle$, and $V_{\veck\veck'}$ is the
in-medium effective pairing interaction.  This effective pairing
interaction has the structure
\bea
\label{eq:CBFpi}
V_{\veck\veck'} &=&  W_{\veck\veck'}+(\vert \ep_{\veck}\vert
+ \vert \ep_{\veck'}\vert )N_{\veck\veck'},  \\
\label{eq:Wdef}
W_{\veck\veck'} &=& \langle \veck \uparrow ,-\veck\downarrow |
\hat W(1,2) | \veck'\uparrow ,-\veck'\downarrow \rangle_a, \\
\label{eq:Ndef}
N_{\veck\veck'} &=& \langle \veck \uparrow ,-\veck\downarrow |
\hat N(1,2) | \veck'\uparrow, - \veck'\downarrow \rangle_a ,
\eea
where the index $a$ implies antisymmetrization.  The two-body operators 
$W(1,2)$ and $N(1,2)$, along with the single-particle energies 
$\ep_{\veck}$, are to be determined from matrix elements 
$H_{mn}=\langle \Psi_m | {\hat H} | \Psi_n \rangle $ and 
$I_{mn} = \langle \Psi_m | \Psi_n \rangle $ of the
Hamiltonian and identity through their natural decompositions
\bea
\label{decomp}
I_{mn} 
&\equiv& \delta_{mn} +  N_{mn},
\\
H'_{mn} &\equiv & W_{ mn} + \frac{1}{2}\left({H'}_{mm}+{H'}_{nn }\right)
N_{ mn},
\eea
where $H' = H - H_{00}$.  

On the assumption that the energy gap is small compared to the Fermi
energy, such that the feedback of pairing on normal-state properties
can be neglected, it is justified to consider one Cooper pair at a
time in analyzing the correlated BCS state \eqref{eq:CBCSdef}.  Upon
imposing this decoupling approximation, the Bogolyubov amplitudes no
longer appear in the gap equation derived by functional minimization
of Eq.~\eqref{eq:HmuN}. This CBF gap equation then becomes identical 
in form to the standard mean-field BCS gap equation, but with the bare 
pairing interaction $v$ replaced by the effective pairing interaction $V$
defined in Eq.~\eqref{eq:CBFpi}, and the single-particle energies 
$\epsilon_\veck$ given by those of the correlated normal ground state.
 
Evaluation of these normal-state inputs to the CBF gap equation, 
predicated on optimal determination of the correlation factor $F_N$,
has involved significant formal and computational developments.
At the level of Jastrow correlations (\ie, having truncated the
series \eqref{eq:JFeen2} for the operator $U_N$ at the two-body term
$n=2$) the obligatory Euler-Lagrange (EL) optimization requires that
the function $u_2(r)$ satisfies
\be
\label{eq:Euler}
\frac{\delta H_{00} }{\delta u_2 }(r)= 0.
\ee
Associated with the resulting energy minimum of the correlated normal
trial ground state $| \Psi_0^{(N)} \rangle$ are a radial distribution 
function $g(r)$ and its Fourier partner, the static structure function 
$S(k)$.

To proceed further and solve Eq.~\eqref{eq:Euler}, a reliable method 
is needed for evaluation of the diagonal and off-diagonal matrix 
elements in the normal-state correlated basis \eqref{eq:NormCBF}.
Initially, cluster-expansion techniques were introduced to calculate
matrix elements in a basis of correlated states of the Jastrow-Feenberg 
type, primarily for the ground-state energy, one-body density matrix, 
and pair distribution functions, but also for perturbative extensions.
In the simple Jastrow case, these are expansions in the number of correlation 
bonds $\eta(r) = f^2(r)-1$, {\it or} the number of correlated bodies.
They are effectively low-density expansions, loosely analogous to 
the wound-parameter or hole-line expansions of Brueckner-Bethe-Goldstone
theory, their terms being given a diagrammatic representation analogous 
to those for imperfect classical gases \cite{Mayer:StatPhys}.  Later, methods 
were developed, originally for the radial distribution 
function $g(r)$, which permitted simultaneous 
resummation of certain important classes of cluster diagrams, in 
particular of nodal ($N$) and non-nodal ($X$) connectivity, and otherwise 
identified by the direct or exchange involvement of their root points 1,2 
[specifically direct-direct (dd), direct-exchange (de), exchange-exchange 
(ee), or cyclic exchange (cc)]; see \cite{Clark1979PPNP,Krotscheck2002book} 
for details.  Application of these resummation techniques to other 
observables culminated in Fermi-hypernetted chain (FHNC) theory 
\cite{FantoniRosati1975,KrotscheckRistig1975} for the analysis of 
the Jastrow-Feenberg correlated normal ground state, subsequently being 
extended to evaluation of off-diagonal as well as diagonal Hamiltonian matrix 
elements \cite{KrotscheckClark1979}.

In combination with EL optimization, the simplest nontrivial
implementation of FHNC resummation that is consistent in the sense of
parquet analysis \cite{Jackson1982,Jackson1985}, named EL-FHNC//0,
incorporates both the random-phase approximation and the
Bethe-Goldstone equation (thus rings and ladders) in a ``collective''
or averaged-GF approximation \cite{Krotscheck2002book,Fan2015}.  
The latter involves treating particle-particle and hole-hole propagation 
in the same average way. Adopting the EL-FHNC//0 approximation, 
the Euler equation \eqref{eq:Euler} takes the form
\be
\label{eq:kEuler}
S(k)\left[1+2{\frac{S_F^2(k)}{t(k)}}{\tilde V}_{ph}(k)\right]^{1/2} 
= S_F(k),
\ee
where $t(k)=k^2/2m$, and $S(k)$ and $S_F(k)$ are respectively 
the static structure functions of the interacting and noninteracting 
systems.  The effective interaction ${\tilde V}_{ph}(k)$ has the
Fourier partner
\bea
V_{ph}(r) &=&  \left[1 + \Gamma_{\rm dd}(r) \right] v(r) + 
 \frac{1}{m}\left|\nabla\sqrt{1+\Gamma_{\rm dd}(r)}\right|^2 
\nonumber \\ &+&   \Gamma_{\rm{dd}}(r)w_{\rm I}(r),
\label{eq:ph},
\eea
where $v(r)$ is the bare iteraction, $\Gamma_{\rm dd}$ (the
FHNC-dressed counterpart of $f^2(r)-1$ in the Jastrow treatment) has
Fourier transform
\be
\label{eq:GFHNC}
{\widetilde \Gamma}_{\rm dd}(k) = \left[S(k)-S_F(k)\right]/S_F^2(k),
\ee
while
\be
{\tilde w}_I(k) = - t(k)
\left[\frac{1}{S_F(k)}-\frac{1}{S(k)}\right]^2
\left[\frac{S(k)}{S_F(k)}+\frac{1}{2}\right]
\label{eq:inducedint}
\ee
is an induced interaction.

The two-body operators $W(1,2)$ and $N(1,2)$ required for evaluation of
the CBF-dressed pairing matrix elements $V_{\veck\veck'}$ of 
Eq.~\eqref{eq:CBFpi} are defined by
\bea
\label{eq:NWloc}
N(1,2) &=& N(r_{12})\, =\, \Gamma_{\rm dd}(r_{12})\,,\nonumber\\
W(1,2) &=& W(r_{12})\,,\quad \tilde  W(k) =
- \frac{t(k)}{S_F(k)}\tilde \Gamma_{\rm dd}(k),
\eea
again in the collective approximation.  
The operator $W(1,2)$ is in practice just the {\it particle-hole interaction},
given in coordinate space by Eq.~\eqref{eq:ph}.  It includes a so-called 
direct interaction consisting of the bare interaction $v(r)$, {\it moderated} 
by dd-dressed two-body correlations, plus a kinetic term caused by the 
deformation of the wave function at short distances.  The induced interaction 
represented by the last term of Eq.~\eqref{eq:ph}, of long range, accounts for 
exchange of virtual phonons, i.e., density fluctuations.\footnote{It should
be noted that since spin-dependent correlations are not present in the assumed
form of the correlation operator $F_N$, the effects of spin-density
fluctuations on the ground-state energy estimate and the pairing gap -- 
known to be a suppression of this, has to be included within 
CBF perturbation theory \cite{ChenClark1993}.}
Finally, the single-particle energies that enter the ``energy-numerator'' 
term in Eq.~\eqref{eq:CBFpi} proportional to $N_{\veck\veck'}$ reduce to
\bea
\label{eq:CBFspect}
\ep_k &=& t(k) -\mu + \frac{\tilde X'_{\rm cc}(k)}{1-\tilde X_{\rm cc}(k)}
+ {\rm const.} 
\eea
where the constant is fixed by the condition $\epsilon_{k_F} = 0$, while 
\bea\label{eq:defX}
\tilde X'_{\rm cc}(k) &=& 
-\frac{n}{\nu} \int d^3r \, e^{i\veck\cdot\vecr} \Gamma_{dd} (r)\ell(k_Fr)
\eea
is a sum of non-nodal diagrams, with $\ell(x) = (3/x)j_1(x)$ denoting
the Slater exchange function, $\nu$ the single-particle degeneracy,
and $j_1(x)$ the spherical Bessel function of the first kind.  The
expression for $\tilde X_{\rm cc}(k)$ in Eq.~\eqref{eq:CBFspect}
follows from Eq.~\eqref{eq:defX} upon replacing $\Gamma_{dd}(r)$ by
$W(r)$.

This last step completes a closed system of equations, starting with 
Eq.~\eqref{eq:kEuler}, that no longer contains any reference to the 
Jastrow correlation function $f(r)=\exp[u_2(r)/2]$.  These equations 
could just as well have been derived in any generic many-body theory, 
including the GF and coupl\-ed-cluster approaches, and especially 
$\cal T$-matrix theory~\cite{Fan2015}.

A concrete implementation of the theory as described above has been
carried out in \cite{FanKrotscheck2017} for the $^1S_0$ pairing
gap in low-density neutron matter using the EL-FHNC//0 approximation
for two simplified NN interactions -- Argonne $V_4'$ and Reid soft core
$V_6$ \cite {Reid1968,Day1981}, both essentially phase-shift
equivalent to Argonne $V_{18}$ in the density range involved.
Earlier calculations within the same framework  were carried out in
\cite{ChenClark1986,ChenClark1993} but implemented only low-order
cluster expansion.

\subsection{Monte Carlo methods }
\label{eq:MC_methods}

Our survey of the many-body methods for microscopic computational
exploration of pairing behavior in nuclear systems would not be
complete without the important class of stochastic approaches based on
Monte Carlo (MC) algorithms. While MC methods have been extensively
applied to the normal (unpaired) state of neutron and nuclear matter
over an extended period~\cite{Gandolfi2009EoS}, the much more
challenging problem of pairing has been addressed in only a handful of
studies during the last decade 
\cite{Fabrocini2005,Gandolfi2008,Gezerlis2008,Gandolfi2009,Gezerlis2010}.
These studies have focused on phase-shift equivalent interactions,
especially the Argonne-Urbana class of potentials. The essence of the
MC method is the solution of the non-relativistic Schr\"odinger
equation using stochastic sampling of configurations, as the system is
advanced in imaginary time.  In practice, an infinite system is simulated 
in a finite box containing a fixed number of particles subject to 
periodic boundary conditions.  Of specific interest for this review 
are the Green Function Monte Carlo (GFMC) and Auxiliary Field Diffusion 
Monte Carlo (AFDMC) algorithms. The latest GFMC computations of bulk 
energy and pairing gaps in nuclear matter have been performed for systems 
of $\sim 60$ nucleons; larger numbers of particles can be accommodated 
in AFDMC simulations~\cite{GandolfiGezerlis2015}.

AFDMC is a special kind of GFMC method in which spin/isospin
configurations are sampled instead of explicitly summed, allowing
extension of the calculation to higher density~\cite{Fabrocini2005}.
The most recent computations of this kind use a fixed-phase
approximation, which resolves the technical difficulties associated
with the presence of a tensor
interaction~\cite{Gandolfi2008,Gezerlis2008,Gandolfi2009}.  This work
also employs the full bare interaction assumed instead of projecting
it on some specific partial-wave channel (\eg, $^1S_0$ for low-density
neutron matter).  Depending on the forms of the starting or trial
correlated superfluid and normal states, the energy difference between
their evolved versions can be under $4\%$~\cite{Gandolfi2009}.  The
starting superfluid state is taken as the product of a state-dependent
Jastrow-type correlation factor and a token superfluid state
consisting of the projection of the BCS state on the $N$-particle
Hilbert space of the system.  For even $N$ the latter is a Pfaffian of
pair wave functions $\phi(ij)$ satisfying prescribed boundary
conditions.  The pair functions are determined from a variational CBF
calculation of the energy expectation value using extended FHNC
techniques.  In the case of odd neutron number, the energy of the
unpaired neutron is chosen to minimize the energy.

The standard Green Function MC (GFMC) method and the simpler
variational MC (VMC) procedure sample only spatial
configurations~\cite{Gezerlis2008,Gezerlis2010,GandolfiGezerlis2015}.
VMC calculations use Monte Carlo integration to minimize the
expectation value of the Hamiltonian, optimizing the trial wave
function.  In the GFMC approach the Schr\"odinger equation is cast in
the diffusion form with respect to imaginary time and the initial
trial wave function is evolved to obtain the lowest energy
eigenstate. As in the AFDMC approach, the starting wave function is
taken to be of Jastrow-Pfaffian form with a fixed number of particles
subject to periodic boundary conditions. The Jastrow part of the wave
function is obtained from a lowest-order con\-strain\-ed-variational
(LOCV) method \cite{Pandharipande1971}.

It should be understood that, of necessity, these methods do not
evolve or reach a state with full BCS pairing correlations, which
would be a state of indefinite particle number residing in Fock space,
but rather its projection onto an $N$ particle subspace.  As is done
in the case of finite nuclei, the energy gap in pure neutron matter is
determined (up to the sign) from the odd-even staggering formula for
odd neutron number $N$, thus
\bea
\Delta (N) = E(N) - \frac{1}{2}\left[
E(N+1) + E(N-1)\right].
\eea

More recent GFMC computations~\cite{Gezerlis2008,Gezerlis2010} predict
gaps which are about $30\%$ smaller than those obtained with AFDMC.
Furthermore, the gaps obtained by the two MC methods are suppressed
compared to the bare BCS result, as is usually the case with the other
methods (SCGF, CBF, etc.) discussed above. One may anticipate that, 
within their error bars, the MC computations faithfully account for the 
strong short-range repulsion of phase-shift equivalent NN interactions.  
Simulations with larger number of particles may provide further insight 
into the accuracy of the extrapolations to infinite matter and the role 
of long-range correlations. 

\begin{figure}[bt]
\begin{center}
\includegraphics[width=0.99\hsize]{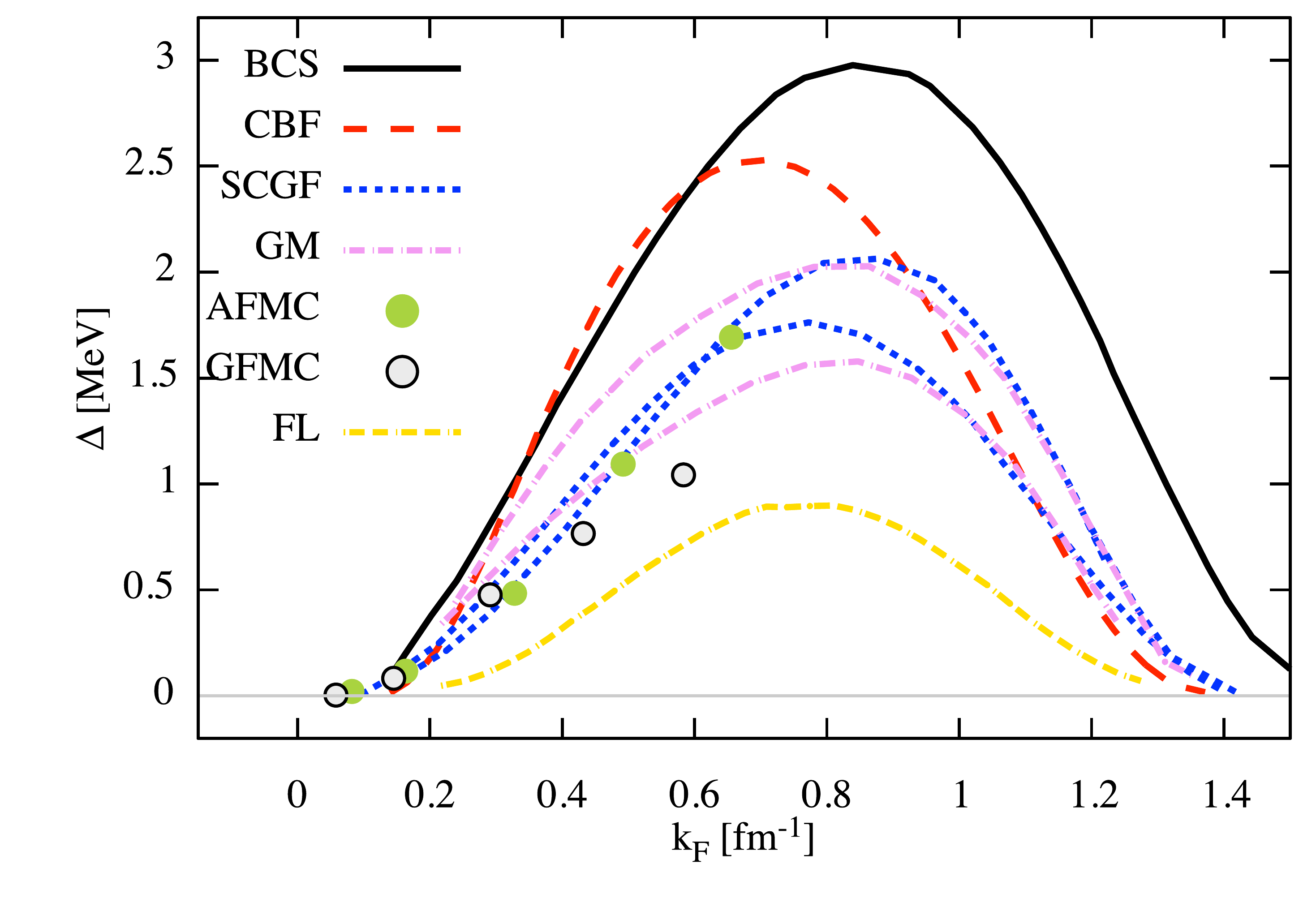}
\caption{ Dependence of the $^1S_0$ pairing gap $\Delta=\Delta(k=k_F)$
  in low-density neutron matter on the Fermi momentum $k_F$ as
  computed using different many-body theories. BCS: solution of the
  BCS gap equation with a free single-particle spectrum. CBF:
  FHNC-optimal correlated basis theory
  ~\cite{FanKrotscheck2017}. SCGF: Self-Consistent Green Functions 
  theory including only BCS and short-range correlations (curve 
  with higher-maximum) and including also long-range correlations 
  (other curve)~\cite{Ding2016}. GM: solutions of the gap equation with
  self-energies derived from a $\cal G$-matrix, with long-range
  correlations either absent (upper curve) or present (lower
  curve)~\cite{CaoLombardo2006}.  AFMC: Auxiliary Field Monte Carlo
  computations~\cite{Gandolfi2008}. GFMC: Green Function Monte-Carlo
  calculations~\cite{Gezerlis2008}.  FL: nucleonic pairing within
  Fermi-Liquid theory that includes long-range polarization
  effects~\cite{WambachAinsworthPines1993}.  BCS, SCGF, and GM results
  were obtained with the Argonne $V_{18}$ interaction, with the GM
  calculation also including a three-nucleon force based on meson exchange.
  Calculations CBF, AFDMC, GFMC used reduced versions of $V_{18}$,
  respectively $V_{4'}$, $V_{8'}$, and $V_4$.  FL used the Reid
  soft-core potential~\cite{Reid1968}.  }
\label{fig:gaps_overview}
\end{center}
\end{figure}
\subsection{Overview of the results}
\label{sec:S_wave_overview}

We close this section with an overview of the results obtained for the
simplest problem, namely the neutron $^1S_0$ pairing gap, discussion
of $^3P_2$--$^3F_2$ pairing being reserved for
Sec.~\ref{sec:higher_partial_waves}.  Figure~\ref{fig:gaps_overview}
collects a selection of results for this gap, all but one of which is
based on a version of the Argonne family of NN potentials, implying
that the observed differences are due primarily to differences between
the many-body methods applied.  All these methods (except the Monte
Carlo approaches, which provide data only in the lower-density domain)
predict a peak value of the gap $\Delta_F$, \ie, $\Delta_\veck$
evaluated for $k= k_F$, close to $0.8~{\rm fm}^{-1}$ (which
  corresponds to the number density $n =0.017$ fm$^{-3}$). The peak
value itself varies in the range 0.8 to 2.5 MeV.  The CBF
\cite{FanKrotscheck2017}, SCGF \cite{Ding2016}, and GM
\cite{CaoLombardo2006} theories predict peak values of the gap within
an interval of 0.5 MeV around a value of the order of 2 MeV. (It
  must be noted here that unlike most treatments of pairing within the
  CBF-variational framework, that of \cite{FanKrotscheck2017}
  incorporates the specific effect of density fluctuations, which
  enhance the gap.  The dominant spin-density fluctuations
  \cite{ClarkKallman1976}, which produce a stronger {\it suppression}
  of the gap, may be estimated within CBF perturbation theory, or by
  the introduction and optimization of long-range spin-dependent
  correlation functions.)    The Fermi-liquid (FL) methods
\cite{WambachAinsworthPines1993} predict $\Delta_F$ values smaller by
about 1~MeV, which is attributed to the suppression of pairing by
spin-density fluctuations. The MC results
\cite{Gandolfi2008,Gezerlis2008} at lower densities are consistent
with the results obtained within non-MC theories, but we recall that
the gap in the MC computations is extracted from the difference in the
energies of the normal and paired states, extrapolated to the
thermodynamical limit, rather than from solution of the gap equation.

Different methods for solving the BCS gap equation for interactions
that are consistent with nucleon-nucleon scattering data lead to
essentially the same result for the gap, provided the high-momentum
states are properly taken into account in the numerical
procedure. Additionally, a number of effective models of the two-body
interaction have been tested on the pairing problem in neutron
matter. These interactions are designed for efficient and accurate
computation of properties of finite nuclei.  Calculations based on
effective interactions such as the purely phenomenological Gogny
interaction or the $V_{\rm low-k}$ potentials which are extracted
  from the phase-shift equivalent realistic interactions produce gaps
in neutron matter that are close numerically to those obtained from
realistic, full (\ie\ un-truncated)
interactions~\cite{SedrakianKuoMuther2003,SunPan2013,Kuckei2003,Ramanan2007}.
The same is true for the more recent chiral potentials with varying
cut-off~\cite{RiosPollsDickhoff2017,Kaiser2005,Drischler2017,Maurizio2014,Finelli2015}.
Particular features of these interactions (e.g.\ localization at small
momenta) are advantageous in many-body approaches that are not well
suited to bare full potentials because of their short-range repulsive
component.

\section{Unconventional pairing and BCS-BEC  crossovers}
\label{sec:Unconv_BCSBEC} 

New classes of superfluid fermionic states arise when the pairing is
between fermions residing on different Fermi surfaces. Such a situation
arises generically in multi-component systems with cross-species
pairing. The simplest example is an electronic superconductor in a spin
polarizing magnetic field that induces an imbalance between the number of
spin-up and down electrons. In nuclear physics we encounter such
a situation when pairing occurs between neutrons and protons in isospin
asymmetric matter or among neutrons (or protons) placed in a strong
magnetic field.

Mathematically, the novelty of such phases is associated with a
non-zero anti-symmetric piece $E_A$ of the quasiparticle spectrum in
Eq.~\eqref{eq:poles}, which by definition requires
$\varepsilon(p) \neq \varepsilon(-p)$, \ie, breaking of the
invariance with respect to reversal of time or spatial symmetry. We
shall refer to such systems below as imbalanced superfluids, a term
that has become common in the theory of cold fermionic atoms, where these
systems can be tested experimentally.

Historically, the studies of imbalanced superfluids began shortly
after the advent of BCS theory in the context of electronic
materials containing paramagnetic
impurities~\cite{Clogston1962,Chandrasekhar1962,Sarma1963}. The effect
of impurity scattering on electrons, on average, was modeled in terms
of an effective magnetic field, which then induces an imbalance
between the spin-up and spin-down electrons.

The initial studies were carried in the weak-coupling formalism, where
the back-reaction of the pairing on the chemical potential of the
system can be ignored. The imbalance was parametrized in terms of the
difference $\delta\mu$ in the chemical potentials of the species,
which led to the following picture for the gap $\Delta$ as a function
of $\delta\mu$~\cite{Clogston1962,Chandrasekhar1962,Sarma1963}.  The
gap is a double-valued function, the upper branch of the two solutions
being a constant $\Delta (\delta\mu) = \Delta (0)$ in the range
$0\le\delta\mu \le \Delta(0)$ and zero beyond the point
$\delta\mu = \Delta(0)$.  The lower branch exists in the range
$\Delta(0)/2\le\delta\mu\le\Delta(0)$, with the gap increasing from
zero at the lower limit to $\Delta(0)$ at the upper limit.  Only the
portion $\delta\mu\le\Delta(0)/\sqrt{2}= \delta\mu_1$ of the upper
branch is stable in the sense that the superconducting state lowers
the ground-state energy of the superfluid~\cite{Sarma1963}.  In the
remaining region of imbalance, the superconducting state is unstable
(Sarma instability).  The maximal value of imbalanced $\delta\mu_1$
sustained by the system is known as Chandrasekhar-Clogston limit.

Imbalanced pairing in infinite nuclear systems naturally became of
interest in the context of $^3S_1$--$^3D_1$ and $^3D_2$ pairing in
isospin asymmetrical nuclear matter~\cite{Alm1993,AlmSedrakian1996},
and the critical temperatures in these channels were computed using
$\cal T$-matrix theory and realistic interactions. The full BCS formulation
was applied at about the same time~\cite{SedrakianAlmLombardo1997},
and subsequently the single-particle energies $\epsilon_p$ were
renormalized within Brueckner theory, resulting in a major reduction
of the gap values and more realistic values of critical isospin
asymmetries~\cite{SedrakianLombardo2000}.

The ground state of an imbalanced superfluid may entail breaking of
global symmetries, notably translational or rotational symmetries, in
some range of parameter space. Breaking of translational invariance
was first proposed and studied independently by
~\cite{FuldeFerrell1964} (FF) and \cite{LO1964} (LO),
(collectively, FFLO), who discovered that the superconducting state
where the Cooper pairs carry a finite center of mass (CM) momentum can
extend to imbalances beyond those restricted by the
Chandrasekhar-Clogston limit. In the weak coupling case, the maximal
value of the difference in the chemical potentials of the species for
the FFLO type of pairing is
$\delta\mu_2 = 0.755 \, \Delta(0)\,[>\delta\mu_1 = 0.707\,\Delta(0)]$.
The condensate predicted by Ref.~\cite{FuldeFerrell1964} assumes
$\Delta(\vecr) = \Delta_0\, {\rm exp}(-i\vecQ\cdot \vecr)$ for the gap
function, where $\vecQ$ is the CM momentum.  Ref.~\cite{LO1964}
explored various lattice types and concluded that the
body-centered-cubic lattice is the most stable configuration near the
critical temperature.  Imbalanced pairing involving finite momentum of
pairs of neutrons and protons in infinite nuclear systems has been
studied in $^3S_1 $--$^3D_1$ and $^3D_2$ pairing channels, both within
$\cal T$-matrix theory~\cite{AlmSedrakian1996} and in extensions of the BCS
theory to account for violation of spatial
symmetries~\cite{Sedrakian2001LOFF,MaoHuangZhuang2009,SteinHuang2012,SteinSedrakian2014}.

Two alternatives to FFLO phase of imbalanced superfluids, proposed
later, find their application in nuclear systems. One involves
deformations of the Fermi surfaces of the fermion species in
population imbalance, with the prospect of improving the ground-state
energy of the paired system compared with the standard configuration
of Fermi spheres
\cite{MuetherSedrakian2002,MuetherSedrakian2003}. Another possibility
is the separation of phases, originally suggested in the context of
cold atomic gases~\cite{BedaqueCaldas2003}, and studied thereafter in
infinite nuclear systems in the $^3S_1$--$^3D_1$
channel~\cite{SteinHuang2012,SteinSedrakian2014}.

Extensive work on imbalanced superfluids was carried out within the
area of ultracold atomic gases starting shortly after the observation
of Bose-Einstein condensates in traps~\cite{BedaqueCaldas2003,StoofHoubiers1996,Combescot2001,SedrakianMurPetit2005,LiuWilczek2003,Forbes2005,Schmitt2014}.
For a review and further references
see~\cite{BlochDalibardZwerger2008}. These systems offer unique
possibilities for testing the physics of imbalanced superfluidity under
controllable conditions in laboratory
experiments~\cite{Partridge2006,Zwierlein2006}.

Furthermore, imbalanced superfluids have been extensively studied
in the context of color superconductivity of cold quark matter in
compact stars, where the three lightest flavors of quarks and three
quark colors make the possible patterns of pairing especially
interesting; for reviews and further references
see~\cite{AlfordSchmitt2008RvMP,AnglaniCasalbuoni2014RvMP,RajagopalWilczek2001}.

\subsection{Formalism}
\label{sec:sd_formalism}

Proceeding to the examination of imbalanced phases in more detail 
in terms of the underlying many-body theory and its application to
nuclear matter, we start with a brief outline of the formalism based 
on the extension of the GF method to imbalanced systems within the
imaginary-time Matsubara formalism (for more details,
see~\cite{SteinSedrakian2014}).

Consider a mixture of neutrons ($n$) and protons ($p$) at some density
and temperature. The GF of the superfluid, written in the
Nambu-Gor'kov basis, is given by
\bea 
\label{eq:props} i\mathscr{G}_{12} =
i\left(\begin{array}{cc} G_{12}^{+} & F_{12}^{-}\\
    F_{12}^+ & G_{12}^{-}\end{array}\right) = \left(\begin{array}{cc}
    \langle T_\tau\psi_1\psi_2^+\rangle
    & \langle T_\tau\psi_1\psi_2\rangle \\
    \langle T_\tau\psi_1^\dagger\psi_2^\dagger\rangle & \langle \tilde T_\tau\psi_1\psi_2^\dagger\rangle
\end{array}\right),\nonumber\\
\eea 
where $G_{12}^{+}\equiv G^{+}_{\alpha\beta}(X_1,X_2)$, etc., and
$X = (t,\vecr)$ is the four-dimensional time-space coordinate.  The
Greek indices label discrete spin and isospin variables.  The
operators in \eqref{eq:props} can be viewed as  bi-spinors, \ie,
$$\psi_{\alpha}=(\psi_{n\uparrow},\psi_{n\downarrow},\psi_{p\uparrow},\psi_{p\downarrow})^T,$$
where the indices $n,p$ label a particle's isospin and
$\uparrow, \downarrow$ label its spin.

The solutions of the Dyson equation for the GF defined in 
\eqref{eq:props} are 
\bea
\label{eq:Gfs1}
G_{n/p}^{\pm} &=&
\frac{ik_{\nu}\pm\epsilon_{p/n}^{\mp}}{(ik_{\nu}-E^+_{\mp/\pm})(ik_{\nu}+E^-_{\pm/\mp})},\\
\label{eq:Gfs2}
F_{np}^{\pm} &=&
\frac{-i\Delta}{(ik_{\nu}-E^+_{\pm})(ik_{\nu}+E^-_{\mp})},\\
\label{eq:Gfs3}
F_{pn}^{\pm} &=&
\frac{i\Delta}{(ik_{\nu}-E^+_{\mp})(ik_{\nu}+E^-_{\pm})},
\eea
where  $ik_{\nu}$ is the Matsubara frequency,
\bea
\label{eq:norm_spec}
\epsilon^{\pm}_{n/p} &=& \frac{1}{2m^*}\left(\veck\pm
  \frac{\vecQ}{2}\right)^2-\mu_{n/p}
\eea
are the normal state spectra of neutrons and protons, with $\mu_{n/p}$
denoting their chemical potentials, $m^*$ their effective mass, and
$\vecQ$ is the center-of-mass momentum of the Cooper pair.  (Note that
the difference between the effective masses of neutrons and protons is
small at the low densities of interest and is neglected.) There are four
branches of the quasiparticle spectrum, which are given by
\be\label{eq:spectrum_Era}
E_{r}^{a} = \sqrt{E_S^2+\Delta^2} + r\delta\mu +a E_A,
\ee
where $a, r \in \{+,-\}$ and 
\bea
\label{eq:E_SA}
E_S =\frac{\vert \vecQ\vert^2/4+k^2}{2m^*}-\bar\mu, \qquad
E_A = \frac{\veck\cdot \vecQ}{2m^*} ,
\eea 
are the symmetrical and anti-symmetrical parts of the spectrum, with
$\bar\mu \equiv (\mu_n+\mu_p)/2$ average of neutron and proton
chemical potentials. Taking $ik_\nu \rightarrow k_0+i0^+$, the GF of
Eqs.~\eqref{eq:Gfs1}--\eqref{eq:Gfs3} are analytically continued to
obtain their retarded counterparts.  The densities of neutrons and
protons are then defined in a standard fashion by
\bea\label{eq:densities_imbalance}
n_{n/p} (\vecQ)&=&\frac{2}{\beta}\int\!\!\frac{d^3k}{(2\pi)^3}\sum_{\nu}
G^+_{n/p}(k_{\nu},\veck,\vecQ)\nonumber\\
&=&2 \int\!\!\frac{d^3k}{(2\pi)^3}
\Biggl[
\frac{1}{2}               
\left(1+\frac{E_S}{\sqrt{E_S^2+\Delta^2}}\right) f(E^+_{\mp})\nonumber\\
&+&\frac{1}{2}             
\left(1-\frac{E_S}{\sqrt{E_S^2+\Delta^2}}\right) f(-E^-_{\pm})
\Biggr],
\eea
where $k = (k_0,\veck)$ is the four-momentum, $\beta$ is the inverse
temperature. 
If the interaction is time-local, the pairing gap is given by 
\bea \label{eq:gap_imbalance}
\Delta(\veck,\vecQ) &=&  \frac{1}{4\beta} \int\!\!\frac{d^3k'}{(2\pi)^3}\sum_{\nu}
V(\veck,\veck')               \nonumber\\
&& \Img  \Bigl[
  F^+_{np} (k'_{\nu},\veck',\vecQ)
+F^-_{np} (k'_{\nu},\veck',\vecQ)  \nonumber\\
&-&F^+_{pn} (k'_{\nu},\veck',\vecQ)
-F^-_{pn} (k'_{\nu},\veck',\vecQ)
\Bigr],
\eea
where $ V(\veck,\veck')$ is the neutron-proton interaction.  This 
interaction could be a bare or effective version, depending on the level 
of approximation; we will illustrate the physical content of the
theory using bare interactions  to establish a benchmark. 

Performing a partial-wave expansion of the expression in
Eq.~\eqref{eq:gap_imbalance} and an energy integration, one arrives at
the gap equation
\bea \label{eq:gap2}
\Delta_l(Q) &=& \frac{1}{4}\sum_{a,r,l'} \int\!\!\frac{d^3k'}{(2\pi)^3}
V_{l,l'}(k,k')
\nonumber\\ &\times&
\frac{\Delta_{l'}(k',Q)}{2\sqrt{E_{S}^2(k')+\Delta^2(k',Q)}}[1-2f(E^r_a)],
\eea
where $V_{l,l'}(k,k')$ is the interaction in the $^3S_1$--$^3D_1$
partial wave and $\Delta^2 =\sum_l \Delta_l^2$.  Note that the magnitude
of the vector $\vecQ$ enters Eqs.~\eqref{eq:densities_imbalance} and
\eqref{eq:gap2} parametrically and should be determined from
minimization of the free energy. Its direction is
chosen by the system spontaneously. This minimum condition leads, in
fact, to an additional equation for $Q$ that should be solved along
with Eqs.~\eqref{eq:densities_imbalance} and \eqref{eq:gap2}.

Quite generally, the ground state of nuclear matter is obtained from
the minimization of the respective free energies of the phases
 \be \label{eq:free} F_S =
E_S-TS_S,\quad F_N = E_N-TS_N, 
\ee
where indices $S$ and $N$ refer to the superfluid and normal phases,
$E$ is the internal energy (statistical average of the system
Hamiltonian), and $S$ is the entropy.

The formalism developed to this point also covers the treatment
of the heterogeneous, phase-separated (hereafter PS) phase proposed 
in \cite{BedaqueCaldas2003} and implemented in the context of nuclear matter in
\cite{SteinSedrakian2014,SteinHuang2012}.  Allowing for separation of
phases implies that we can choose to maximize pairing by having
isospin-symmetrical superfluid domains, with all the excess neutrons
accommodated in normal regions.  Then we already have all the necessary 
ingredients for evaluating the free energy of such a state, using the
simple construction 
\bea \label{eq:free_mixed}
\mathscr{F}(x_f,\alpha) = (1-x_f) F_S(\alpha = 0) 
+ x_f F_N(\alpha \neq 0), \,\, (\vert \vecQ \vert= 0),\nonumber\\
\eea
where $x_f$ is the filling fraction of the unpaired component.  Here
$\alpha$ is the isospin asymmetry parameter defined by 
\bea\label{eq:alpha-def}
\alpha = \frac{n_n -n_p}{n_n+n_p} .
\eea
  In the superfluid
phase ($S$), one has by definition $n_n^{(S)}=n_p^{(S)} = n^{(S)}/2$. 
In the unpaired phase ($N$), the densities of neutrons and protons need 
not be equal and are assigned the values $n_{n/p}^{(N)}$. 
Thus, the net densities of neutrons/protons per unit volume are given 
by $n_{n/p} = (1-x_f)n^{(S)} + x_fn_{n/p}^{(N)}$.

In the preceding discussion of imbalanced phases we considered a
particular realization of the FFLO phase with single plane-wave
modulation of the gap parameter in space, which corresponds to the
original FF phase. This phase in fact breaks only the rotational
symmetry along the direction of the vector $\vecQ$, but in many
respects this phase is a representative for other realizations of the
FFLO phases.

To summarize this subsection, we have surveyed the formalism for imbalanced
superfluids and their realization in superfluid nuclear matter in which
four possible distinct phases can arise.  These can be classified
 in terms of $Q$, $\Delta$, and $x_f$ as follows $(Q \equiv \vert \vecQ\vert)$
\bea\label{eq:phases}
\left\{
\begin{array}{llll}
Q = 0,  &\Delta \neq 0, & x_f = 0,& \textrm{BCS phase,}\\
Q \neq 0, & \Delta \neq 0, &x_f = 0,&\textrm{FFLO phase,} \\
Q = 0,  &\Delta \neq 0, & x_f \neq 0,&\textrm{PS phase,}\\
Q = 0, & \Delta = 0, & x_f = 1, &\textrm{unpaired phase.} \\
\end{array}
\right. 
\eea
The competition between these phases is decided on the basis of 
minimization of the ground state energy.

These four phases of fermionic matter are conceived for two species of
fermions having spherically symmetric Fermi surfaces.  In
Sec.~\ref{sec:dfs_phase} below we will amend our discussion with yet
another phase that breaks the spatial symmetries by deformations of
the Fermi surfaces away from spherical
shape~\cite{MuetherSedrakian2002,MuetherSedrakian2003}.

\subsection{Homogeneous phase}
\label{sec:homogeneous}

Equations~\eqref{eq:densities_imbalance} and \eqref{eq:gap_imbalance} have
to be solved simultaneously in general. In weakly coupled superfluids,
Eq.~\eqref{eq:densities_imbalance} can be evaluated with the normal
state spectrum by setting $\Delta = 0$, in which case it decouples from
Eq. \eqref{eq:gap_imbalance}. This approximation is invalid for
strongly coupled systems, in particular for nuclear systems at very
low densities.

Once the solutions are found, the free energy can be evaluated and a
specific phase can be assigned to a given temperature and density.
\begin{figure}[t]
\centering 
\includegraphics[width=0.8\hsize]{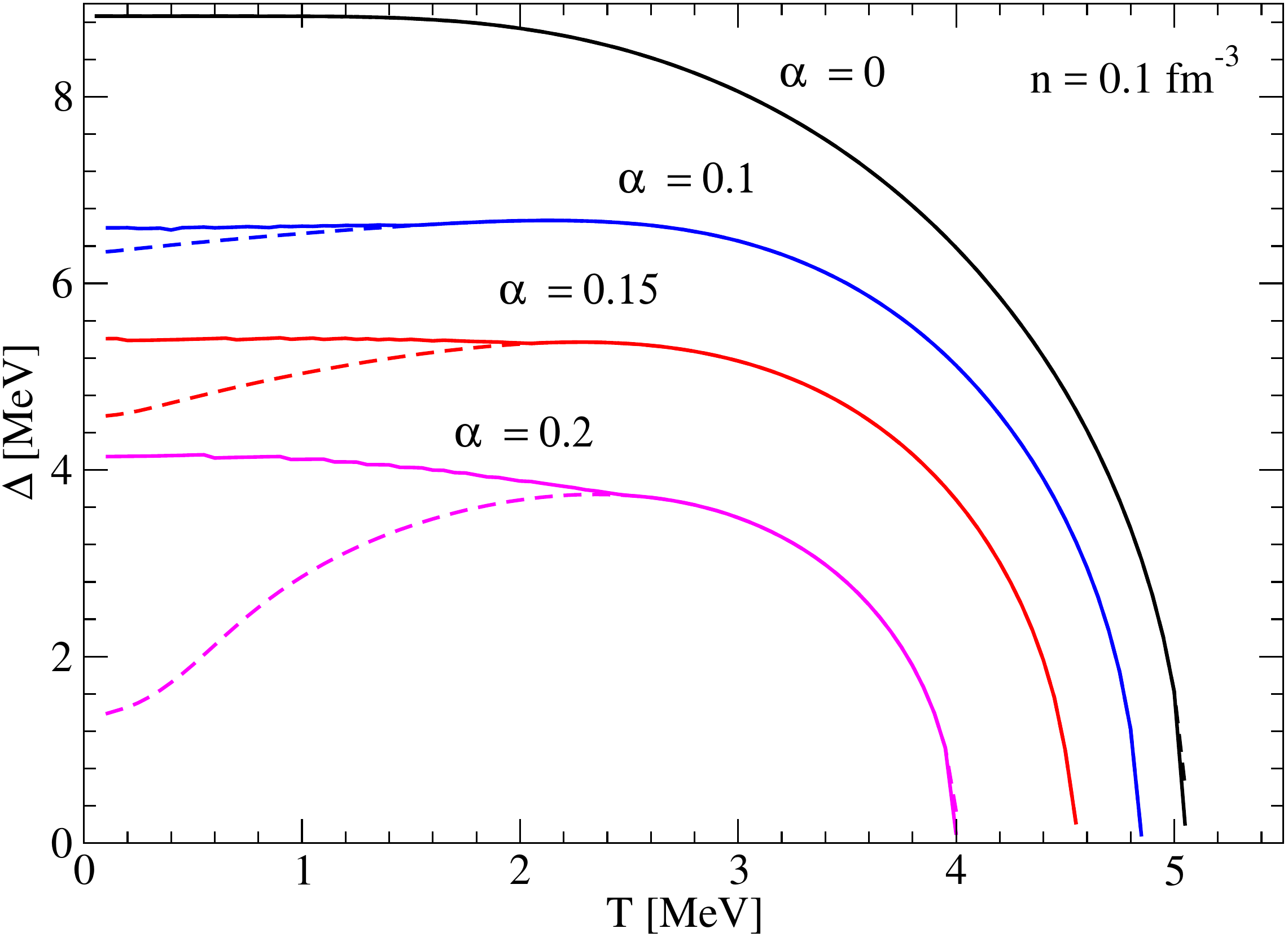}
\caption{Dependence of the pairing gap on temperature in the 
homogeneous phase (dashed lines) and FFLO phase (solid lines) for nuclear
matter density $n = 0.1$ fm$^{-3}$. The curves are
labeled by the values of isospin asymmetry $\alpha$.
}
\label{fig:homogeneous}
\end{figure}
We start our discussion with the simplest case, the homogeneous
imbalanced superfluids. Their realization depends essentially on the
difference in the chemical potentials of the components,
$\delta \mu =\mu_n-\mu_p \ge 0$, assuming a neutron excess.  A
sufficiently large $\delta \mu$ value will disrupt pairing because
fermions lying on different Fermi surfaces cannot overlap to form
Cooper pairs, due to the lack of shared phase-space.  Finite
temperature can counteract the disruptive effect of $\delta\mu$
because it increases the ``diffuseness'' of the Fermi surfaces and
hence the phase-space overlap between the paired fermions. This
physics is illustrated in Fig.~\ref{fig:homogeneous}, where we see
that the gap has a maximum as a function of temperature as a
consequence of the interplay of two effects: the disruption by
$\delta\mu$ and phase-space expansion by temperature.  For large
enough asymmetries there exists a lower critical temperature
$T_c^*$~\cite{SedrakianAlmLombardo1997,SedrakianLombardo2000} (not
shown in the figure).  Note that a similar phenomenon (referred to as
``re-entrance'') has also been observed in the context of small
superconducting systems, where in the case of an odd number of
particles the gap increases with $T$ near
$T = 0$~\cite{BalianFlocard1998}.  Furthermore, this type of
phenomenon arises in systems with spin-zero pairing, when two fluids
(nuclei and nuclear matter, in the nuclear context) occupy different
volumes see~\cite{Margueron2012,PastoreMargueron2013,BelabbasLi2017}.

It should be mentioned that the ``anomalous'' behavior of the BCS gap below the
temperature corresponding to the maximum gap (see
Fig.~\ref{fig:homogeneous}) leads to a number of anomalies in
thermodynamic quantities, such as negative superfluid density or an
anomalous jump in the specific heat, which can be a signature of a
(metastable) low-temperature homogeneous
phase~\cite{SedrakianPolls2006}.

We also observe in Fig.~\ref{fig:homogeneous} that at temperatures
close to its critical value, homogeneous imbalanced superfluids show
the same dependence of the gap on temperature; consequently, the
thermodynamics in this regime is analogous to that in the ordinary BCS
case.  Computation of the free energy of the homogeneous imbalanced
phase and its comparison to that of other phases listed in \eqref{eq:phases}
shows that it is preferred in a domain of temperatures adjacent to the
critical temperature, where the disruptive effects are
small~\cite{SteinHuang2012,SteinSedrakian2014}.

\subsection{FFLO phases}
\label{sec:fflo}
\begin{figure}[t]
\centering 
\includegraphics[width=0.8\hsize]{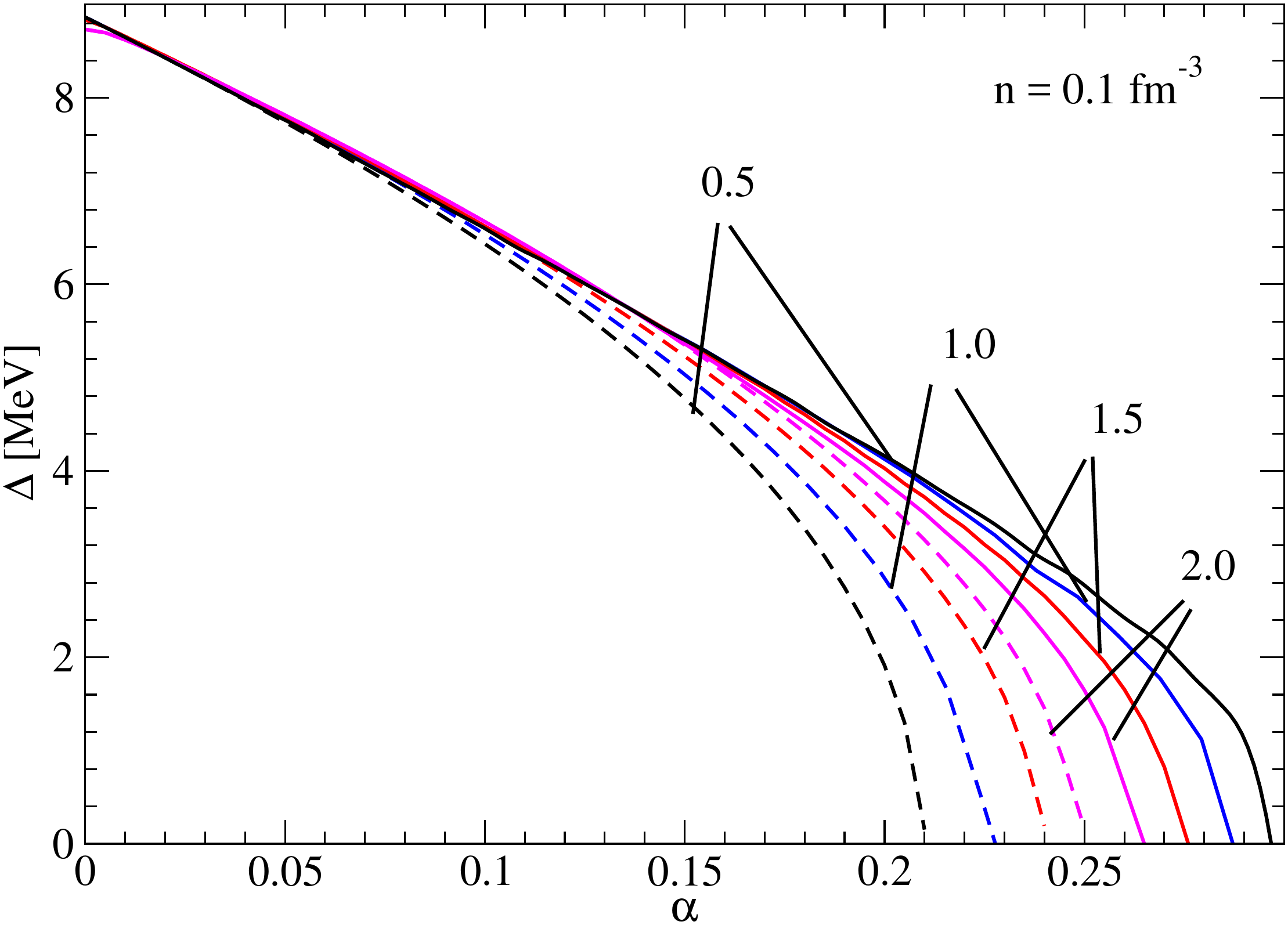}
\caption{Dependence of the gap on isospin asymmetry in the 
homogeneous phase (dashed lines) and FFLO phase (solid lines) for several 
temperature values indicated in the plot. The 
nuclear-matter density is fixed at $n = 0.1$ fm$^{-3}$. 
}
\label{fig:gap_alpha}
\end{figure}
Our next example is the Fulde-Ferrell (FF) phase of nuclear matter paired in the
$^3S_1$--$^3D_1$ channel. The physics of this phase can be understood
by observing that the finite momentum $\vecQ$ affects the spectrum of
particles in a twofold manner: there is a shift in the symmetric part
of the spectrum $E_S \to E_S + (Q^2)/8m^*$ and moreover $E_A = 
\pm (\veck\cdot \vecQ)/2m^* \neq 0$. Thus, in the FF phase there is a positive 
increase in the quasiparticle kinetic energy $\propto Q^2$, which 
disfavors it relative to the BCS state.  However, the anisotropic 
term $\propto \veck \cdot \vecQ$ changes the phase-space overlap of 
the fermions and promotes pairing in certain directions. Clearly, the 
FF phase is stabilized when the increase in the kinetic energy loss 
caused by moving the condensate is overcome by the gain in the potential 
energy of pairing due to the increase in the phase-space overlap. 

The mechanism leading to a stable FF phase significantly affects the
low-temperature behavior of the imbalanced superfluid as illustrated
in Fig.~\ref{fig:homogeneous}.  It is seen that the FF and homogeneous
solutions coincide above a certain temperature, but there is a
bifurcation at low temperatures.  The high-temperature segment
corresponds to the BCS state, with the temperature dependence of the
gap given by the standard asymptotic behavior
$\Delta(\alpha) \sim [T_c(\alpha)(T_c(\alpha)-T)]^{1/2}$, where
$T_c(\alpha)$ is the (upper) critical temperature.  The quenching of
the BCS gap (dashed lines) discussed above is replaced in the FF phase
by solutions which are self-similar to the BCS solutions with
$d\Delta(T)/dT \le 0$~\cite{HeJinZhuang2006}.  Also, the anomalies in
the thermodynamic quantities found in the homogeneous phase are absent
in the FF state~\cite{JinHeZhuang2007}.

Figure~\ref{fig:gap_alpha} illustrates the competition between the FF
phase and the homogeneous BCS phase by showing the dependence of the
gap on asymmetry for several constant temperatures.  There exist 
two segments for each temperature: the low-$\alpha$ segment where 
both phases predict the same $\alpha$ dependence, and the large-$\alpha$ 
segment, where the gap values for the FF phase systematically extend
to larger $\alpha$ values.   The small-$\alpha$ region is
characterized by linear dependence of the gap on $\alpha$; the
large-$\alpha$ asymptotic behavior is
$\Delta(\alpha)\sim \Delta_{00} \left(1-\alpha/\alpha_c\right)^{1/2}$,
where $\alpha_c$ is the critical asymmetry characteristic of a given 
phase. 

\begin{figure}[tb] 
\begin{center}
\includegraphics[width=0.7\hsize]{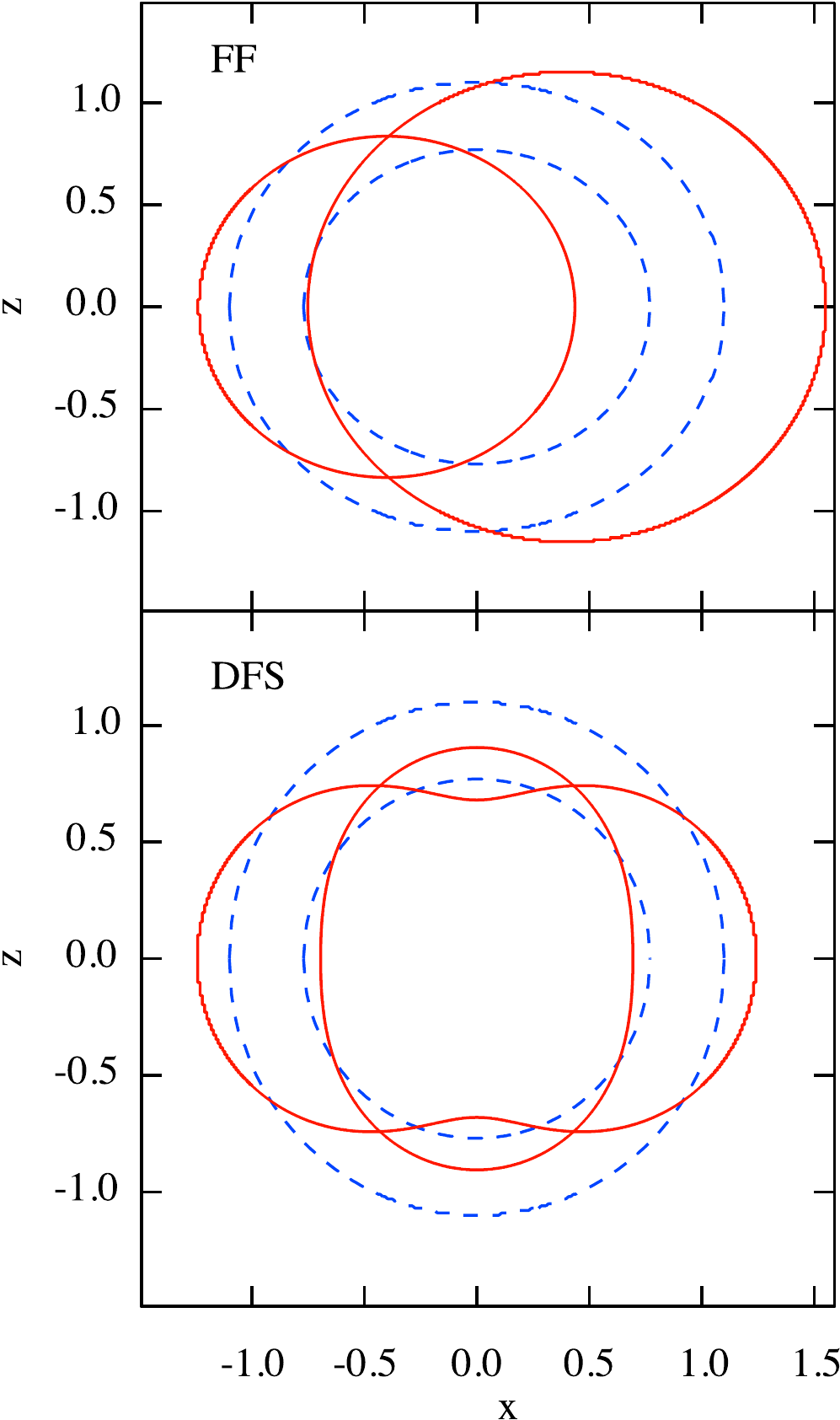}
\caption{Illustration of projected Fermi surfaces (solid traces) in
  the FF (upper panel) and DFS (lower panel) phases.  The concentric
  (dashed) circles represent projections of spherical Fermi surfaces
  of neutrons and protons under imbalance in the $x$-$z$ plane. In the
  case of the FF phase $\mu_n=25$ and $\mu_p=12$ MeV with
  $\vert \vecQ\vert = 0.4$ fm$^{-1}$; for the DFS phase $\mu_n=32$ and
  $\mu_p=10$ MeV with $\delta\epsilon=0.7$.}
\label{fig:ellips}
\end{center}
\end{figure}

\subsection{Deformed Fermi surface phase}
\label{sec:dfs_phase}

The deformed Fermi surface (DFS) phase restores pairing correlations in
imbalanced systems via deformations of the Fermi surfaces of the two
fermion species away from the perfectly spherical
shape~\cite{MuetherSedrakian2002,MuetherSedrakian2003,SedrakianMurPetit2005}.
The agent of these deformations could be the non-central component of
the interaction which, in principle, should be already present in the
unpaired state, but can be strongly magnified by the superconducting
state. The deformations can be spontaneous, as conjectured
in~\cite{MuetherSedrakian2002,MuetherSedrakian2003,SedrakianMurPetit2005},
in the sense that the original Hamiltonian is O(3) symmetric, but the
ground state breaks this symmetry down to a subgroup, for example to
O(2).  To explore whether deformations lead to improvements of the
ground state energy of the system, it is useful to consider spontaneous
deformations which are parametrized as
\begin{eqnarray}
\label{exp}
E^{a,a'}_r=  E^{a}_r + a'\epsilon_2 P_2(\cos\theta),
\end{eqnarray}
where $a' = \pm$, the $E_r^a$ are given by Eq.~\eqref{eq:spectrum_Era},
$P_2(x)$ is the $n=2$ Legendre polynomial, and $\theta$ is the
angle formed by the particle momentum and the direction of the
spontaneous breaking of rotational symmetry. 

The gap equation has been solved and the free energy computed for a
deformation parameter defined as the relative deformation of the two
Fermi surfaces, $\delta\epsilon = (\epsilon_{2,+}-\epsilon_{2,-})/2$.
This parameter is the analog of the total momentum $\vecQ$ in the
analysis of the FF phase.  Computations for nuclear matter and cold
atoms show that, in a certain domain of asymmetries, the energy is
minimized for non-zero $\delta\epsilon$, \ie, there is a stable
minimum corresponding to a state with deformed Fermi surfaces of the
components. 

The Fermi surfaces in the FF and DFS phases are illustrated in
Fig.~\ref{fig:ellips} along with those in an imbalanced homogeneous
phase.  The intersections of the Fermi surfaces of the two fermionic
species in the cases of the FF and DFS phases reveal the mechanisms of
the phase-space overlap between the components and the enhancement of
pairing correlations achieved in these phases.

As already indicated in Sec.~\ref{sec:interface}, cold-atom systems
offer a playground for testing the theoretical ideas that emerged in the
studies of imbalanced system in various contexts. An interesting
extension, which we will not discuss in this review, is the study of
imbalanced superfluids in periodical external potentials created by
optical lattices; for a review see~\cite{Kinnunen2018}.

\subsection{BCS-BEC transition}
\label{sec:BCS_BEC_intro}

\begin{figure}[t]
\begin{center}
\includegraphics[width=0.9\hsize]{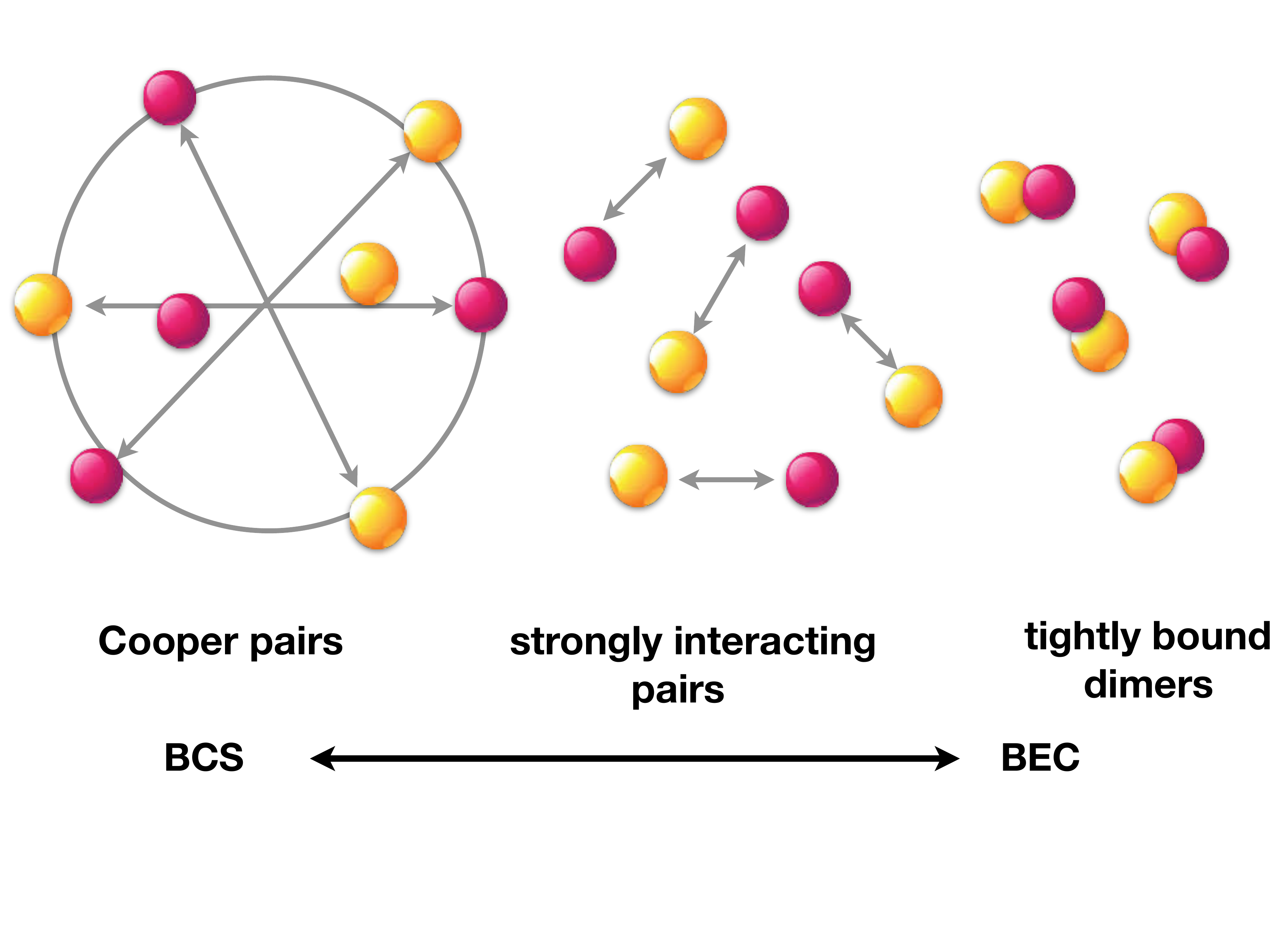}
\caption{Illustration of the BCS-BEC transition. Left: Weakly 
  coupled BCS state characterized by loosely bound Cooper pairs. 
  Center: Intermediate coupling, with strongly interacting pairs. 
  Right: Tightly bound dimers (deuterons) forming a BEC.}
\label{fig:bcs_bec}
\end{center}
\end{figure}

Weakly coupled BCS superfluids form Cooper pairs with characteristic
size of the order of the coherence length, which is much larger than
the interparticle distrance. The pairs are weakly bound, as the scale
of binding energy set by the gap $\Delta$ is much smaller than the
Fermi energy. It was conjectured long ago that under gradual decrease
of density, such BCS superfluids will smoothly evolve into a BEC of
tightly bound bosonic dimers, with a size much smaller than the
interparticle distance, in what amounts to a strong-coupling
limit~\cite{Nozieres1985,Eagles1969}. See Fig.~\ref{fig:bcs_bec} for
an illustration. This conjecture has been confirmed in experiments on
cold atomic gases, where the coupling parameter can be manipulated via
tuning the magnetic field to a Feshbach resonance and thus
effectively changing the magnitude of the scattering length and its
sign~\cite{Zwierlein2006,REGAL2007}.
High-density isospin-symmetric nuclear matter would form a weakly
coupled BCS condensate in the dominant $^3S_1$--$^3D_1$ channel. The
BCS-BEC crossover in such a condensate can be achieved by diluting the
system, in which case the reduction of the density of states will
eventually lead to a transition to a BEC state, which in this case is
a condensate of
deuterons~\cite{Alm1993,BaldoLombardoSchuck1995,SteinSchenll1995,SedrakianClark2006PRC,Huang2010,JinUrban2010}. This
transition occurs smoothly without changes in the condensate wave
function; hence it is a crossover in the proper sense.  The BCS-BEC
crossover is also expected in imbalanced systems, in particular in
isospin-asymmetric nuclear matter, unless the pairing is completely
disrupted by the mismatch in the Fermi surfaces of protons and
neutrons~\cite{LombardoNozieres2001,MaoHuangZhuang2009,SteinHuang2012,ShangZuo2013,SunPan2013,SteinSedrakian2014,Stein2014JP}. In
this case phase transitions can be encountered, \ie, the condensate
wave function does not evolve smoothly across the BCS-BEC crossover.
Therefore, it is more appropriate to speak about a BCS-BEC transition
rather than a crossover.  The straightforward modification of the
original theory of ~\cite{Nozieres1985} involves adaptation to a
gaseous mixture of neutrons and deuterons in the strong-coupling
low-density limit. A more subtle issue is the emergence of phase
transitions between various phases of imbalanced superfluids discussed
above as one moves from weak to strong coupling. Note that
  the straightforward application of the mean-field BCS approach to
  the problem of BCS-BEC crossover fails quantitatively in the
  intermediate coupling regime, as it does not include
  pair-fluctuation corrections~\cite{Nozieres1985,Strinati2018}. 

The evidence for isospin-singlet pairing in nuclear phenomenology is
scarce. However, it has been conjectured that large enough nuclei may
feature spin-aligned $np$ pairs, based on recent experimental studies
of the excited states in $^{92}$Pd~\cite{Cederwall2011} as well as
Hartree-Fock-Bogolyubov computations of large
nuclei~\cite{FriedmanBertsch2007}.  Intermediate energy heavy-ion
collisions produce large amounts of deuterons in final states, which
could be an asymptotic state reached once the initially formed BCS
condensate in the $^3S_1$--$^3D_1$ channel crosses over to a BEC of
deuterons~\cite{BaldoLombardoSchuck1995}.  The measured deuteron
distributions are well described by simple statistical models,
therefore there is no direct experimental evidence of condensation of
deuterons in heavy-ion collisions.  It has been also speculated that
deuteron condensates can be formed in the dilute nuclear matter found
in supernova and hot proto-neutron-star matter at sub-saturation
densities; see for
example~\cite{Sumiyoshi2008,HeckelSchneider2009,HempelSchaffner2010,RadutaGulminelli2010,OertelFantina2012,PaisChiacchiera2015,GulminelliRaduta2015,Clark2016JP,Typel2010,WuSedrakian2017}.

The emergence of a BEC in the isospin-singlet channel at asymptotically
low densities is straightforward because in the vacuum there is a
bound state in this channel -- the deuteron. In contrast, the BEC limit
in the isotriplet $^1S_0$ pairing channel is not obvious. Nevertheless,
the unusually large scattering length in this channel for
neutron-neutron scattering, $ a_{n} \simeq -19$ fm, suggests
traces of a BEC in neutron-rich systems such as the halo nuclei or
neutron matter in compact
objects~\cite{Matsuo2006,MargueronSagawa2007,Isayev2008,Kanada-Enyo2009,AbeSeki2009,SunToki2010,RamananUrban2013,Salasnich2011,SunSun2012,SteinSedrakian2016neutron}. We
will address this problem below in Sec.~\ref{sec:spin_pol_NM}.

\subsubsection{BCS-BEC transition in the balanced case}
\label{sec:bcs_bec_balanced}

Consider first the basics of the BCS-BEC crossover in the
$^3S_1$--$^3D_1$ pairing channel for the case of isospin symmetrical
nuclear matter.  The equations that are solved in this case for the
densities and the gap are respectively \eqref{eq:densities_imbalance}
and \eqref{eq:gap2} in the symmetrical limit.
\begin{figure}[t]
\begin{center}
\includegraphics[width=1.\hsize]{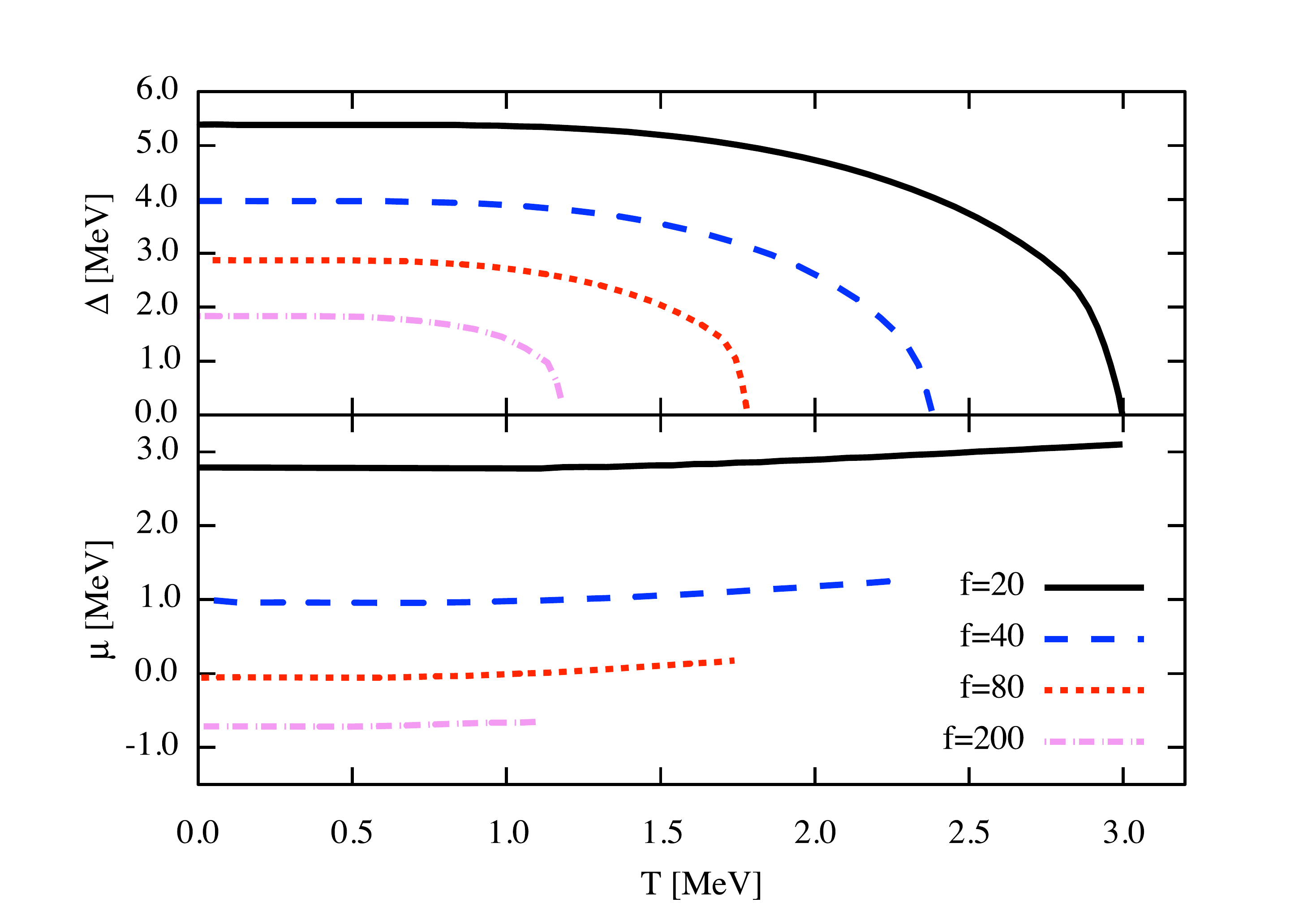}
\caption{ Upper panel: Pairing gap $\Delta(p_F)$ as a function of
  temperature $T$ at different values of the parameter $f = n_0/n$,
  defined in the text. Lower panel: associated chemical potential
  $\mu$ at the same values of $f$. The quoted $f$ values translate
  into the values of the diluteness parameter $na_{np}^3 = $ 1.26, 0.63,
  0.32, and 0.13, for the neutron-proton scattering length
  $a_{np} = 5.41$~fm.  }
\label{fig:bcs_bec_sym}
\end{center}
\end{figure}
Results from simultaneous solution of the gap and density equations
\cite{SedrakianClark2006PRC} are plotted in Fig.~\ref{fig:bcs_bec_sym}
at fixed values of the ratio $f = n_0/n$. (We recall that $n_0 = 0.16$
fm$^{-3}$ is the saturation density of symmetrical nuclear matter.)
The low- and high-temperature asymptotics of the gap function can be
fitted by the BCS-like relations:
$\Delta(T)=\Delta(0)-[2\pi c_1\Delta(0)T]^{1/2}\, {\rm
  exp}(-\Delta(0)/T)$
for $T\to 0$ and $\Delta(T) = 3.06\, c_2 [T_{c} (T_{c} -T)]^{1/2}$ for
$T\to T_{c}$, where $T_c$ is the critical temperature and $c_1$, $c_2$
are adjustable parameters. Note that the critical exponent remains
unchanged and equal the mean-field value 1/2. The values of the parameters yielding a
fit, $c_1 \simeq 0.2$ and $c_2\simeq 0.9$, deviate from the
predictions $c_1=c_2=1$ of BCS theory.  As a consequence, the value of
the ratio $\Delta(0)/T_{c} $ deviates from the BCS prediction of 1.76.
Clearly, the discrepancy depends on the inverse density measure $f$
and reflects the breakdown of the weak-coupling \textit{Ansatz}.

The transition from the BCS to the BEC regime can be traced in terms of 
several characteristic quantities. One such parameter is the 
ratio $ \Delta(0)/\vert \mu\vert $.  Using this measure, we can now 
infer from Fig.~\ref{fig:bcs_bec_sym} that the strong-coupling regime
characterized by $\Delta \gg \mu$ sets in for $f\ge 40$. For small
values of $f~(\sim 20)$, where $\Delta \sim \mu$, the system is in 
the transition region intermediate between BCS and BEC.  A second 
signature of the crossover from weak to strong coupling is the sign 
of the chemical potential of nucleons. Indeed, it changes sign 
for $f \simeq 80$, which is somewhat below the crossover density 
between weak-coupling and strong-coupling regimes deduced above. A 
third method, rather appealing physically, is direct comparison 
of the size of Cooper pairs, taking the ratio of the coherence length 
$\xi$ to the interparticle distance $d\sim n^{1/3}$.  In the BCS 
limit, one has by definition $\xi \gg d$; conversely, in the BEC 
limit $\xi \ll d$.  We use this criterion below in identifying the 
transition parameters. 

Finally, we note that in the context of dilute atomic gases, domains of 
weak and strong coupling are distinguished by the parameter $n|a|^3$, 
where $a$ is the scattering length. In nuclear matter, the strong-coupling 
regime was assigned to $f\ge 40$, which translates to 
$na^3 \simeq 0.6 < 1$ if we use $a_{np} = 5.41$~fm for the $n$-$p$ 
scattering length.  Thus one may conclude that symmetrical nuclear
matter is indeed in the strong coupling-regime at low densities. 

Another interesting feature shown in Fig.~\ref{fig:bcs_bec_sym} is the
asymptotic value of the chemical potential, $\mu =-1.1 $ MeV at
$f \to \infty$ or $n \to 0$. Its value is just half the binding energy
of the deuteron in free space. Formally, this result can be verified
by transforming the gap equation into an eigenvalue problem, in which
case it becomes a Schr\"odinger equation for a two-body bound state
described by the anomalous correlation function, with the chemical
potential as its energy eigenvalue.

We conclude that the BCS condensate of Cooper pairs in the $^3S_1$--$^3D_1$ 
state evolves into a BEC of deuterons under dilution of nuclear matter. 
The crossover is smooth, taking place without change of symmetry of 
the many-body wave function in the case of the isospin symmetrical nuclear
matter~\cite{Alm1993,BaldoLombardoSchuck1995,SteinSchenll1995,SedrakianClark2006PRC,Huang2010,JinUrban2010,Strinati2018}.

\subsubsection{BCS-BEC transition in the imbalanced case}
\label{sec:bec_bec}

How does the physics of the BCS-BEC crossover change under imbalance
between the populations of fermionic spe\-cies that pair?  As we have
seen, the condition of imbalance introduces some new and
unconventional phases in the BCS limit, and it is natural to ask about
their counterparts (if present) in the strong-coupling limit. This
problem has been addressed recently in a series of
papers~\cite{SteinHuang2012,SteinSedrakian2014,Stein2014JP}, in which
the equations for the gap and densities
[Eqs.~\eqref{eq:densities_imbalance} and \eqref{eq:gap2}] were solved
in a framework that provides for description of both the BCS phase and
its low-density BEC counterpart, as well as two unconventional phases
which may arise within a range of isospin asymmetries.  The FFLO phase
was chosen as a representative for phases with broken space symmetries
and the collection of phases was supplemented by the heterogeneous
phase in which the normal fluid and superfluid occupy separate spatial
domains.

Before discussing the phase diagram containing these phases, we survey
the intrinsic properties of the BCS-BEC transition\footnote{While it
  is well-established that one deals with a crossover in the proper
  sense in the case of balanced systems, the imbalance does change the
  nature of transition. Therefore, we will use transition instead of
  crossover when dealing with imbalanced systems.} under isospin
imbalance~\cite{LombardoNozieres2001,SteinHuang2012,SteinSedrakian2014,Stein2014JP}.
These properties include primarily the Cooper-pair wave function, the
occupation probabilities of particles, the coherence length, and the
quasiparticle spectra. Their quantitative study provides additional
physical insight and understanding of how the system evolves from weak
coupling to strong coupling under isospin asymmetry.  We note that in
the case of phase separation, the only non-trivial phase is the
isospin symmetrical BCS phase. Therefore, its intrinsic features,
apart from heterogeneity, are identical with those of the standard BCS
theory and hence need not be addressed separately.

Recall that in ultracold atomic gases the imbalance is achieved by
trapping different amounts of atoms in different hyperfine states, and
the transition is achieved by varying their effective interaction
strength via the Feshbach mechanism. In contrast, in an extended
nuclear system, a BCS-BEC transition is induced by variation of its
density and the isospin asymmetry is fixed by the minimization of the
energy or the initial conditions, as e.g. in nuclear collisions. As a
result, the pairing interaction strength changes, in accord with
changes in the relevant energies for in-medium scattering of two
nucleons set by the Fermi energy of the system.  In consonance, the
density of states changes.  The BCS-BEC transition in the nuclear
system is therefore governed by the combination of these two effects.
In contrast to ultracold atoms, it cannot be manipulated at will.
\begin{figure}[tb]
\begin{center}
\includegraphics[width=8.4cm,height=5.9cm]{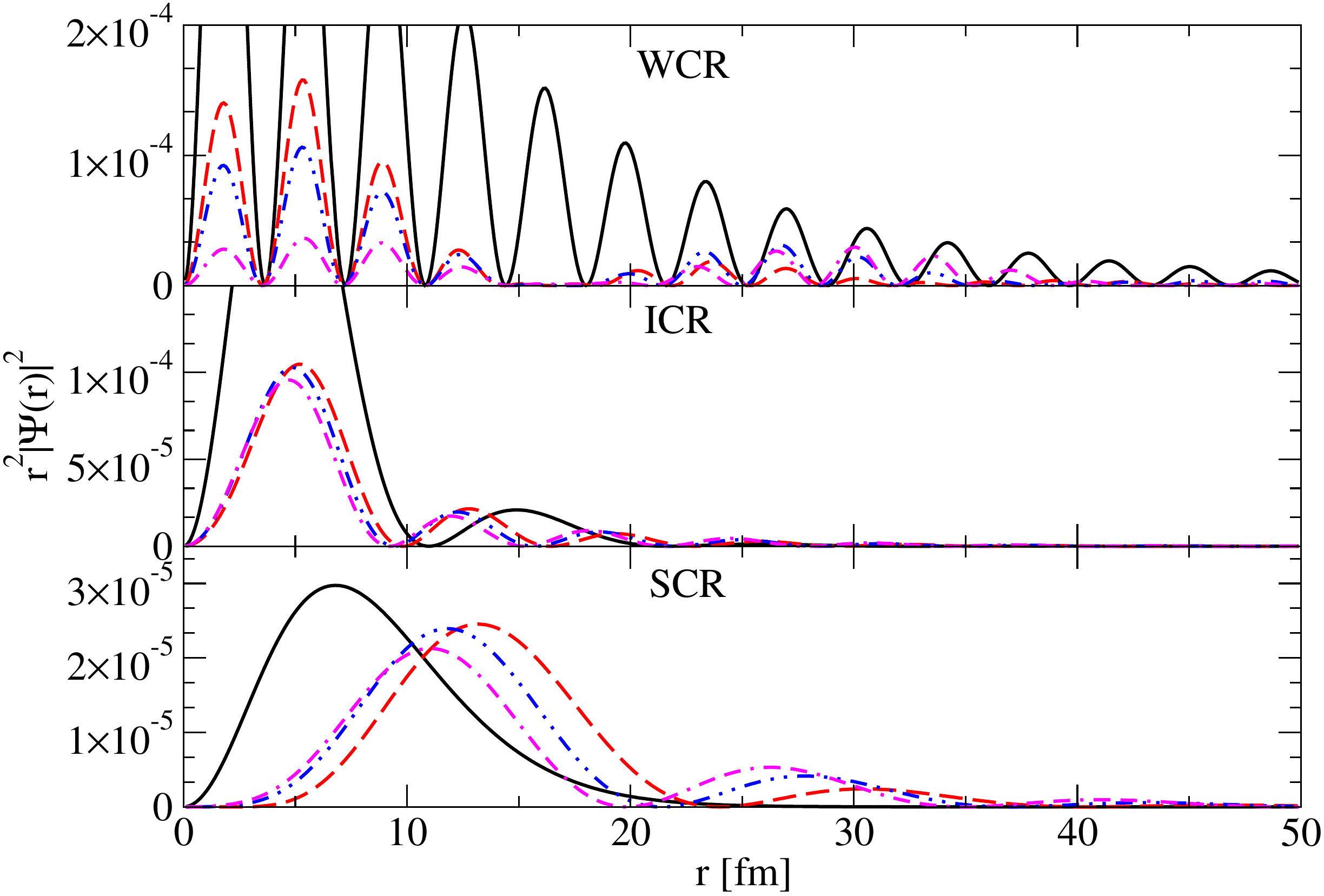}
\caption{ Typical dependence of $r^2|\Psi(r)|^2$ on $r$ in the three
  coupling regimes, weak coupling (WCR), intermediate coupling (ICR),
  and strong coupling (SCR), evaluated for asymmetries $\alpha = 0.0$
  (black solid line), 0.1 (red dashed line), 0.2 (blue dash-double-dotted
  line), 0.3 (magenta dash-dotted line).}
\label{fig:r_Psi}
\end{center}
\end{figure} 

To set the stage, we extract the kernel of the gap equation \eqref{eq:gap2}
\bea 
\label{eq:kernel}
K(p) &\equiv& \sum_{a,r}\frac{1-2f(E^a_r)}{4\sqrt{E_{S}(p)^2+\Delta^2(p,Q)}},
\eea
where we recall that $f(E^a_r)$ is the Fermi distribution function and
$E^a_r$ and $E_S(p)$ are given by Eqs. \eqref{eq:spectrum_Era} and 
\eqref{eq:E_SA}.
Physically, $K(p)$ is the momentum-space wave function of the Cooper
pairs, because it obeys a Schr\"odinger eigenvalue equation in strong
coupling.  In terms of its con\-fi\-gura\-tion-space image, we may write the
wave function of a Cooper pair as
\bea \label{eq:Psi} \Psi(\vecr) = \sqrt{N} \int
\frac{d^3p}{(2\pi)^3}
[K(\vecp,\Delta)-K(\vecp,0)]e^{i\vecp\cdot\vecr}. 
\eea 
Here $N$ is a constant determined by the standard normalization of 
a wave function to unity, and the value $K(\vecp,0)$ of the kernel in the normal 
state is subtracted to regularize the integral, which is otherwise 
divergent.  It is useful also to define the quantities 
\be \label{eq:xi}
\langle
r^2\rangle = \int d^3r\, r^2 \vert \Psi(\vecr)\vert^2, 
\quad 
\xi_{\rm rms} = \sqrt{\langle r^2\rangle},
\ee
where $\xi_{\rm rms}$ is the coherence length, \ie, the spatial
extension of a Cooper pair. This definition can be contrasted to 
the weak-coupling BCS analytical formula 
$\xi_a =  k_F/(\pi m^* \Delta)$. The root-mean-square
definition \eqref{eq:xi} allows one to extend the notion of the
coherence length into the strong-coupling regime; therefore, it can be
compared to the mean interparticle distance $d = (3/4\pi n)^{1/3}$
in the entire range of the BCS-BEC transition.

Figure \ref{fig:r_Psi} shows the integrand of $\langle r^2 \rangle$ in
Eq.~\eqref{eq:xi} as a function of radial distance $r$ at densities
representative for the three coupling regimes involved in the BCS-BEC
transition.  At densities corresponding to the weak-coupling regime
(labeled WCR, $\log_{10}\, n/n_0 =  -0.5$),  this
function (as well as the wave function $\Psi(r)$ itself, not shown)
has a well-defined oscillatory form of period $2\pi/k_F$, which
persists for multiple tens of fm.  The behavior of such a state is
commensurate with the long-range order inherent to BCS picture, where
the spatial extension of pairs, measured by the coherence length
$\xi$, is much greater than the interparticle distance $d$.  In
intermediate- and strong-coupling regimes (ICR and SCR, $\log_{10} n/n_0=
-1.5$ and $-2.5$, respectively), the wave
function becomes concentrated at the origin, possibly showing a few
oscillations indicative of the transition between limiting cases.
This behavior is descriptive of particles well-localized in space, a
distinctive characteristic of the BEC regime.

\begin{figure}[tb]
\begin{center}
\includegraphics[width=8.4cm,height=5.9cm]{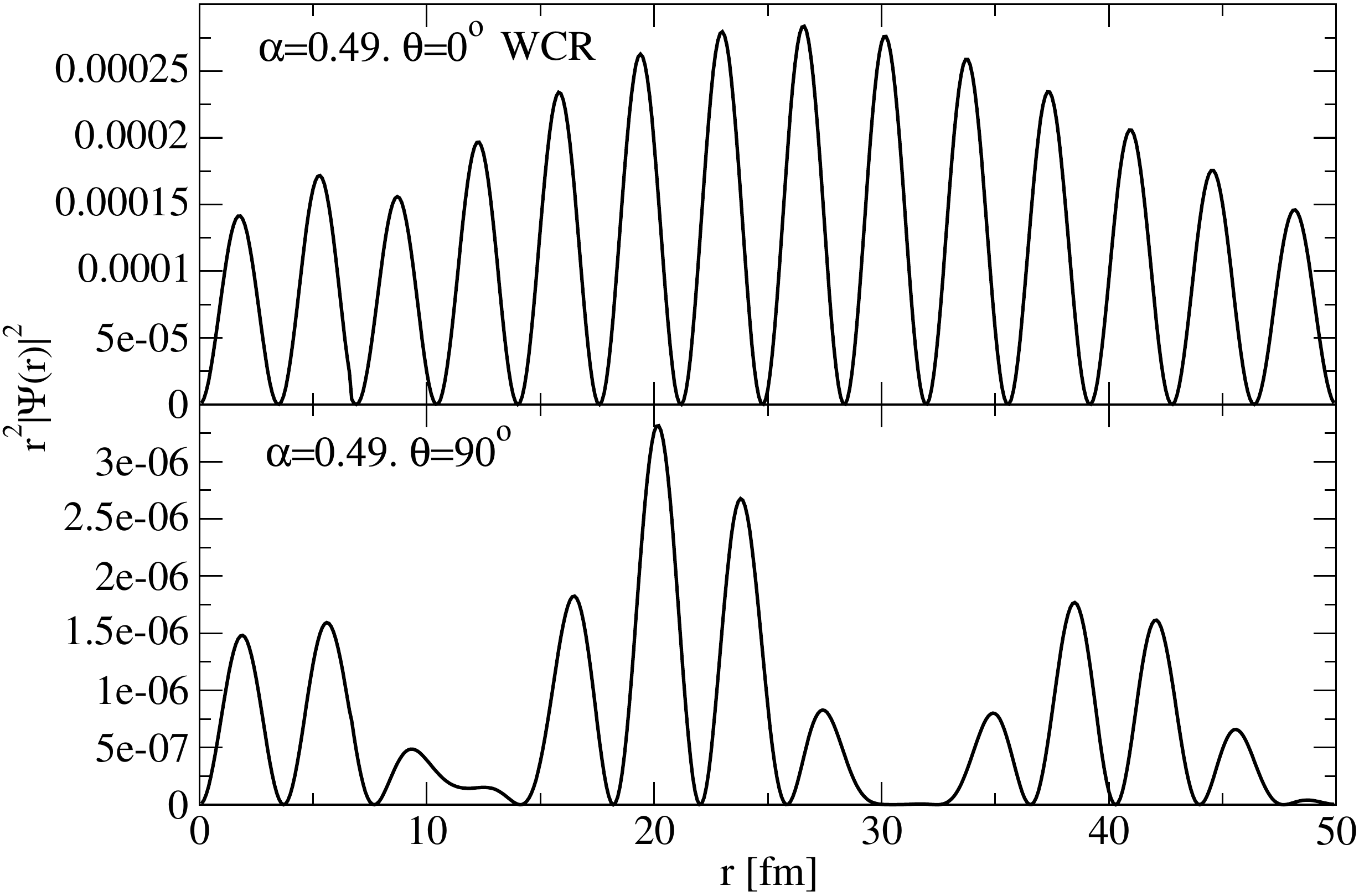}
\caption{Dependence of $r^2|\Psi(r)|^2$ on $r$ in the WCR
  coupling regime for $\theta = 0^o$ (upper panel) 
and $\theta = 90^o$ (lower panel) for the indicated asymmetry
 at which the FF phase is the ground state.}
\label{fig:Psi_loff2}
\end{center}
\end{figure}
In Fig.~\ref{fig:Psi_loff2} we show the same quantity $r^2|\Psi(r)|^2$
as in Fig. \ref{fig:r_Psi}, but 
in the FF phase at two different angles $\theta$, evaluated for an
asymmetry $\alpha = 0.49$ ($\delta \mu = 6.45$ MeV), where this phase
is the ground state of the matter with $\Delta = 1.27$ MeV and
$Q = 0.4$ fm$^{-1}$.  At $\theta = 0$ the perfect oscillatory behavior
of the BCS case is intact, with a slight modulation due to 
non-zero $Q$.  At $\theta = 90^{\rm o}$, the amplitude of
the oscillations is modulated by a second oscillatory mode with period
$2\pi/Q \sim 20$ fm,  in addition to the first mode having  period
$2\pi/k_F$. In the FF phase and for $\theta = 0^{\rm o}$, the term
$\propto \cos \theta $ renders the quasiparticle spectrum and 
therefore the Cooper pair wave function close to that expected 
from ordinary BCS theory.  In contrast, for
$\theta = 90^{\rm o}$, the term $\propto \cos \theta $ is zero and
marked differences are seen, notably damping of the
amplitude of oscillations.

\begin{figure}[tb]
\begin{center}
\includegraphics[width=8.4cm,height=6cm]{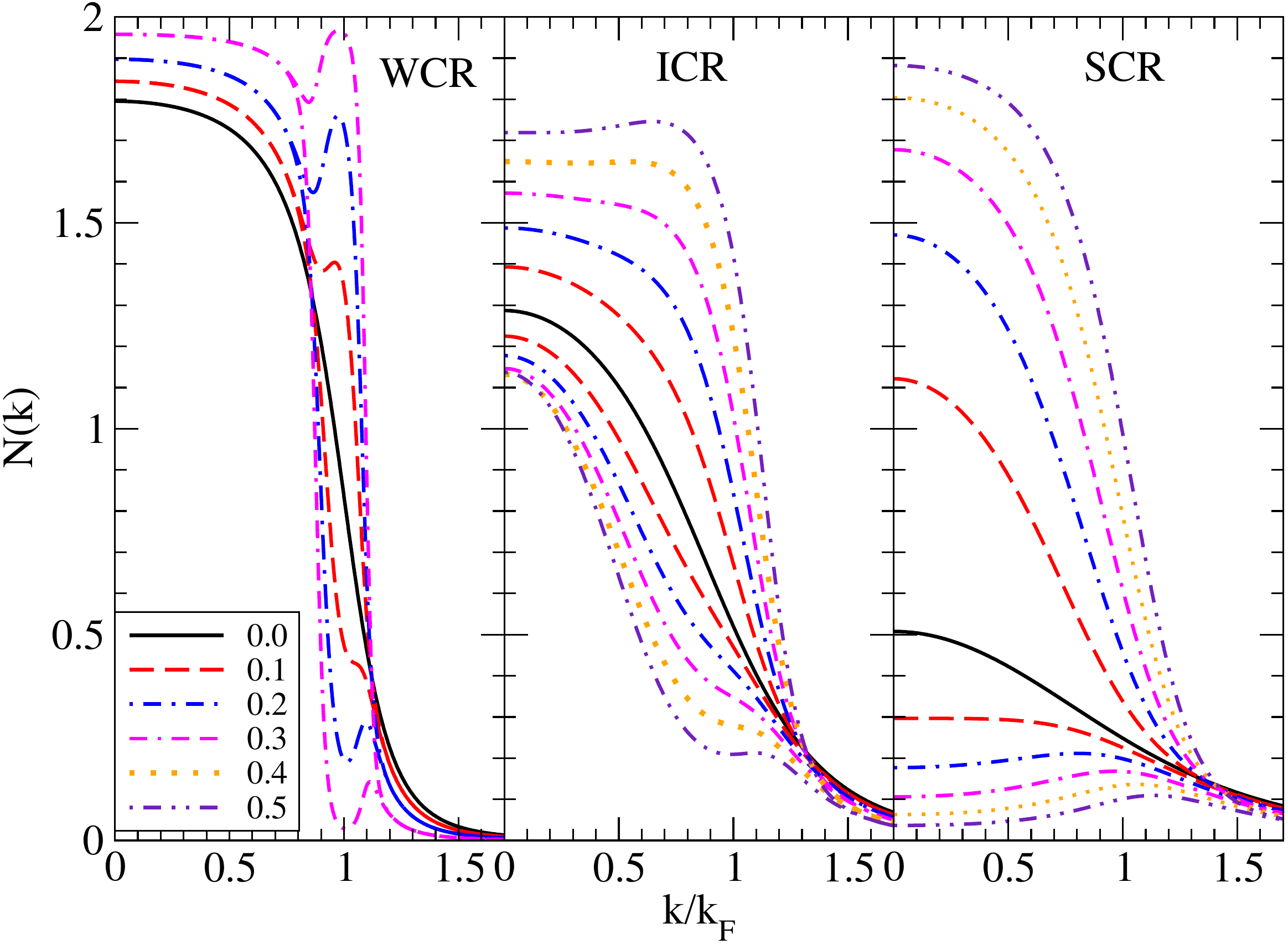}
\caption{ Dependence of the neutron and proton occupation numbers on
  momentum $k$ (in units of Fermi momentum) at densities
  $\log_{10} n/n_0= -0.5, -1.5$ and $-2.5$ corresponding to the three
  coupling regimes WCR, ICR, and SCR, respectively, for a range of
  asymmetries $\alpha$ indicated in the legend. The boundaries of the
  exclusion regions seen in the WCR (left panel) are smooth due to the
  non-zero value of temperature. }
\label{n_p_dens}
\end{center}
\end{figure}

Also of central interest are the occupation numbers $N_{n/p}(k)$ of
proton and neutron states, which are identified as the integrands of
Eq.~\eqref{eq:densities_imbalance}.  At zero temperature and in
unpaired matter, the functions $N_{n/p}(k)$ are discontinuous at the
Fermi surface.  Numerical results for balanced and imbalanced
superfluids are shown in the three coupling regimes of interest in
Fig.~\ref{n_p_dens}. A key feature of this figure that is universal
for imbalanced superfluids and nuclear systems is the appearance of a
``breach''
\cite{LiuWilczek2003,GubankovaLiu2003,Forbes2005,GubankovaSchmitt2006}
or ``blocking region''~\cite{LombardoNozieres2001} for large
asymmetries.  These designations refer to the entire expulsion of the
minority component (in this case the protons, $N_p=0$) from a region
around the Fermi momentum of the balanced system, accompanied by
maximal occupancy ($N_n/2=1$) of the majority component (here the
neutrons).  Examination of $N_{n/p}(k)$ in the FF phase for different
angles shows that for small enough $\theta$ the breach disappears and
the occupation numbers resemble each other in shape.  This reflects
the fact that for certain directions the effects of asymmetry are
mitigated by the non-zero $Q$.

The ICR (middle panel) is characterized by loss of the Fermi character
of the occupation numbers and vanishing of the breach. In addition,
for large enough $\alpha$, the occupation number of the minority
component becomes non-monotonic.  In the SCR (rightmost panel) one is
dealing with a BEC of strongly coupled pairs, in which the minority
component is reorganized at larger asymmetries into a distribution in
which the modes are populated starting from a certain nonzero value.
Consequently, the Fermi sphere of the weakly coupled BCS condensate
transforms into a shallow shell structure in the strongly coupled BEC.

Quasiparticle spectra provide further insight into the nature of
superfluid states.  In the balanced case, one finds a dispersion
relation with a minimum $E^+_{+} = E^+_{-} = \Delta$ for $k= k_F$ [the
spectra with lower $\pm$ indices being degenerate; see
Eq.~\eqref{eq:spectrum_Era}].  For non-zero asymmetries one has
$E^+_{\pm} = E_S \pm \delta\mu$, which induces a shift in the
minima. For protons the spectrum becomes {\it gapless}: no energy is
required to create excitations of two modes (say $k_1$ and $k_2$) for
which the dispersion relation intersects the zero-energy axis. This
phenomenon is referred as {\it gapless
superconductivity}~\cite{Abrikosov:Fundamentals}.  The momentum
interval $k_1\le k\le k_2$ is in fact where the ``breach'' in the
occupation of the minority component exists.

Finally, in the SCR, the balanced limit corresponds simply to a gas of
deuterons, and the dispersion relation has a minimum at the origin
that corresponds to the (average) chemical potential, which in the low
density limit tends to the value $-1.1$ MeV, as discussed above.
Imbalance changes the position of the average chemical potential
downwards and separates the quasiparticle spectra by an amount
$\delta\mu$.  Because there is unique minimum, the dispersion relation
crosses zero only once at a finite $k$. 

Upon introducing the FF phase, if one again considers different values
of the angle $\theta$, it turns out that for $\theta = 0^o$, two of
the four branches of quasiparticle spectra closely resemble the
spectrum of the ordinary BCS phase.  For large $\theta \le 90^{\rm o}$,
the dispersion relations supported by this phase are close to those of the
imbalanced BCS case, which implies strong suppression of pairing.
This behavior again points to the key mechanism by which the FF phase
enhances pairing -- the restoration of pairing correlations through an
improved overlap between the Fermi surfaces of neutrons and protons in
certain directions.

\subsection{Toward a complete phase diagram}
\label{sec:phase_diagram}

A central problem in the physics of imbalanced many-fermion systems 
is the concrete realization of their phase diagram in the parameter 
space spanned by the density (or in cold-atom physics by the scattering 
length), the temperature, and the degree of imbalance.  While the 
details of the phase diagram will certainly depend on the specifics 
of the interactions (contact vs. finite range, purely central or 
complicated by tensor and spin-orbit components), the ge\-ne\-ric 
structure of the phase diagram should be universal.  It is also 
expected to exhibit universal features across diverse systems 
including cold atomic gases, nuclear systems, and dense quark 
matter.

\begin{figure}[tb]
\begin{center}
\includegraphics[width=0.9\hsize]{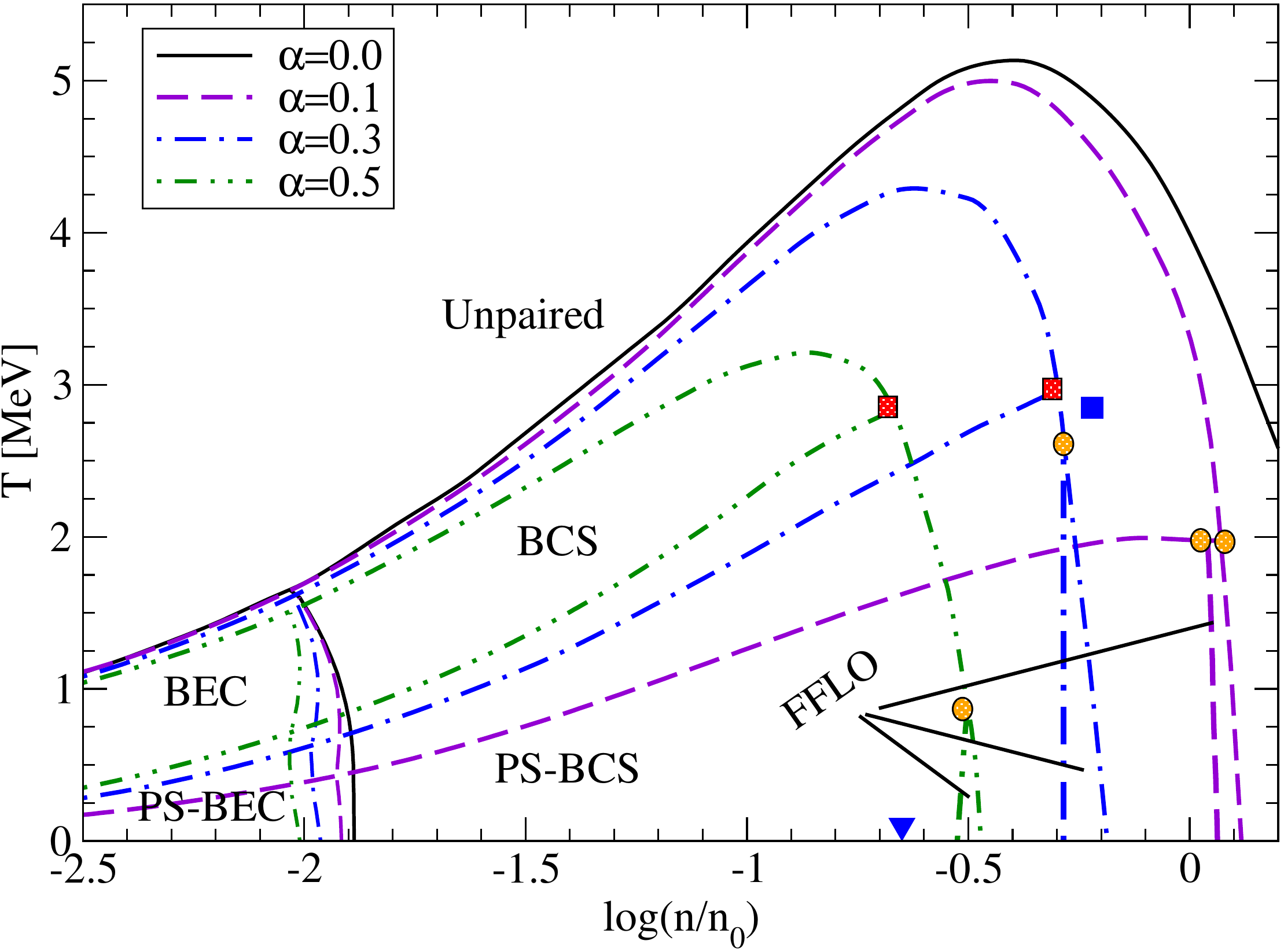}
\caption{Phase diagram of dilute nuclear matter in the
  temperature-density plane for several isospin asymmetries $\alpha$,
  where the density is normalized by the nuclear saturation density
  $n_0 $. For each fixed $\alpha$ there are two
  tri-critical points, the point bordering the FFLO phase being
  always a Lifshitz point~\cite{Hornreich1980}. In the special case
  $\alpha = 0.255$, these tri-critical points merge into a
  tetra-critical point for $\log_{10}(n/n_0) = -0.22$ and $T = 2.85$ MeV
  (square dot).  The FFLO phase completely disappears to the left of
  the point $\log_{10}(n/n_0) = -0.65$ and $T=0$ (shown by the
  triangle) for $\alpha = 0.62.$ The transition from BCS pairing to
  BEC, identified by the change of sign of the chemical potential
  $\bar\mu$, occurs on the vertical lines located around $\log_{10}(n/n_0)
  =-2$.  }
\label{fig:phasediagram}
\end{center}
\end{figure}

Results of a detailed study of the phase diagram of the imbalanced 
systems presented by generally asymmetric nuclear matter, which 
admits the four phases listed in Eq.~\eqref{eq:phases} and discussed 
above, were reported in a series of 
two papers~\cite{SteinHuang2012,SteinSedrakian2014}.  The resulting
phase diagram is shown in Fig.~\ref{fig:phasediagram}. The phases 
are arranged in the temperature-density plane, and the phase 
boundaries have been computed for several values of isospin asymmetry 
$\alpha$.
The  ge\-ne\-ric  structure of the phase diagram is as follows.  (a) Above 
the critical temperature $T_{c0}(n)$ for the normal-to-superfluid 
phase transition at $\alpha =0$, the nuclear matter is in the unpaired
phase. (b) At low temperatures, high densities, and moderate to large
asymmetries, the FFLO phase forms the ground state in the triangular
regions indicated in Fig.~\ref{fig:phasediagram}.  (c) Moving to
stronger couplings (lower densities), one finds the domain of phase
separation (PS) at sufficiently low temperatures. (d) The ordinary BCS
phase with isospin asymmetry intervenes at higher temperatures. 

As seen in Fig.~\ref{fig:phasediagram}, the extreme low-density
(strong coupling) limit features two counterparts of the weakly
coupled phases: first, the BCS phase at intermediate temperatures
evolves into the BEC phase of deuterons; second, the PS-BCS phase
transforms into the PS-BEC phase, in which the superfluid domains
contain a BEC of deuterons.  These transitions are indicated in the
figure by phase boundaries, although we should stress that the BCS-BEC
transition and the PS-BCS to PS-BEC transition are smooth crossovers.
The transition to the normal state and the phase transitions between
the superfluid phases are generally of second order and are indicated by
thin solid lines in Fig.~\ref{fig:phasediagram}.  The only exception
 is the  PS-BCS to FFLO phase transition, which is of first
order. We further
notice that at finite asymmetry there is a locus where three of the
four phases meet, corresponding to a tri-critical point. For each
fixed $\alpha$ there are two tri-critical points. Of these, one is a
Lifshitz point, since one of the adjacent phases represents a modulated
phase~\cite{Hornreich1980}. We observe also that at low temperatures,
the ordinary BCS-BEC crossover, which is a smooth crossover to an
asymptotic state corresponding to a mixture of a Bose condensate of
deuterons and a gas of excess neutrons, is replaced by a new type of
transition in which the fragmented superfluid contains a deuteron BEC
surrounded by a phase containing neutron-rich unpaired nuclear matter.

\subsection{Spin polarized neutron matter}
\label{sec:spin_pol_NM}

Another important scenario in which unconventional nuclear
superfluidity arises is spin-polarized neutron matter in strong
magnetic fields.  Strongly magnetized neutron stars, known as
magnetars, are characterized by surface fields of order
$B\sim 10^{15}$~G~\cite{ThompsonDuncan1995,TurollaZane2015} and may
feature fields that are larger by factors of a few in their
interiors~\cite{Chatterjee2015,Bocquet1995}. 
Magnetic fields of this magnitude can
suppress the pairing of neutrons and protons in the $S$-wave
state~\cite{SinhaSedrakian2015,SteinSedrakian2016neutron}, but the
mechanisms of suppression for charged and neutral condensates are
different. The proton $S$-wave pairing is quenched because of the
Landau diamagnetic currents of protons induced by the field; this
happens once the Larmor radius of a proton in the magnetic field becomes of
the order of the coherence length of the proton condensate. The
neutron pairing is suppressed when the $S$-wave neutron gap becomes of
the order of the Pauli-paramagnetic interaction of the neutron spin
with the magnetic field $B$.  The magnitude of this interaction is
$\vert \tilde {\mu}_N \vert B$, the neutron spin magnetic moment being
given by $ \tilde \mu_N = g_n (m_n /m_n^*)\mu_N$, where $g_n = -1.91$
is the neutron $g$ factor, $m_n^*$ its effective mass, and
$\mu_N= e \hbar /2m_nc$ the nuclear magneton.

Thus, the physics of neutron matter in strong magnetic fields would
parallel that of the $^3S_1$--$^3D_1$ condensate discussed in
Sec.~\ref{sec:Unconv_BCSBEC}, with the paramagnetic interaction
playing the role of the isospin asymmetry.  This possibility was
anticipated for the FFLO phase~\cite{Sedrakian2001LOFF}, and the
phase-separated state of neutron matter has been investigated in
detail in~\cite{GezerlisSharma2012}.  More recently, signatures of the
BCS-BEC crossover in spin-polarized neutron matter and the emergence
of dineutron correlations in the presence of a magnetic field have
received attention~\cite{SteinSedrakian2016neutron}, generalizing the
previous studies of this phenomenon in unmagnetized neutron
matter~\cite{Matsuo2006,MargueronSagawa2007,Isayev2008,Kanada-Enyo2009,AbeSeki2009,SunToki2010,Salasnich2011,SunSun2012}.

The critical field of unpairing of $S$-wave superfluidity in neutron
matter is of great phenomenological interest for the physics of magnetar
crusts.  The magnitude of the magnetic field $B$ in the crust and outer-core
regions of magnetars cannot be determined directly from observations.
One may anticipate that their interior fields are somewhat larger than
the surface fields $B\sim 10^{15}$~G based on the modeling of
magnetar equilibrium figures. Some magnetar models suggest that strong
toroidal $B$-fields are confined to the crust. Therefore, if local
fields are larger than the critical field for unpairing, neutron 
superfluidity will be absent.  Computation of the critical 
field~\cite{SteinSedrakian2016neutron} indicates that at a
temperature $T=0.05$~MeV characteristic of neutron stars, it
is of order $10^{16}$ G at the base of the inner crust, \ie, 
at $\log_{10}(n/n_0) = -3$, and increases up to $10^{17}$~G 
for densities one order of magnitude larger.

\begin{figure}[tb]
\begin{center}
\includegraphics[width=0.9\hsize]{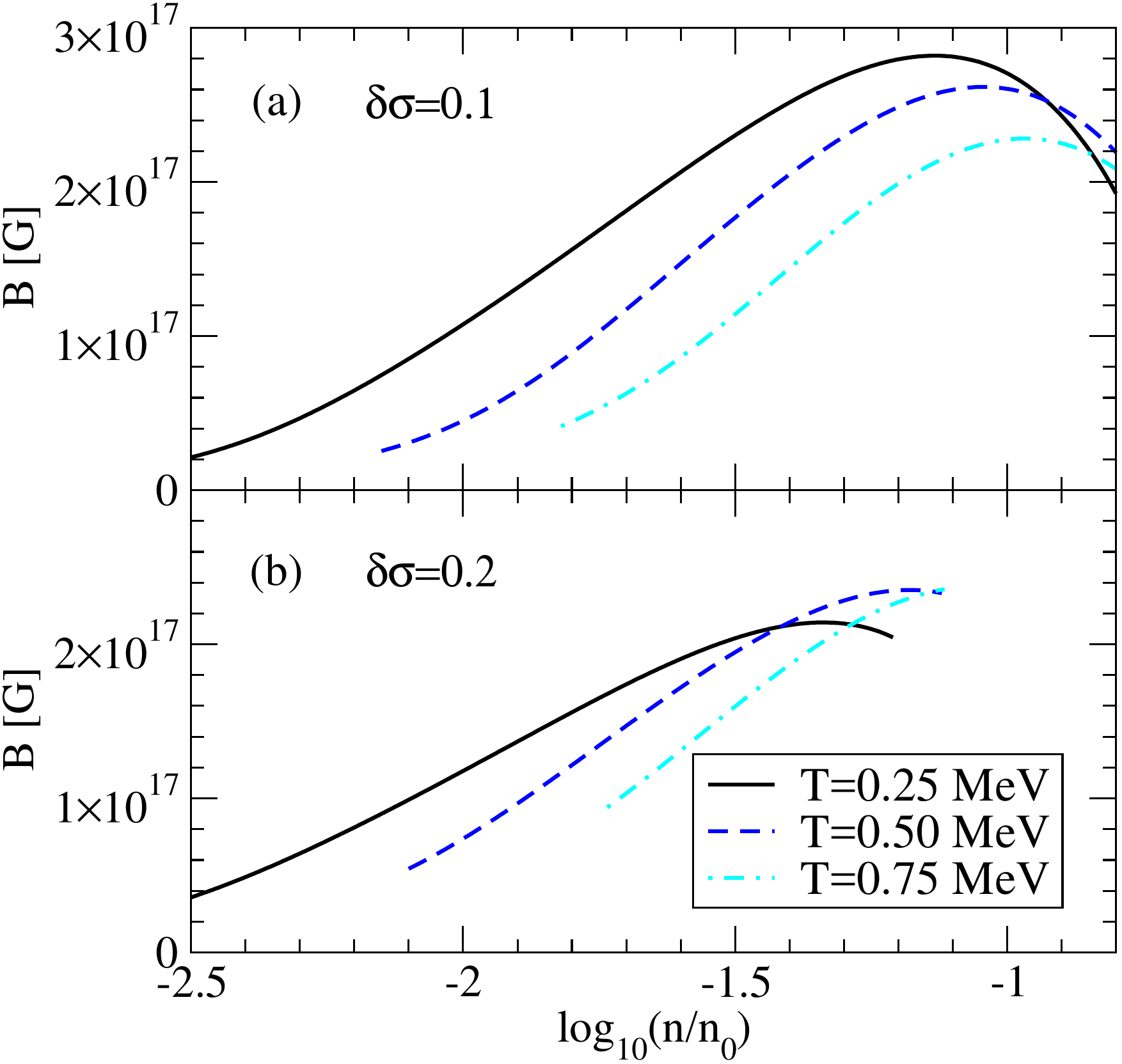}
\caption{Magnetic field required to create a specified spin
  polarization $\delta\sigma$ plotted as a function of density (normalized
  by $n_0$), at two fixed values of polarization, $\delta\sigma = 0.1$
  (panel a) and 0.2 (panel b), and for temperatures $T=0.25$ (solid
  line), 0.5 (dashed line), and 0.75 (dash-dotted line) MeV.}
\label{fig:spin_pol}
\end{center}
\end{figure}

Thus, in contrast to the analogous case of asymmetric nuclear matter, 
the interest in spin-polarized neutron matter lies primarily in 
the dependence of pairing on the magnetic field, rather on the 
spin polarization {\it per se}.  Figure~\ref{fig:spin_pol} shows the
magnitude of the field needed to generate a prescribed polarization in
neutron matter; it is seen that at low densities lower magnetic fields
are required.  It is also to be noted that the magnetic field required to 
produce a specified polarization increases with decreasing temperature. The
combined effect of variation of the gap and the polarization with 
density produces critical magnetic fields that are maximal at about
$n/n_0 = -1$, as discussed above.

\begin{figure}[t]
\begin{center}
\includegraphics[width=0.9\hsize]{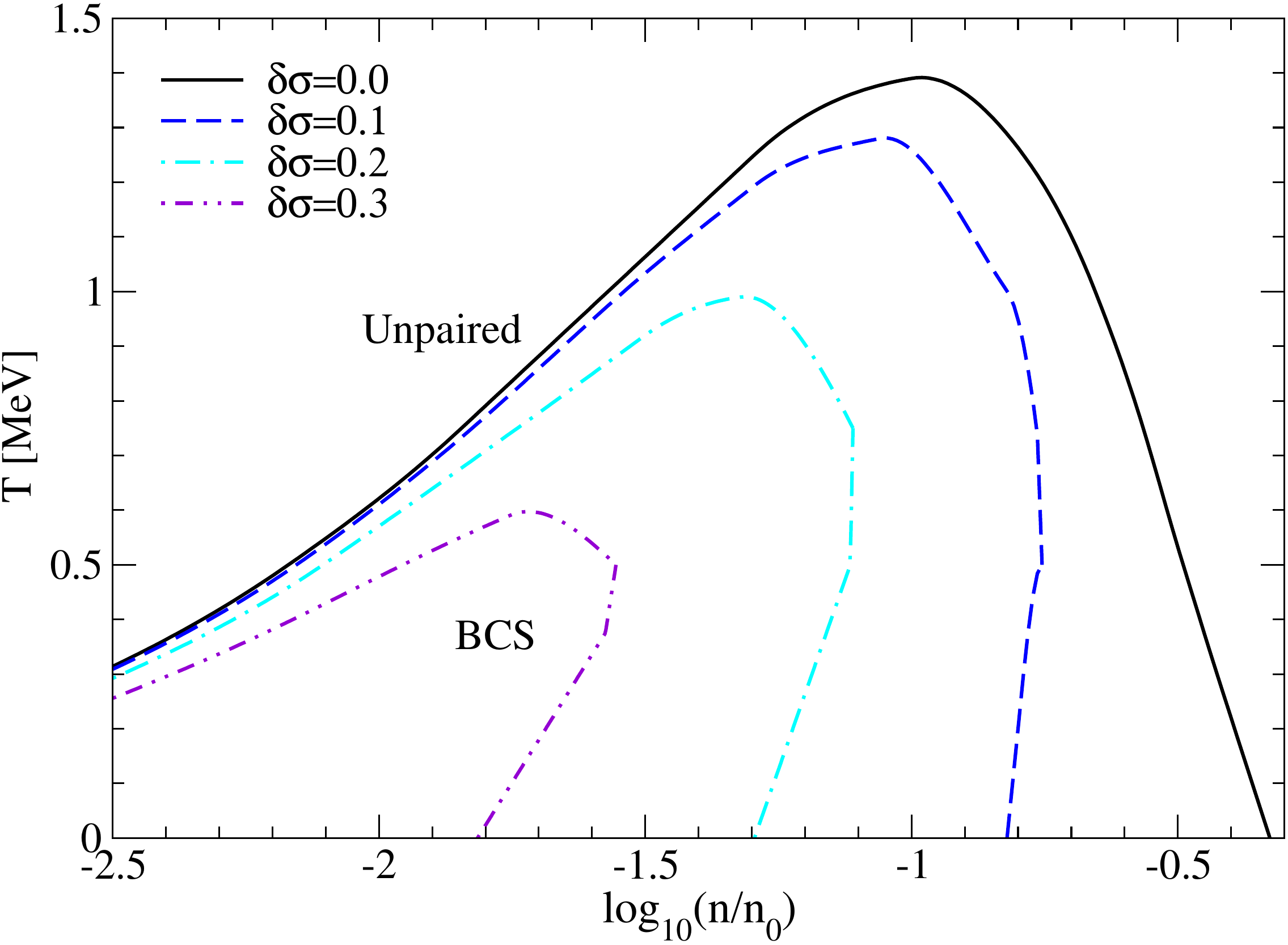}
\caption{ Phase diagram of neutron matter in the temperature-density
  plane for several values of the spin polarization $\delta\sigma$, showing
  the spin-asymmetric BCS phase and the unpaired phase at low and 
  high temperatures respectively. Note that the phase-separation lines
  have double-valued character for $\delta\sigma \neq 0$. 
}
\label{fig:neutron_phase_diagram}
\end{center}
\end{figure}

The phase diagram of neutron matter at fixed spin polarization,
defined as $\delta\sigma =
(n_{n\uparrow}-n_{n\downarrow})/(n_{n\uparrow}+n_{n\downarrow})$ with
$n_{n\uparrow/n\downarrow}$
being the densities of spin-up and down neutrons,  is displayed 
in Fig.~\ref{fig:neutron_phase_diagram}. It resembles the phase diagram
of asymmetric nuclear matter shown in Fig.~\ref{fig:phasediagram}, but
it contains only the BCS and unpaired phases.   The possibility of the 
FFLO phase filling the low-temperature and high-density pockets formed 
by the critical lines for $\delta\sigma \neq 0$ requires further study. 

Analysis of the principal intrinsic features of the spin-polarized
neutron condensate --  which include the Cooper-pair wave function,
occupation numbers, and quasiparticle spectra -- shows that their
behavior runs parallel to those of asymmetrical nuclear matter,
already discussed in some detail.  In particular, one finds
~\cite{SteinSedrakian2016neutron} that the Cooper-pair wave functions
and the function $r^2\vert \Psi(r)\vert^2$ exhibit oscillatory
behavior characteristic of long-range order, the wave vector
of the oscillations being $2\pi/k_{Fn}$, where $k_{Fn}$ is the 
neutron Fermi wave number.  The quasiparticle occupation numbers 
likewise display a breach around the Fermi momentum $k_F$, which 
is most pronounced in the high-density and low-temperature limit 
where the matter is highly degenerate. Furthermore, the feature 
of gapless superfluidity is again observed in this case: 
at large polarizations, the energy spectrum of the minority-spin 
particles crosses the zero-energy level, where modes can be excited 
without any energy cost.
\begin{table}[t]
\begin{tabular}{ccccccc}
\hline 
$\textrm{log}_{10}\left({n}/{n_0}\right)$ & $k_{Fn}$ &
                              $\Delta$
  & $m^\ast/m$ & $\mu_n$ & $d$ & $\xi_{\rm rms}$   \\
 & [fm$^{-1}$] & [MeV] & & [MeV] & [fm] & [fm]   \\
\hline 
   $-1.0$ & 0.78 & 2.46 & 0.967 &12.94 & 2.46 & 4.87 \\
$-1.5$ & 0.53 & 1.91 & 0.989 & \,\,\,5.65 & 3.61 & 3.55 \\
$-2.0$ & 0.36 & 1.07 & 0.997 & \,\,\,2.49 & 5.30 & 2.36 \\
\hline\\
\end{tabular}
\caption{ Parameters of the $^1S_0$ condensate for $T=0.25$ MeV 
  and $\delta\sigma =0$ at selected values of the total particle density 
  $n$ (in units of $n_0$).  Other 
  table entries are the Fermi momentum $k_F=(3\pi^2 n)^{1/3}$, 
  pairing gap $\Delta$, effective mass (in units of the bare mass), 
  chemical potential $\mu_n$, interparticle distance $d$, and 
  coherence length $\xi_{\rm rms}$. 
}
\label{table_neut}
\end{table}

As argued above, neutron matter exhibits the features of a BCS-BEC
crossover, although at asymptotically low densities two neutrons are
not bound. Evidence of this transition is clearly seen in 
Table~\ref{table_neut} by comparing (a) the first row (high-density 
entry) showing $\xi_{\rm rms}/d > 1$ and $\Delta/\mu \ll 1$, which 
is characteristic to the BCS phase, with the third row (low-density
entry), where $\xi_{\rm rms}/d < 1$ and $\Delta/\mu \sim 1$. 
The behavior at low density is interpreted as a precursor state
to a (non-existent) dineutron BEC.

\section{Astrophysical manifestations of 
pairing in neutron stars}
\label{sec:astro}

\subsection{Pairing patterns in neutron stars}
\label{sec:pairing_patterns}

So far we have concentrated on the microscopic physics of superfluidity 
in extended nuclear systems in a general setting, over broad ranges of 
density and  temperature.  Our
next task is to adapt these considerations to the conditions
prevailing in neutron stars.  The matter in neutron stars is
characterized by conserved charges, specifically baryon number and
electrical charge. Also, at the microscopic level, the interior of a
neutron star is in approximate weak equilibrium.  This
condition, combined with charge conservation, determines the phase and
composition of neutron-star matter at any given depth in the star
(cf.~Sec.\ref{sec:isospin_asym}).\footnote{More detailed expositions
of the composition and structure of compact stars can be found,
e.g., in the texts \cite{ShapiroTeukolsky1983,Glendenning_book} 
and \cite{weber_book} and recent review
articles~\cite{Blaschke2018,Oertel2017RvMP,2004Sci...304..536L,Page2013,Sedrakian2007PrPNP,AlfordSchmitt2008RvMP}.}
\begin{figure}[t]
\begin{center}
\includegraphics[width=9cm,height=7cm]{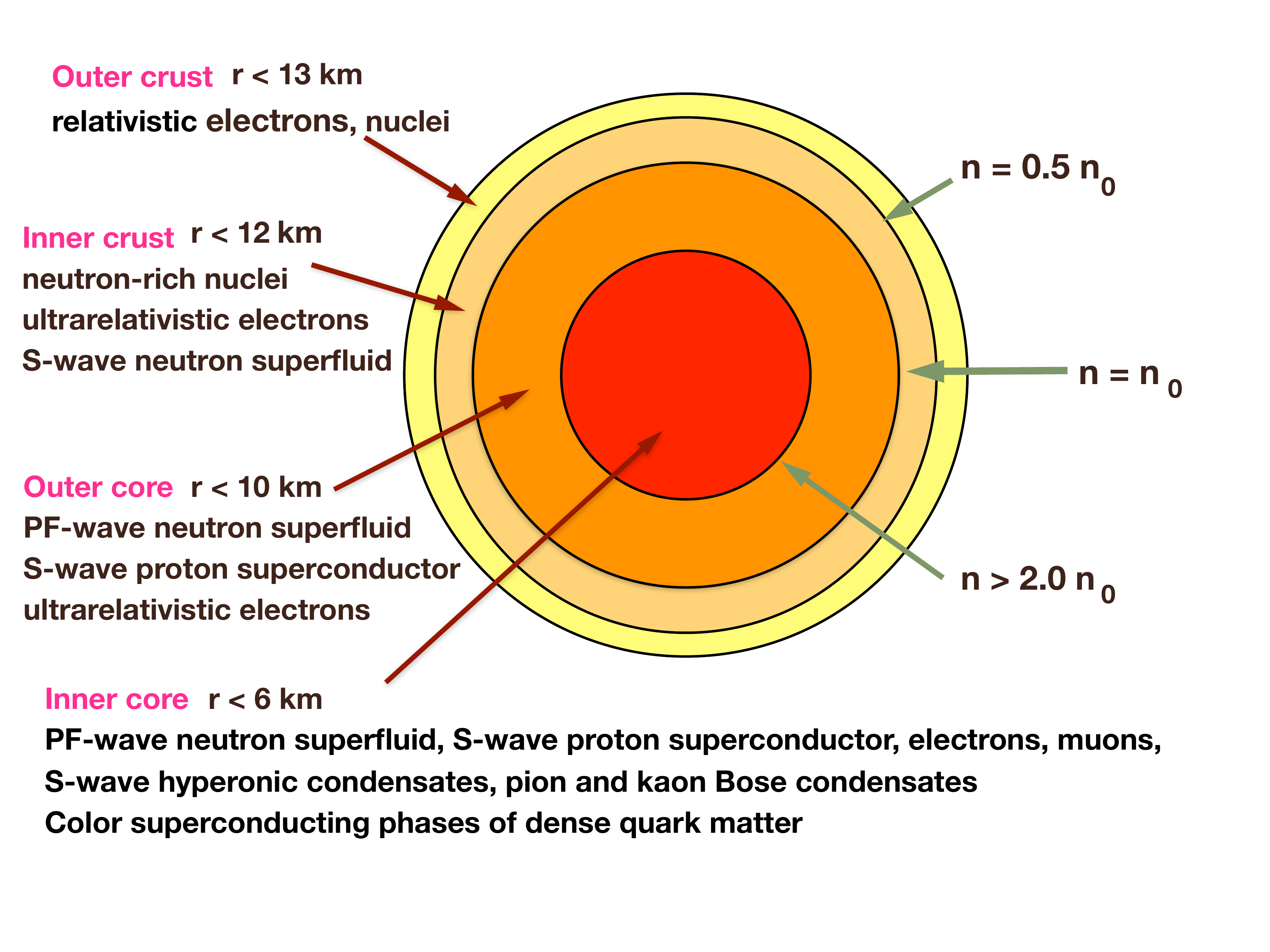}
\caption{Schematic interior of a $M = 1.4 M_{\odot}$ mass neutron
  star. Values of the radial coordinate $r$ enclosing each region of
  star are indicated along with the transition density in units of
  $n_0$. The particle content and possible condensates of each phase
  are indicated as well; note that in the inner core hadronic and quark
  phases are mutually exclusive in any given unit of the volume,
  although they can co-exists in a form of a mixed phase. (The figure
  is not to scale, low-$\rho$/large-$r$ domains being strongly
  expanded.)  }
\label{fig:NS_map}
\end{center}
\end{figure}

Figure~\ref{fig:NS_map} shows a schematic cross-section of the
interior of a neutron star of mass $M=1.4 M_{\odot}$. Among the
multitude of possible phases occurring at different densities, the
locations of nucleonic superfluid and superconducting phases are
indicated in order of increasing depth in the star, along with
possible hypernuclear condensates and color-superconducting phases 
in the inner core of the star.

We now discuss briefly the key phases that exist inside a neutron
star, beginning just below its surface and moving toward the center.
We will assume that matter is in its lowest energy state. This
  might not be the case in a number of contexts, such as in the case of
  accreting neutron stars.  At densities $\rho \simeq 10^6$ g
cm$^{-3}$, neutron-star matter is fully ionized, being composed of
ions of $^{56}$Fe and relativistic electrons.  Such matter, much like
that in a white dwarf but involving heavier isotopes, solidifies below
a melting temperature $T_m\sim 10^9-10^{10}$~K, such that we
anticipate a solid phase in mature neutron stars.  We note that this
solid phase may have a very thin blanket (several cm in total) made up
of lighter elements including H, He, etc., in ionized, atomic, or
molecular form.  The composition of this enveloping blanket can, in
principle, be extracted from observations of thermal radiation
from the surface of the star.

Under the constraints of neutrality and equilibrium, the matter
becomes more neutron-rich as the depth and hence the density
increases.  The outer crust of the star, which spans the density range
$10^6 \leq \rho \leq 10^{11}~{\rm g~cm}^{-3}$, is made up of a sequence
of nuclei, their neutron fraction increasing with depth, a
characteristic sequence being $^{62}$Ni,
$^{86}$Kr,
$^{84}$Se,
$^{82}$Ge,
$^{80}$Zn,
$^{124}$Mo, $^{122}$Zr, $^{120}$Sr, and their neutron-rich isotopes.
The lattice formed by these nuclei may not be perfect and may contain 
nuclei with mass numbers different from those predicted for the ground 
state. Nucleonic superfluidity in the outer crust exists inside
the individual bound finite nuclei and can be described using standard 
methods, such as Hartree-Fock plus BCS or Hartree-Fock-Bogolyubov 
theories~\cite{PearsonPhysRevC2011,Grill2011,PastoreMargueron2013,Goriely2016,Pastore2017JPhG}. 

At a density $\rho \simeq 4\times 10^{11}$ g cm$^{-3}$, neutrons drip
out of the nuclei and start filling continuum states. Consequently a
degenerate neutron gas occupies the space between the nuclear
clusters.  The resulting phase, which features neutron-rich nuclei
immersed in an electronic background and a dilute neutron gas,
occupies the inner crust of a neutron star and extends in density up
to half the saturation density of symmetrical nuclear matter,
$\rho_0 = 2.8\times 10^{14}$ g cm$^{-3}$.  Here one finds sequences of
nuclei that are neutron-rich isotopes of Zr and Sn, which have proton
numbers $Z=40$ and $50$ respectively.  As a rule, the mass number of
the nuclei increases with density and lies in the range
$100 \le A \le 1500 $. At the bottom of the inner crust the spherical
nuclei are replaced by aspherical ones which form the so-called
``pasta'' phases of neutron stars, which were first proposed
in~\cite{Ravenhall1983PhRvL,Oyamatsu1984PThPh,Hashimoto1984PThPh}, and
further studied in~\cite{Lorenz1993PhRvL,Pethick1995,WatanabeIida2000}. Further
significant advances on the structure and transport properties of these phases 
were achieved in the past decade mainly on the basis of either
molecular dynamics simulations or (time-dependent) Hartree-Fock
density functional
approaches~\cite{Newton2009PhRvC,Schneider2013PhRvC,Horowitz2015PhRvL,Schuetrumpf2015,Fattoyev2017PhRvC,Caplan2018PhRvL,Schuetrumpf2014PhRvC,Schuetrumpf2016,Schuetrumpf2014}.
There exist some parallels between the pasta phases and terrestrial
liquid crystals~\cite{Pethick1998PhLB,Kobyakov2014PhRvL}.  We will not
discuss these interesting topics here, see the
reviews~\cite{Caplan2017RvMP,Blaschke2018}.  Since the unbound
low-energy neutrons tend to fill a Fermi sphere and their interaction
in the $^1S_0$ channel is attractive, they form a superfluid, which is
the main object of applications of the theories discussed in
Sec.~\ref{sec:Methods}.  The neutron condensate in the inner crust
plays a fundmental role in theories of neutron-star
cooling~\cite{Page2013,Yakovlev2001,weber_book,YakovlevPethick2004} as
well as in theories of their rotational dynamics (for a recent review
and further references see ~\cite{Haskell2017}).

The inner crust terminates with a first-order phase transition at the
point where the clusters merge together to form a continuum. The new
phase, occupying the outer core of the star, is a fluid mixture of 
neutrons ($n$), protons ($p$), and electrons ($e$), along with
muons ($\mu$) appearing at somewhat higher densities. 

The actual phase structure of matter that exists in the densest part
of the core ($\rho > 2 \rho_0$) is uncertain. A number of conjectured
phases have been explored.  One possibility is the appearance of
hyperons in matter, which has attracted much attention in recent years. The
mechanism driving the onset of hyperons is the Pauli exclusion
principle: once the Fermi energies of neutrons and electrons
(including their rest mases) become of the order of the in-medium
masses of $\Sigma^{\pm,0}$, $\Lambda$, or $\Xi^{\pm,0}$ hyperons,
their formation becomes energetically more favorable, with increasing
density, than further increase in the Fermi energies of the neutron
and electron constituents.  If the hyperons experience mutual
attractive interactions, they will form condensates by the same BCS
mechanism that operates for non-strange baryons -- a possibility that
will be considered below.

Depending on the equation of state of the matter, the central densities in
the most massive neutron stars can lie in the range 5-10 times $\rho_0$,
and deconfinement of quarks becomes a plausible outcome.
Deconfinement may set in after the hyperons appear in matter or even 
before, depending on the unknown density for the deconfinement transition. 
We will not discuss quark color superconductivity in this review, although
it may have profound implications for neutron-star phenomenology; the 
interested reader is referred to the reviews~\cite{AlfordSchmitt2008RvMP,AnglaniCasalbuoni2014RvMP,RajagopalWilczek2001}.
Medium-modification of meson properties may also lead to their
Bose-Einstein condensation and, consequently, their superfluidity.  
For pions, such condensation can arise through an instability of 
the particle-hole nucleonic excitations in the medium having 
pion quantum numbers.  The interplay between pion condensation 
and nucleonic pairing has been covered extensively in the literature; see the
reviews~\cite{1981PThPh..65.1333T,1993PThPS.112...27T}.

Turning to the superfluid phases within neutron stars, we first
concentrate on neutron and proton condensates at low (partial)
densities (respectively in the inner crust and outer core), for which
more reliable computations can be made.  We recall the uncertainties
involved in the many-body methods outlined in Sec.~\ref{sec:Methods},
in particular the fact that many-body calculation of pairing gaps are
usually simplified by adopting the decoupling approximation, \ie, by
computing the single-particle energies in the normal state.

\begin{figure}[t]
\begin{center}
\includegraphics[width=8.5cm,height=7cm]{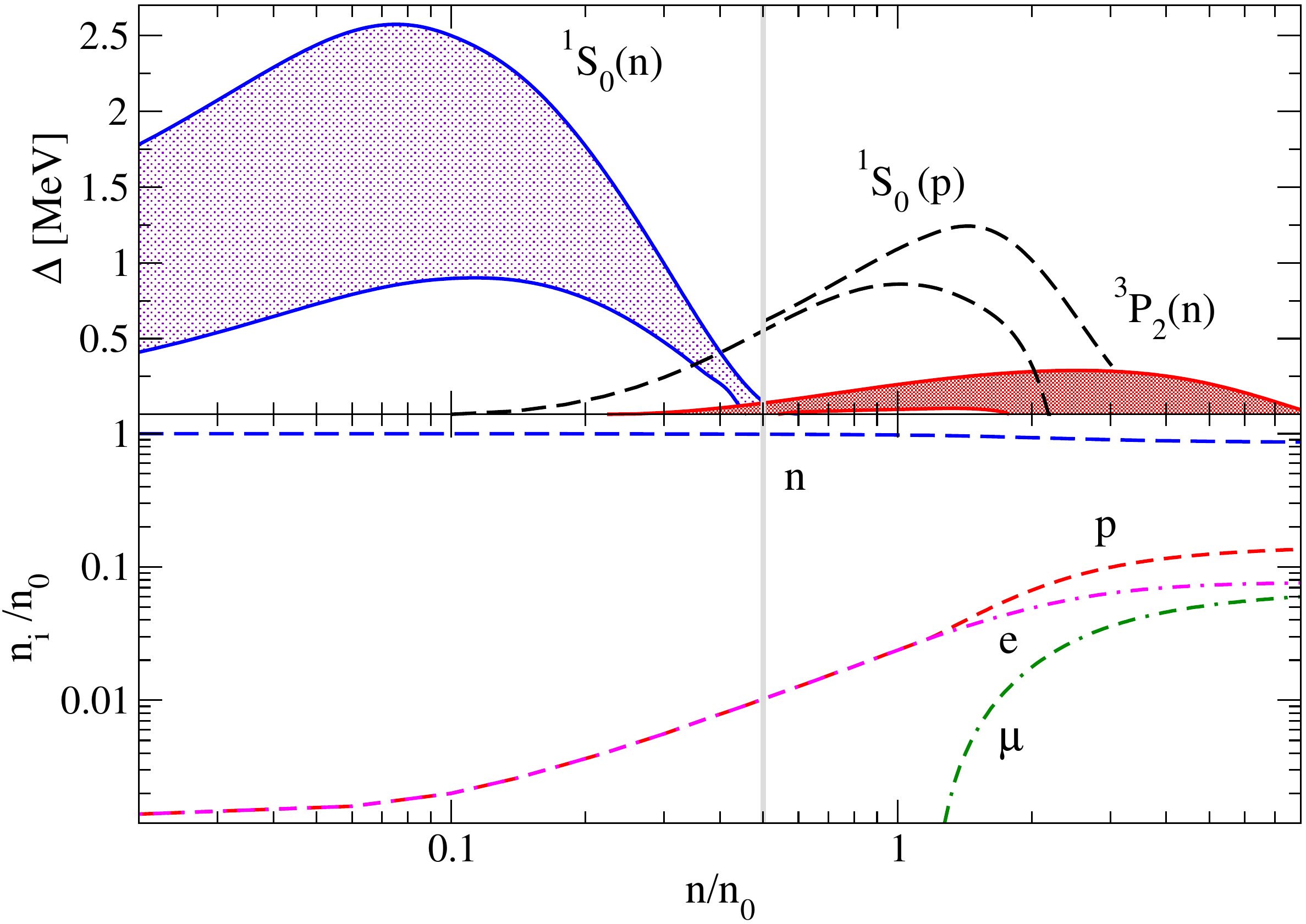}
\caption{ Upper panel: Dependence of pairing gaps for neutrons
  ($^1S_0$ and $^3P_2$--$^3F_2$ channels) and for protons ($^1S_0$
  channel) on baryonic density in units of nuclear saturation density.
  The range of the neutron $S$-wave gap has an upper boundary given by
  the result of~\cite{FanKrotscheck2017}. Its lower boundary is given
  by that of \cite{WambachAinsworthPines1993}. The range of the proton
  gap has an upper and lower boundaries given by the results of
  \cite{BaldoSchulze2007} and \cite{BaldoCugnon1992}.  The neutron
  $^3P_2$--$^3F_2$ gap range has a lower and upper boundary given by
  the results of \cite{BaldoElgaroy1998_Pwave} and \cite{Ding2016}
  respectively.  Note that the neutron pairing gaps were obtained
    for pure neutron matter, whereas those for protons were obtained
    in $\beta$-equilibrated neutron star matter.  Lower panel:
  Composition of the core of a neutron star at $T=0$, as constructed
  from a relativistic density functional~\cite{ColucciSedrakian13},
  which was used to map the pairing gaps to the net baryon density $n$
  of neutron star matter.  Note that this composition differs
    from the ones used to compute the proton pairing gaps, but the
    error introduced by this discrepancy should be insignificant
    compared to the uncertainties arising from the pairing force. The
  particle abundances $n_i/n$ with $i\in \{n, p, e, \mu\}$ are
  normalized by the net baryon density $n$, which is measured in units
  of the nuclear saturation density $n_0$. The vertical line shows the
  crust-core transition density.  }
\label{fig:npe_diagram_gaps}
\end{center}
\end{figure}
Figure~\ref{fig:npe_diagram_gaps} displays the neutron and proton
pairing gaps as functions of the baryon density, for the composition
shown in the same figure. The bands for neutrons are chosen to show
the range of reasonable values with boundaries corresponding to actual
computations; the upper boundary corresponds to the CBF calculations
of \cite{FanKrotscheck2017}, whereas the lower boundary corresponds to
Fermi-liquid computations of \cite{WambachAinsworthPines1993}.
Neutron $S$-wave superfluidity occurs at lower densities corresponding
to the neutron-star crust. It is described essentially by the result
for pure neutron matter, because at these densities the protons are
confined in the nuclear clusters. However, the coupling between the
neutron fluid and crustal phonon modes or band structure induced by
the lattice can alter the value of the gap (see
Sec.~\ref{sec:collective_modes} for a discussion).  The $^1S_0$
neutron gap closes in the vicinity of the crust-core interface. In the
outer core, where the proton density becomes comparable to the neutron
density in the inner crust, the protons also pair in the $^1S_0$
state, with a gap of order of 1~MeV. The upper and lower
boundaries for protons are shown for computations based on the
BCS theory with single-particle renormalization from Brueckner
theory. The upper boundary is obtained when the three-body (3B or 3N)
force is neglected in the solution of the gap equation and the
equation of state of matter~\cite{BaldoCugnon1992}, whereas the lower
boundary is obtained when the three-body forces are included in
both~\cite{BaldoSchulze2007}.  Neutrons in the core pair with much
smaller gaps of order 100 keV.  The gap obtained in SCGF theory
\cite{Ding2016} provides the lower boundary. The upper boundary is
based on BCS theory with screening corrections and single-particle
spectrum derived from Brueckner theory~\cite{BaldoElgaroy1998_Pwave}.
Stronger suppression by polarization effects due to spin dependence of
the effective interaction in neutron matter has been found in
\cite{SchwenkFriman2004}. The $P$-wave and $P$-$F$ wave gap computations
will be assessed in the next subsection.
\begin{figure}[bt]
\begin{center}
\includegraphics[width=0.99\hsize]{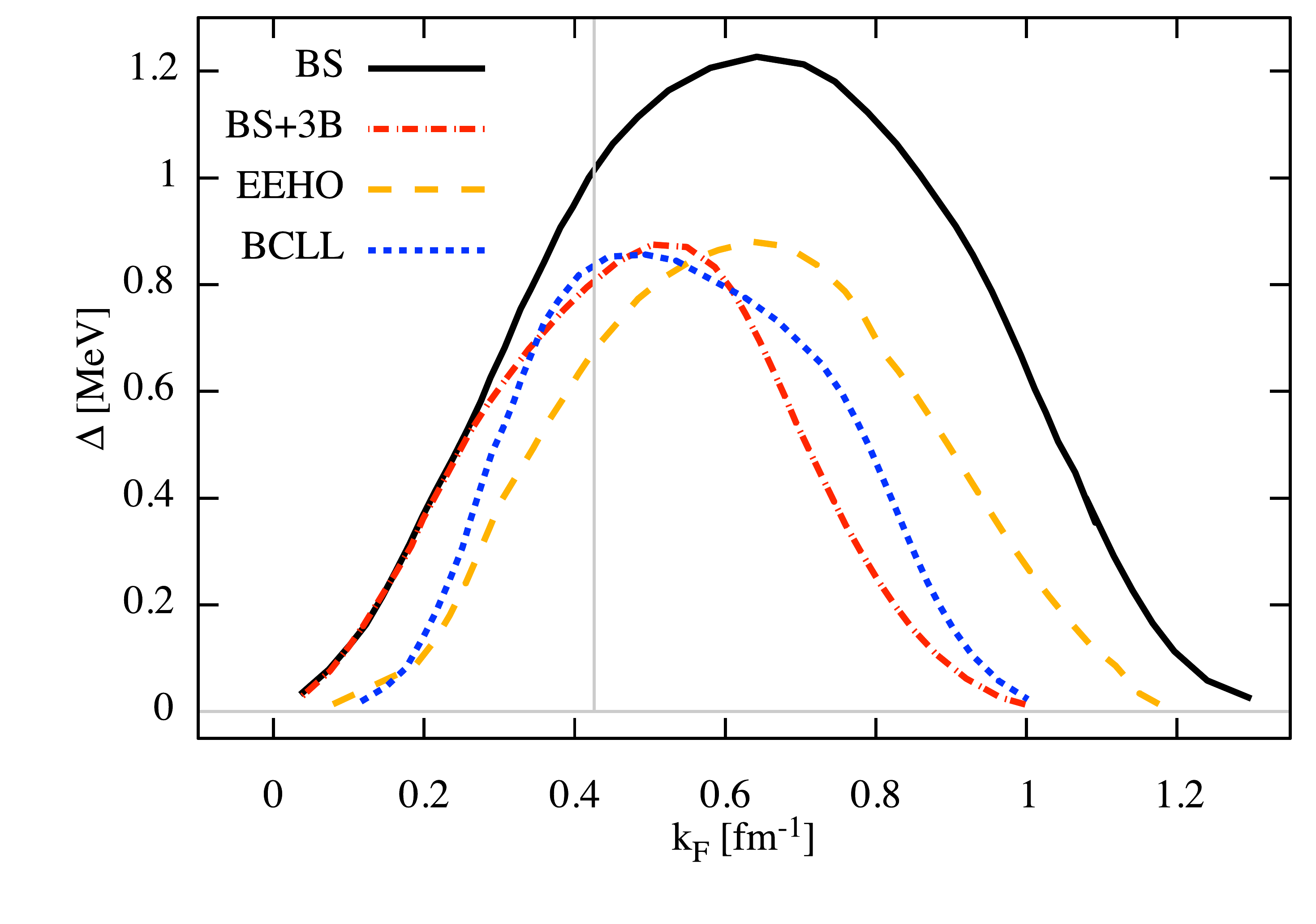}
\caption{Dependence of the $^1S_0$ pairing gap of protons on their
  Fermi momentum from BCS theory with phase-shift equivalent interactions: 
  the BS result is based on the Argonne $V_{18}$ two-body interaction, whereas 
  BS+3B includes in addition an adjusted version of the Urbana UIX three-body 
  interaction~\cite{BaldoSchulze2007}; the EEHO result was obtained with the 
  Bonn two-body interaction~\cite{Elgaroy1996_Swave};
  BCLL refers to an earlier computation with the Argonne two-body
  interaction~\cite{BaldoCugnon1992}. The vertical line indicates the
  Fermi momentum corresponding to the crust-core transition. }
\label{fig:gaps_protons_overview}
\end{center}
\end{figure}

Figure~\ref{fig:gaps_protons_overview} shows the pairing gaps of
protons obtained with the phase-shift equivalent Argonne
\cite{BaldoCugnon1992,BaldoSchulze2007} and Bonn
potentials~\cite{Elgaroy1996_Swave}. The computations were carried in
the BCS approximation by using the bare two-body or two- plus
three-body interactions. The single-particle energies in all
computations were obtained from the Brueckner-Hartree-Fock theory of
nuclear matter. These were then used to obtain the effective mass of
the quasiparticles in the gap equation.  The results BS and BCLL were
obtained with the Argonne $V_{18}$ two-body interaction; the BS+3B
calculation also includes a version of the Urbana UIX three-body
interaction. It is seen that the three-body forces reduce the gap by
$25\%$. In addition, the wave-function renormalization
  specified by Eq.~\eqref{eq:renormalization} and polarization effects
  are taken into account in Ref.~\cite{BaldoSchulze2007}.  The
absence of proton superconductivity, if confirmed by future
computations, would have profound implications for the physics of
compact stars. We also note that in all models the proton pairing gap
attains its maximum (almost) at the crust-core boundary; this may have
some interesting implications for the {\it type} of proton
superconductivity (\ie, type-II vs.\ type-I) throughout the core of
the star and, consequently, for the formation of flux tubes vs.\
superconducting domains across the core~(see \cite{SinhaSedrakian2015}
and references therein).

Clearly, the dependence of gaps on the density of stellar matter could
strongly depend on the underlying model that provides the equation of
state and the particle fractions.  Nevertheless, the general
arrangement of the pairing gaps depicted in
Fig.~\ref{fig:npe_diagram_gaps}, such as the transition from $S$- to
$P$-wave neutron pairing around the crust-core interface, and larger
proton $S$-wave than neutron $P$-wave gap in the core, is rather robust.

So far our discussion has focused on pairing in homogeneous and
isotropic neutron or proton matter.  As discussed above, the matter in
neutron star crusts is a multi-component system consisting of neutron
fluid, nuclear clusters, and background electrons. Since both neutron
matter and nuclear clusters contain correlated Cooper pairs of
fermions, it is useful to address these pairings in a unified
manner. Indeed, the distinction between the clusters and the neutron
fluid is appropriate at low densities, but as the density increases
the surface of clusters is blurred and the transition to neutron
matter becomes smooth. It is then more appropriate to consider the
{\it pairing amplitudes} of neutrons or protons within a unit cell of
the nuclear lattice. We now discuss the static pairing properties of
such cells, leaving their collective excitations and dynamics to
Secs.~\ref{sec:collective_modes} and \ref{sec:dynamics}.  This problem
can be addressed at a number of levels of sophistication ranging from
DFT-based approaches to fully microscopic Hartree-Fock-Bogolyubov
theories. The early work of Ref.~\cite{1973NuPhA.207..298N} on neutron star
crusts, which is often used as a benchmark for the equation of state
and composition of the inner crust of a neutron star, does not take
into account pairing correlations. These correlations were later
incorporated using density functionals that contain terms coming from
pairing correlations~\cite{BulgacYu2004IJMPE,Fayans2000NuPhA}.
Application of such a functional carried out for a Wigner-Seitz (WS)
cell in a neutron star's inner crust~\cite{BaldoSaperstein2007EPJA}
showed that the introduction of pairing correlations does not change
the net energy of the system significantly, but changes the number $Z$
of protons bound in a cluster and the size of the WS cell~\footnote{In
  the WS approximation one replaces the unit cell of the lattice by a
  spherical cell of volume $1/n_N$, where $n_N$ is the number density
  of nuclei.  The number of electrons in a cell is equal $Z$ (the
  charge of the nucleus) and their density over the cell is assumed to
  be constant. }.  Specifically, $Z=20,\, 24,\, 26$ values appear
instead of $Z=40$ and 50 determined in the unpaired case.  The main
uncertainty in these calculations involves the polarization
corrections to the pairing gap in neutron matter and the associated
contribution to the density functional. The pairing gap $\Delta(r)$ as
a function of the radius $r$ of the WS cell shows a pronounced minimum
at the surface of the nucleus. In the large-$r$ limit the gap is
larger than it is inside the cluster at low to intermediate densities,
but this disparity is strongly reduced as the densities of the outside
neutron gas and the cluster become comparable. The relative magnitudes
of the pairing gaps within the cluster and outside are qualitatively
consistent with the predictions of the local-density approximation.

  In the weak-coupling limit, a semiclassical approach has been applied
  to the pairing in neutron star crusts within a given WS
  cell~\cite{Vinas2010,SchuckPhysRevLett2011}. In this approach, the
  pairing is enhanced at the surface of the nucleus, with asymptotic
  behaviors at small and large $r$ showing the same trends as above.  A
  semiclassical method that includes shell effects -- namely the
  extended Thomas-Fermi plus Strutinsky integral method -- was applied
  to assess the role of proton pairing~\cite{Pearson2015PhRvC}.  It
  was found that pairing acts to smooth out the proton shell effects,
  without significantly changing the energetically favored value of
  $Z$, which was found to be close to 40.

  At the microscopic level, the Hartree-Fock-Bogolyubov theory
  \cite{Goodman1981NuPhA} has been applied extensively in the past
  decade to determine the composition and pairing of crustal matter in
  neutron stars, as well as its specific heat, using the WS
  approximation;
  see~\cite{SandulescuPhysRevC.70,Sandulescu2004PhRvC,Margueron2011,PastoreMargueron2013,Pastore2015PhRvC,Monrozeau2007,Pastore2015MNRAS}
  and references therein for the earlier work.  The key features
  related to the behavior of the relative magnitude of the gaps in the
  continuum and the cluster bound state discussed above are reproduced
  in this case as well. It has also been suggested that the observed
  depression/enhancement of the gap value at the surface of the
  cluster can be attributed to the manner in which the Dirichlet-Neumann mixed
  boundary conditions are imposed at the boundary of the WS
  cell~\cite{Pastore2015PhRvC}. Specifically, either even-parity wave
  functions and the first derivative of odd-parity wave functions vanish
  at the boundary, or visa versa. Because the physical results should
  not depend on the choice of the boundary condition, the observed
  disagreement should be attributed to the limitations of the WS
  approach.

  The validity of the WS approximation was tested in BCS calculations
  of pairing using the band theory of solids and ideas of anisotropic
  multi-band superconductivity assuming a body-centered cubic
  lattice~\cite{Chamel2010PhRvC}. In this case Bloch boundary
  conditions are imposed on the wave functions of neutrons. The
  neutron pairing gap, averaged over all continuum states, was found
  to be reduced due to the presence of inhomogeneities
  (clusters). However, the value of the pairing gap on the Fermi
  surface is only weakly affected by the band structure and can be
  well approximated by the corresponding value of the pairing gap in
  uniform matter, if an appropriate average neutron density is
  used. The amplitude of the neutron pairing gap is found to be smooth
  in this case, which suggests that the rapid change in its value
  close to the cluster's surface found in alternative approaches is
  associated with the WS approximation.  

\subsection{Pairing in higher partial waves}
\label{sec:higher_partial_waves}

From the preceding overview of neutron-star structure it is clear
that, if present, $^3P_2$--$^3F_2$ pairing in neutron matter is
phenomenologically important, since the neutron fluid in the core of
the star occupies a large volume fraction.  We now review the analyses
and computations involving this version of triplet odd-parity pairing,
as it involves a number of new aspects relative both to $^1S_0$ and
$^3S_1$--$^3D_1$
pairing~\cite{BaldoCugnon1992,BaldoElgaroy1998_Pwave,ZverevClarkKhodel2003,SchwenkFriman2004,Maurizio2014,Papakonstantinou2017,KhodelClark2004,KhodelClark2001,KhodelKhodelClark2001,SrinivasRamanan2016}.
In this case there can be competition between states involving various
projections $M$ of the orbital angular momentum $L$ (or total angular
momentum $J$). Additional complications arise from a dominant 
spin-orbit interaction and the tensor coupling of the $^3P_2$ 
partial wave to the $^3F_2$ state.

To solve the gap equation in the uncoupled $P$-wave channel, one starts 
with the expansion of the pairing interaction in partial waves,
\be
 V(\vecp,\vecp') =
  4\pi \sum_L (2L+1) P_L({\hat{\vecp}\cdot\hat{\vecp}'}) V_L(p,p'),
\ee
where the $P_L$ are  Legendre polynomials, and an associated 
expansion of the gap function 
\begin{equation}
  \Delta(\vecp) =
  \sum_{L,M} \sqrt{\frac{4\pi}{2L+1}} Y_{LM}(\hat \vecp) \Delta_{LM}(p)
\end{equation}
in spherical harmonics $Y_{LM}$, where $L$ is the orbital quantum
number and $M$ the corresponding magnetic quantum number.  It is
apparent that the non-linearity of the gap equation couples the
various gap components of $\Delta_{LM}(p)$. Regarding the $M$ 
dependence, this problem is usually simplified by performing an 
angle average within the denominator of the kernel of the gap 
equation, by focusing on real solutions and replacing
$\sqrt{\varepsilon^2(\vecp)+\Delta^2(\vecp)}$ by
$\sqrt{\varepsilon^2(p)^2+D^2(p)}$, where the ``angle-averaged'' gap
is given by
\begin{equation}
D^2(p) \equiv\int \frac{d\Omega }{4\pi} \, \Delta^2(\vecp) =
\sum_{L,M} \frac{1}{2L+1} [\Delta_{LM}(p)]^2 
\end{equation}
and $\varepsilon(p)$ is a single-particle energy in the normal state.

With this approximation the angular integrals are trivial, and 
one finds a one-dimensional gap equation for the $L$-th 
component of the gap: 
\begin{equation}
\label{eq:Delta_L}
  \Delta_L(p) = - \int_0^\infty
\frac{dp' p'}{\pi} \frac{V_L(p,p')}{\sqrt{ \varepsilon(p')^2 +
  \sum_{L'} [\Delta_{L'}(p')]^2 }}
  \Delta_L(p').
\end{equation}
In a first approximation one may neglect the terms in the sum on the
right-hand side of \eqref{eq:Delta_L} having $L'\neq L$, based on the
common assumption that a specific pairing channel is dominant over the
density range concerned. This assumption gains credence from the argument
in Sec.~\ref{sec:various_pw} relating densities (more precisely $k_F$
ranges) to in-medium collision energies, and hence to the dominant NN
phase shift.  However, a ``specific pairing channel'' may involve
coupling of different $L$ states through spin dependence of the
interaction. In the present case there is a substantial coupling to the
$^3F_2$ wave due to the tensor components of the pairing force, which
must be included to obtain quantitative results.  Thus the gap
equation to be solved takes the form of coupled equations for $P$- and
$F$-wave components of the gap, \ie,
\bea
 \left( \begin{array}{c} \Delta_L \\ \Delta_{L'} \end{array} \right)
=  \int_0^\infty\!\!\! \frac{dp' p'^2}{\pi E(p')} 
 \left( \begin{array}{rr}
  -V_{LL} & V_{LL'} \\ V_{L'L} & -V_{L'L'} 
 \end{array} \right)
 \left(\begin{array}{c} \Delta_L \\ \Delta_{L'} \end{array}\right),\nonumber\\
\eea
where $E(p) = \sqrt{ \varepsilon^2(p) +D^2(p)}$ and 
$D^2(p)\equiv[\Delta_{L}(p)]^2 + [\Delta_{L'}(p)]^2$, with $L=1$ and $L'=3$.
This coupling is analogous to that in the $^3S_1$--$^3D_1$ channel 
discussed in Secs.~\ref{sec:various_pw} and \ref{sec:Tmatrix}.

\begin{figure}[bt]
\begin{center}
\includegraphics[width=0.99\hsize]{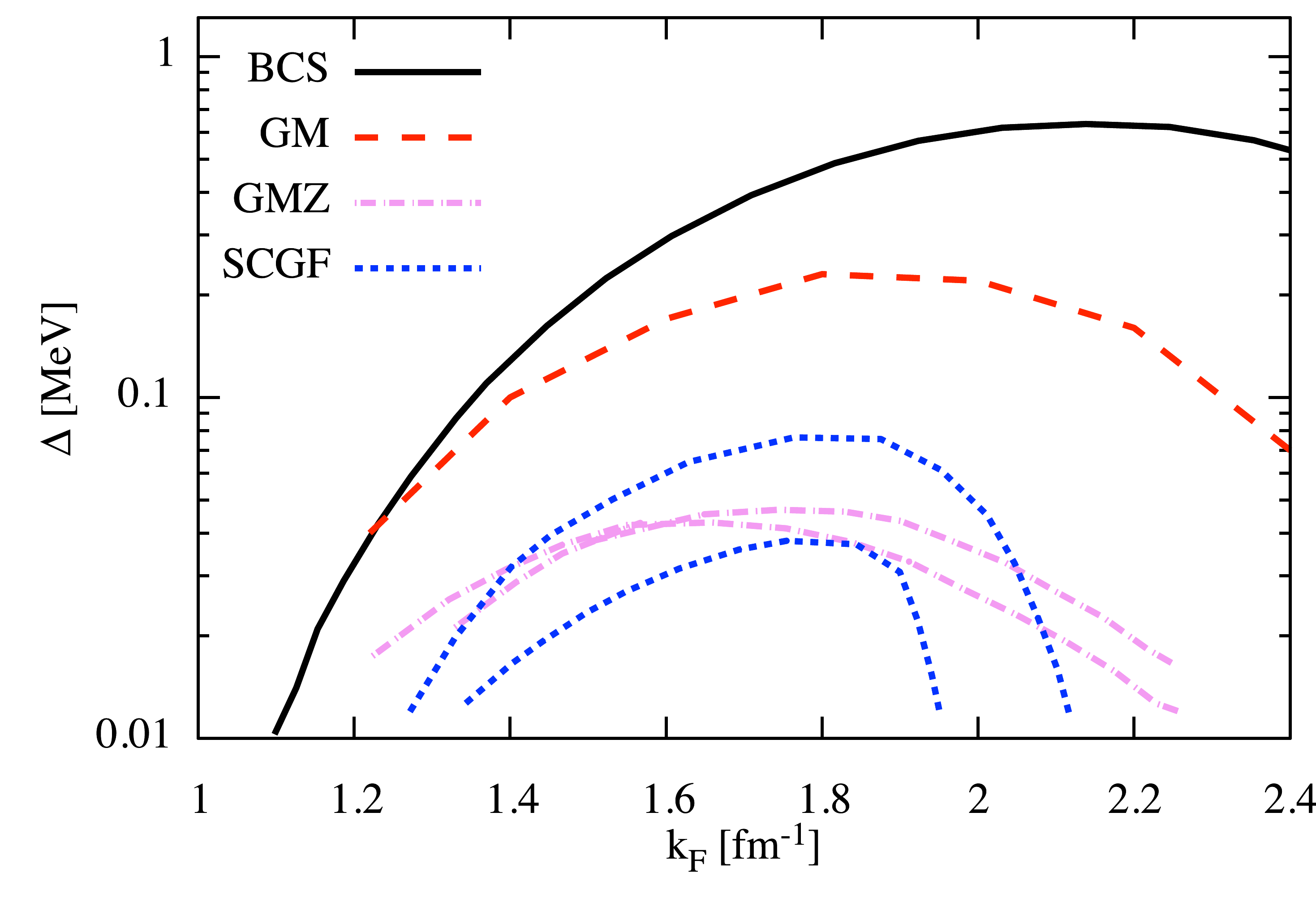}
\caption{ Dependence of the $^3P_2$--$^3F_2$ pairing gap in neutron
  matter on the Fermi momentum in various many-body theories. BCS: the
  solution of the BCS gap equation with free single particle
  spectrum. GM: solution of the gap equation with self-energies
  derived from $\cal G$-matrix~\cite{BaldoElgaroy1998_Pwave}. GMZ: uses the
  same theory as GM, but accounts for wave-function renormalization of
  the quasiparticle spectrum (lower maximum curve) plus three-body
  forces (higher maximum curve)~\cite{DongLombardZuo2013}. SCGF:
  self-consistent GF computations including only
  short-range correlations (higher-maximum curve) and both short- and
  long-range correlations (lower maximum curve)~\cite{Ding2016}. The
  results BCS, GM and SCGF were obtained with the Argonne
  $V_{18}$ potential; the GMZ result, with the Bonn-B potential.}
\label{fig:gaps_overview_P_wave}
\end{center}
\end{figure}

Figure~\ref{fig:gaps_overview_P_wave} collects a selection of results
for the $P$-wave gap.  Quantitative
understanding of $P$-wave pairing, or more precisely $^3P_2$--$^3F_2$
pairing, is further complicated by the presence of three-body forces,
which play an increasingly important role at the high baryon densities
of the outer-core region of a neutron star.  (At the lower densities
where $^1S_0$ neutron pairing dominates, the three-body force can be
safely neglected.) In the case of phase-shift equivalent NN
potentials, models of the three-nucleon (3N) force in use have been
constrained by the physics of light nuclei, as is the case for the
Urbana UIX 3N interaction~\cite{Pudliner1997}.  This model has been
used in conjunction with the Argonne $V_{18}$ NN interaction to
estimate the $^3P_2$--$^3F_2$ pairing
gap~\cite{ZuoCuiLombardoSchulze2008}.  However, a readjustment of the
parameters of the UIX interaction, rendering it less repulsive, was
required to guarantee agreement with the empirical nuclear-matter
saturation properties within Brueckner-Hartree-Fock theory, which in
turn was used to obtain the single-particle spectrum entering the gap
equation~\cite{ZhouBurgio2004}.  Within this scheme, the triplet gap
was found to be slightly reduced from the result for $V_{18}$ alone,
to a maximum value of the order 0.5~MeV. (We note that upon neglecting
the single-particle renormalization, the assumed 3N interaction causes
an enhancement of the gap.)
The Bonn-B meson-exchange phase-shift equivalent potential including
both two- and three-body terms has also been used to estimate the effect 
of 3N interactions on $^3P_2$--$^3F_2$ pairing~\cite{DongLombardZuo2013}.  
In this study the maximum of the gap is reached at about 0.6 MeV, but it 
is attained at slightly higher densities (higher neutron $k_F$ values).  
Upon including the effect of wave-function renormalization, as specified 
by Eq.~\eqref{eq:renormalization}, the gap was suppressed by an order 
of magnitude.

Importantly, the $^3P_2$--$^3F_2$ gap in neutron matter has also been 
evaluated for chiral two-nucleon (2N) and 3N interactions~\cite{Maurizio2014}.  
Assuming a free single-particle energy spectrum, the 3N force was found 
to produce an enhancement in third order of the chiral expansion.  
Introduction of single-particle renormalization leads to a moderate 
suppression of the gap, to a maximal value of about 0.4 MeV.  In another 
recent exploratory study~\cite{Papakonstantinou2017}, strong sensitivity 
of the $^3P_2$ gap to the choice of the 3N interactions was demonstrated 
and the mandatory consistency between the 2N and 3N forces was emphasized.
This numerical study differs from most others in that the 3N
pairing interaction is not simulated by a density-dependent effective
2N interaction.

In summary, these and other similar computational efforts have shown
the importance of including 3N interactions for accurate determination
of the pairing gaps in the $^3P_2$--$^3F_2$ coupled channel.  To this
end, consistent extrapolations of the NN and 3N interactions from nuclear 
saturation to higher densities, such that these forces are properly 
constrained by empirical data, are imperative, together with microscopic 
many-body theories that are reliable at high density.

This brings us to a remaining aspect of the problem of spin-triplet pairing 
in high-density matter that demands further attention, namely the influence 
of correlations that are not included in BCS theory. Long-range correlations 
have been found to induce a strong suppression of the gap when non-central 
Landau interactions are used in conjunction with the weak-coupling 
formula~\cite{SchwenkFriman2004}. A more recent computation~\cite{Ding2016} 
indicates that the suppression of pairing by long-range correlations 
is counteracted by enhancement due to short-range correlations, \ie, 
these two factors tend to compensate each other.

At high densities, isospin-symmetrical nuclear matter should pair in
the $^3D_2$ channel, by forming isospin-singlet pairs, the attractive
interaction in this channel being stron\-ger than that in the $^3P_2$
channel, as seen in Fig.~\ref{fig:phaseshifts}.  In much  the same
way as one expects a transition from $^3S_1$--$^3D_1$ pairing to
$^1S_0$ pairing with rising isospin asymmetry, nuclear matter at still
higher densities should undergo a transition from $^3D_2$ pairing to
$^3P_2$--$^3F_2$ pairing as isospin asymmetry increases from zero to
larger values.  However, existing calculations~\cite{AlmSedrakian1996}
have shown that a small imbalance in isospin populations already
destroys $D$-wave pairing, so it can be realized only in nearly
symmetrical nuclear matter.

\subsection{Hyperonic pairing}
\label{subsec:H_pairing}

The inner core of a neutron star may contain a hyperonic component,
because the rise of neutron and electron energies with density
can be compensated by the onset of lower-energy hyperons. Just as  was
the case with protons, a relatively low fraction of hyperons implies that
they will pair in the $^1S_0$ channel if there is a sufficiently attractive
component available in their interaction at low energies~\cite{BalbergBarnea1998,Takatsuka2002,TakatsukaNishizaki2006,VidanaTolos2004,WangShen2010,XuWuRen2014,RadutaSedrakian2017,Takatsuka_XiPairing}.

We can exclude from the outset the possibility of hyperon-nucleon
pairing and pairing between non-identical hyperons, due to the
imbalance between the chemical potentials of baryon components of the
core (however, see~\cite{ZhouSchulze2005}). In contrast to the purely
nucleonic case~\cite{SteinSedrakian2014,SteinHuang2012,Clark2016JP},
additional imbalance will arise because of the substantial disparity
in the masses of different hyperons as well the large difference
between the masses of nucleons and hyperons. It cannot be ruled out
that the Fermi surfaces of non-identical particles may, for some
density, be close enough to support cross-species pairing, but this
can only occur in a rather limited density range and depends
significantly on the underlying model of the composition of matter.

Before discussing the pairing gaps in hypernuclear matter, we need to
address the ambient composition and the single-particle energies of
hyperons. The complication here is that the non-relativistic theories
we have thus far considered fail to account for the measured two-solar
mass pulsars~\cite{Balberg1997,Baldo2000,Burgio2011,RijkenSchulze2016}.
Relativistic covariant density-functional (hereafter DF) theory allows
for modeling the properties of hypernuclear matter consistent with
the astrophysical constraint on masses of hypernuclear compact stars,
as well as with laboratory constraints on the depths of potentials in
(hyper)nuclear matter.\footnote{The fundamentals of covariant density
functional theory for nuclear systems are discussed, for example, in
\cite{Glendenning_book,weber_book,SerotWalecka1997,Meng2016}. Recent
  applications of covariant density functional theory to hypernuclear
  matter are reviewed
  in~\cite{Chatterjee2016EPJA,Oertel2017RvMP}.}  This is achieved
by starting from phenomenological Lagrangians with parameters that are
adjusted {\it a posteriori} to satisfy the available astrophysical and
laboratory constraints.  Adopting relativistic DF theory, the
single-particle energies of hyperons (collectively denoted $\rY$ below)
are expressed as
\begin{equation}
\label{eq:spe_Y}
E^\rY(k)=\sqrt{ k^2+m_\rY^{*2}}+g_{\omega \rY} 
\omega+g_{\phi \rY} \phi+g_{\rho \rY} \tau_{3\rY} \rho+\Sigma_R,
\end{equation}
where $\Sigma_R$ represents the rearrangement term entering the models
with density-dependent couplings, $\mu_\rY=E^\rY(k_F)$ stands for the
chemical potential, and
$m_\rY^*=m_Y-g_{\sigma \rY} \sigma - g_{\sigma^*\rY} \sigma^*$ is the
Dirac effective mass of species $\rY$. Here $\sigma$, $\sigma^*$,
$\rho$, $\omega$ and $\phi$ refer to the mesonic fields, while the
$g_{\alpha \rY}$ with $\alpha \in (\sigma, \sigma^*, \rho, \omega)$
are the hyperon-meson couplings.

A method for computing the pairing gaps in relativistic DF theories
that has been validated in studies of finite nuclei within
relativistic Hartree-Fock-Bogolyubov theory is based on solution of
the non-relativistic BCS equation for a given two-nucleon potential
using single-particle energies and particle composition determined by
the relativistic DF method~\cite{1991ZPhyA.339...23K}.  While there is
a clear inconsistency in treating the unpaired matter and pairing
correlations in different theories, this approach is close in 
spirit to the decoupling approximation widely applied in the
non-relativistic theories discussed previously.

\begin{figure}
\begin{center}
\includegraphics[angle=0, width=0.89\columnwidth]{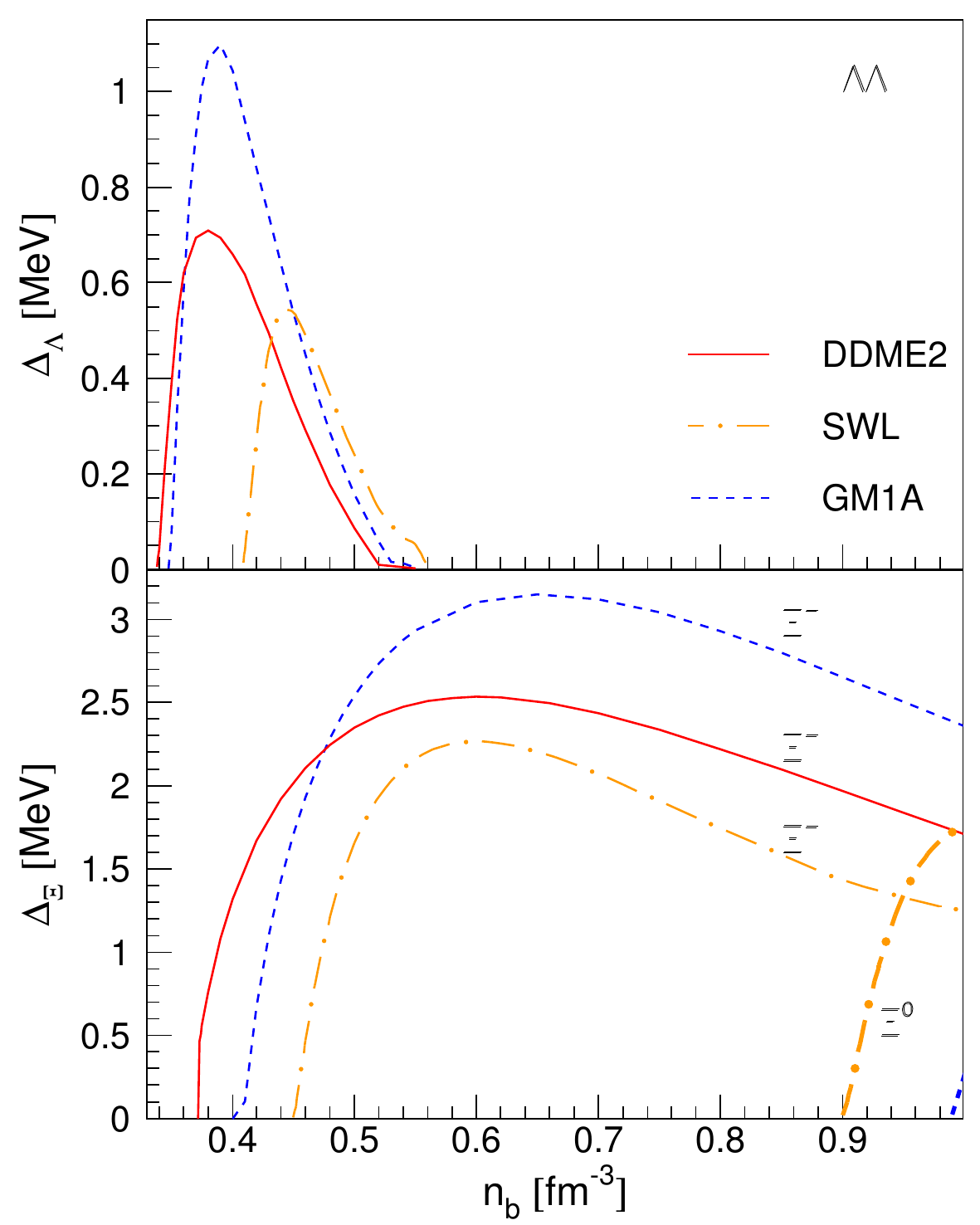}
\end{center}
\caption{Dependence of $^1S_0$ pairing gaps in hyperonic matter on
  baryonic number density $n_b$ for medium-dependent single-particle
  energies and composition computed according to DDME2 (solid), GM1A
  (dashed), and SWL (dash-dotted) models. The upper panel refers to
  the $\Lambda$ gaps; the lower panel to the $\Xi^{-}$ (thin lines)
  and $\Xi^{0}$ (thick lines) gaps \cite{RadutaSedrakian2017}.  For
  DDME2 and GM1A functionals, $\Xi^{0}$ hyperons do not appear in the
  density range shown in the figure.  }
\label{fig:Gaps_nB}
\end{figure}
In the BCS approximation, it becomes straightforward to solve the gap 
equation for the pairing of hyperons of a given type Y, which can be 
written as 
\begin{equation}
\Delta_{\rm Y} (k)=-\frac{1}{4 \pi^2} \int dk' k'^2 \frac{V_{\rm YY}(k,k') 
\Delta_{\rm Y}(k')}{\sqrt{\left[E^{\rm Y}(k')-\mu_{\rm YY} \right]^2
+\Delta_{\rm Y}^2(k')}},
\label{eq:gap_Y}
\end{equation}
where $E^{\rm Y}(k)$ is the single-particle energy of hyperon Y
given by Eq.~\eqref{eq:spe_Y}. In this approximation, the pairing
interaction is given by the YY interaction in the $^1S_0$ channel, 
\begin{equation}
  V_{\rm YY}(k,k')= 4 \pi \int dr r^2 j_0(k r) V_{\rm YY}(r) j_0(k' r),
\label{eq:Vkk_Y}
\end{equation}
where $j_0(k r)=\sin(kr)/(kr)$ is the spherical Bessel function of
order zero and $V_{\rm YY}(r)$ is the YY interaction in coordinate
space. 

Recently, hypernuclear pairing in compact stars was addressed in the
context of their cooling \cite{RadutaSedrakian2017} using the strategy
outlined above. For the $\Lambda\Lambda$ pairing interaction, the
configuration-space parameterization of the ESC00
potential~\cite{Rijken2001} proposed by \cite{FilikhinNPA2002} was
adopted. For the $\Xi^{-}\Xi^{-}$ and $\Xi^{0}\Xi^{0}$ interactions, a
model presented in \cite{Garcilazo2016} was selected, specifically the
one that corresponds to the Nijmegen Extended Soft Core potential
ESC08c~\cite{Rijken:2013wxa}.  The\-se potentials were chosen to maximize
the attraction in the respective channels, in order to obtain an upper
bound on the pairing gap within BCS theory.

Figure~\ref{fig:Gaps_nB} shows the dependence of the pairing gaps for
$\Lambda$ and $\Xi^{-,0}$ hyperons on the baryon
density~\cite{RadutaSedrakian2017}.  Because this involves input for
the composition of matter and self-energy effects, the results
depend on the chosen DFs, which are labeled as
DDME2~\cite{FortinPRC2016}, GM1A~\cite{Gusakov_MNRAS2014}, and
SWL~\cite{Spinella-PhD}.  It is interesting that the $\Lambda$ pairing
is restricted to densities $n_b\le 0.55$ fm$^{-3}$. Accordingly, at
higher densities there may exist regions of unpaired $\Lambda$ matter
in the most massive stars, provided there is no significant attraction
that causes pairing in higher partial waves.  The $\Xi^{-}$ hyperons
remain paired up to the highest densities considered in
\cite{RadutaSedrakian2017}.

\subsection{Overview of neutrino radiation from compact stars}

Theoretical modeling of the thermal evolution (cooling) of neutron
stars tests neutron-star interior composition predicted by microscopic
theories of dense matter.  Such models are confronted with
observations of the X-ray emission from the surfaces of nearby neutron
stars.  The cooling evolution is roughly divided into three
stages~\cite{Pehick1992RvMP,Page2013,Yakovlev2001,weber_book,Sedrakian2007PrPNP,YakovlevPethick2004}: (a)
There is an initial transient stage during which the core temperature
drops to about 0.1  MeV$\approx 1.16 \times 10^9$~K
the subsequent thermal history of the star does not depend on this
initial stage. (b) The neutrino cooling era lasts for
$t\lesssim 10^5$yr, the main cooling mechanism being neutrino
radiation from the stellar interior. This stage is crucial for
theoretical predictions of the surface temperatures of the observed
thermally emitting stars.  The importance of studies of the thermal
history of neutron stars lies in its strong dependence on the neutrino
emission rates from dense matter during the neutrino cooling
era. These rates, in turn, depend crucially on the particle content
and superfluidity of the interior components.  (c) The photon cooling
era $t \gtrsim 10^5$~yr is dominated by radiation of photons from the
surface of the star and heating due to the dissipation of rotational
and magnetic
energy~\cite{PetrovichReisenegger2015,SchaabSedrakian1999}.

In this subsection, we briefly review the main processes responsible for
neutrino radiation. More detailed surveys can be found in
\cite{Page2013,1996NuPhA.605..531S,Sedrakian2007PrPNP,Yakovlev2001,SchmittShternin2017,Pehick1992RvMP}. 
In particular, we will focus on the effects of
superfluidity on these processes as background for a more detailed
discussion of microscopic calculations of neutrino emission rates from
superfluid matter in the next section.

The various processes or reactions generating neutrinos can be
classified according to the number of fermions
involved~\cite{ShapiroTeukolsky1983,Pehick1992RvMP}.  The rationale of this
classification is that each degenerate fermion participating in such a reaction
introduces a factor $T/\ep_F$, which is a small parameter when the
temperature is much less than the Fermi energy involved.  Thus, the
processes of leading order in powers of $T/\ep_F$ are given by
\bea \label{eq:urca}
&&n \to p + e + \bar\nu_e , \quad p + e \to n + \nu_e , \\
\label{eq:brems}
&&N  \to N  + \nu_f +\bar\nu_f  \qquad ({\rm forbidden}),
\eea 
where $N\in (n,p)$ refers to a nucleon, $\nu$ and $\bar\nu$ to
neutrino and antineutrino, and index $f=e,\, \mu,\, \tau$ to neutrino
flavors.  In dealing with the quasiparticle states of nucleons having
an infinite lifetime, the rates of these reactions are constrained
kinematically. The second process \eqref{eq:brems}, known as
neutral-current neutrino pair bremsstrahlung, is forbidden by energy
and momentum conservation.  The Urca reaction \eqref{eq:urca} is
kinematically allowed in the matter under $\beta$-equilibrium for
proton fractions 
$Y_p \ge 11-14\%$~\cite{Boguta1981,LattimerPrakashHaensel}.  The
processes having two baryons in the initial (and final) states
obtained from \eqref{eq:urca} and \eqref{eq:brems} by adding a
nucleon, \ie,
\bea \label{eq:mod_urca}
&&N + n \to N + p + e + \bar\nu_f , \\
\label{eq:mod_brems}
&&N + N' \to N + N' + \nu_f +\bar\nu_f ,
\eea 
are allowed kinematically but are suppressed by an extra factor
$(T/\ep_F)^2$.  Estimates of emissivity (power of energy radiated per
unit volume) for the three relevant processes above are
$\varepsilon_{\rm Urca} \sim 10^{27}\times T_9^6$ for reaction
\eqref{eq:urca},
$\varepsilon_{\rm mod.~Urca} \sim 10^{21} \times T_9^8$ for reaction
\eqref{eq:mod_urca} and its inverse, and
$\varepsilon_{\rm \nu\bar\nu} \sim 10^{19}\times T_9^8$ for reaction
\eqref{eq:mod_brems}. Here $T_9$ is the temperature in units $10^9$ K.
It is seen that the modified Urca process is significantly
  less effective than the direct Urca process due to the phase-space
  restriction introduced by the additional two baryons involved in the
  first reaction. However, close to the Urca threshold indicated above the rate
  of the modified Urca process is strongly enhanced due to the pole
  structure of intermediate state propagator connecting the weak and
  strong vertices in this process~\cite{Shternin2018}, i.e., there is a
  smooth transition from modified to direct Urca process as the proton
  fraction increases. In addition to these nucleonic processes,
bremsstrahlung by electrons scattering off nuclei and impurities in
the crust contributes to the neutrino radiation, but we will not
discuss these mechanisms as they are independent of the baryonic
superfluidity~\cite{KaminkerPethick1999}.

Hyperon featuring matter will emit neutrinos via the hyperonic Urca
processes~\cite{PrakashPrakash1992}
\begin{eqnarray} 
\label{eq:UrcaLambda}
\Lambda &\to& p + l  + \bar\nu_l,\\
\label{eq:UrcaSigmaminus}
\Sigma^- &\to& \left(\begin{array}{c} n  \\
                       \Lambda \\
\Sigma^0
\end{array} \right) + l + \bar\nu_l,\\
\label{eq:UrcaXiminus}
\Xi^- &\to& \left(\begin{array}{c} \Lambda  \\ 
                                                         \Xi^0
                        \\
\label{eq:UrcaSigmazero}
\Sigma^0   \end{array} \right) + l + \bar\nu_l,\\
\label{eq:Xizero}
\Xi^0 &\to& \Sigma^+ + l  + \bar\nu_l,
\end{eqnarray}
where $l$ stands for a lepton, either electron or muon, and
$\bar\nu_l$ is the associated antineutrino.  The hyperonic Urca
thresholds are much lower than those for nucleons. Once the 
relative abundances of hyperons exceed a few percent, the hyperonic
Urca processes are kinematically allowed.  Consequently, the reactions
\eqref{eq:UrcaLambda}-\eqref{eq:Xizero} will operate provided the
relevant species of hyperons are present in matter.  Since the hyperon
abundances increase rapidly once they become energetically favorable,
the corresponding threshold densities practically coincide with the
onset densities for hyperons.

For completeness, we point out here that pion BEC will radiate 
via the reaction
\bea\label{eq:pions}
 {\cal U } \to {\cal U } +  e^- + \bar\nu_e,
\eea
where ${\cal U}$ denotes here the basic fermion which is a coherent linear
combination of proton and neutron.  Similar reactions act also in a
kaonic BEC.  The corresponding emissivities are large compared to
those of the baryonic processes above, but the BEC of pions and kaons
is not guaranteed to occur in neutron stars; their discussion is
beyond the scope of this
review, see Refs.~\cite{Migdal1990,Tatsumi2013,Muto2003}.  Finally, once
deconfinement takes place, the quark cores of neutron stars will
radiate neutrinos predominantly through the quark counterparts of the
Urca
process~\eqref{eq:urca}~\cite{AlfordSchmitt2008RvMP,AnglaniCasalbuoni2014RvMP,RajagopalWilczek2001}.

Neutrino radiation is suppressed once superfluid phases are formed in
neutron stars, exponentially at asymptotically low temperature by a
Boltz\-mann fac\-tor ${\rm exp}[-\Delta_{ \rm max}(0)/T]$ for processes of
the Urca type \eqref{eq:urca}, where $\Delta_{\rm max}(0)$ is the
largest of the gaps of nucleons. A more detailed discussion of this
suppression will be given below in
Sec.~\ref{sec:Urca_superfluid}. Similarly, the processes
\eqref{eq:mod_urca} and \eqref{eq:mod_brems} are suppressed at low
temperatures by factors
${\rm exp}\{-[\Delta_N(0)+\Delta_{N'}(0)]/T\}$, where $N$ and $N'$
label the pair of initial (or final) baryons.

Somewhat counterintuitively, superfluidity opens a new channel of
neutrino emission, which is due to the process of neutrino-pair
bremsstrahlung via the neutral-current Cooper pair-breaking and
formation (PBF) process, which can be written schematically as 
\begin{eqnarray}
\label{eq:N_PBF}
N+N\to {\cal C}+ \nu_f + \bar \nu_f, 
\end{eqnarray}
where ${\cal C}$ stands for a nucleonic Cooper pair.  The neutrino
emission by these reactions was computed initially neglecting vertex
corrections~\cite{FlowersRuderman1976,KaminkerHaensel1999,
  VoskresenskiSenatorov1986}, 
which were considered later in a series of
works~\cite{LeinsonPerez2006,SedrakianMuther2007PRC,KolomeitsevVoskresensky2008,KolomeitsevVoskresensky2010,SteinerReddy2009,Sedrakian2012PRC}.

Although at low temperatures the PBF processes are again suppressed
exponentially, they are very effective in cooling neutron stars at
temperatures not far below the critical temperature $T_c$ of a
relevant nucleonic
condensate~\cite{Page2009,SchaabVoskresensky1997}. These processes are
operative also in hyperonic condensates, as discussed below.

\subsection{Pair-breaking and formation processes}
\label{sec:PBF}

We now turn to a more detailed description of the PBF processes
introduced in the previous section.  The initial calculations, which
did not include vertex corrections, led to the conclusion that the
neutrino emission via neutral vector currents is large compared to
that via axial vector currents
~\cite{FlowersRuderman1976,KaminkerHaensel1999,VoskresenskiSenatorov1986}.
More recent work has shown that the vertex corrections substantially
suppress the emission via vector currents while they leave the axial
vector emission
unaffected~\cite{LeinsonPerez2006,SedrakianMuther2007PRC,KolomeitsevVoskresensky2008,KolomeitsevVoskresensky2010,SteinerReddy2009,Sedrakian2012PRC}.
Accordingly, the axial vector current emission is the dominant one.
The physical basis of the strong suppression of the vector
  response function is the conservation of baryon number. The axial
  current is not conserved; hence the neutrino radiation in this
  channel is not suppressed.

The low-energy neutral weak-current interaction Lagrangian describing
the interaction of the neutrino field $\psi$ and baryonic current
$j_{\mu}$ is given by
\be {\cal L}_W
= -\frac{G_F}{2\sqrt{2}} j_{\mu}\bar\psi \gamma^{\mu}(1-\gamma^5)\psi,
\ee 
where $G_F$ is the Fermi coupling constant. The baryon current 
for each $B$-baryon is 
\be j_{\mu} = \bar \psi_B
\gamma_{\mu}(c^{(B)}_V-c^{(B)}_A\gamma^5) \psi_B , \ee 
where $\psi_B$ is the quantum field of the baryon and $c^{(B)}_V$ and
$c^{(B)}_A$ are its vector and axial vector couplings, respectively.
\begin{figure}[t]
\begin{center}
\includegraphics[angle=0, width=0.99\columnwidth]{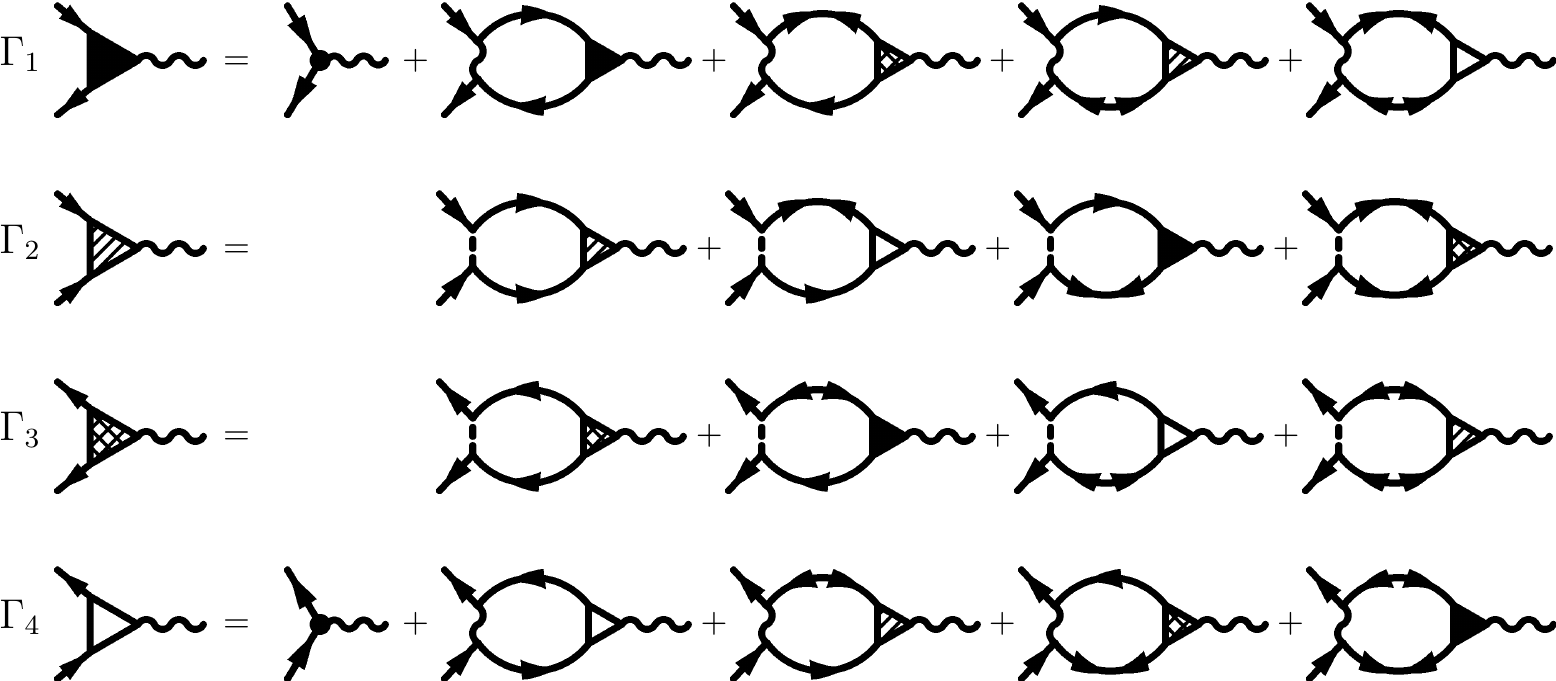}
\end{center}
\caption{ Coupled integral equations for the three-point
  weak-interaction vertices in a superfluid.  The ``normal''
  GF for particles (holes) are shown as lines with single arrows
  directed from left to right (right to left).  The lines with two
  arrows correspond to the ``anomalous'' GF $F$ (two incoming
  arrows) and $\Fd$ (two outgoing arrows).  The ``normal'' vertices
  $\Gamma_1$ and $\Gamma_4$ are shown as full and empty triangles,
  respectively.  The ``anomalous'' vertices $\Gamma_2$ and $\Gamma_3$ 
  are shown as hatched and shaded triangles, respectively. Horizontal wavy 
  lines represent the low-energy GF  of the $Z^0$ gauge boson. 
  Vertical dashed lines stand for the particle-particle interaction 
  $v_{pp}$; wavy lines for the particle-hole interaction $v_{ph}$.  }
\label{fig:Ladders_PBF}
\end{figure}
The rate at which neutrinos are radiated from matter (the neutrino
emissivity) can be obtained either by using the optical theorem in
finite-temperature field theory ~\cite{VoskresenskiSenatorov1986} or
directly from the kinetic equation for neutrinos formulated in terms
of real-time GF~\cite{SedrakianDieperink2000}.  Both methods
lead to the following expression for the neutral-current neutrino pair
bremsstrahlung emissivity
\bea
\label{eq:emiss_pbf} 
\varepsilon_{\rm \nu\bar\nu}&=& - 2\left(
  \frac{G_F}{2\sqrt{2}}\right)^2\int\!  d^4q~ g(\omega)\omega
\sum_{i=1,2}\int\!\frac{d^3q_i}{(2\pi)^32 \omega_i} \nonumber\\
&\times&\Img[
L^{\mu\lambda}(q_i)\,\Pi_{\mu\lambda}(q)]\delta^{(4)}(q-\sum_iq_i),
\eea 
where $q_i = (\omega_i,\vecq_i)$, with $i=1,2$, are the neutrino
momenta, $g(\omega) = [{\rm exp}(\omega/T) - 1]^{-1}$ is the Bose
distribution function, $\Pi_{\mu\lambda}(q)$ is the retarded
polarization tensor of baryons, and 
\bea L^{\mu\nu}(q_1,q_2) &=& 4
\Big[q_1^{\mu}q_2^{\nu}+q_2^{\mu}q_1^{\nu}- (q_1\cdot q_2)g^{\mu\nu}
\nonumber\\
&&-i\epsilon^{\alpha\beta\mu\nu}q_{1\alpha}q_{2\beta}\Big]
\eea 
is the leptonic trace.  Here the emissivity is defined per
neutrino flavor, \ie,  the full rate of neutrino radiation through weak
neutral currents is larger by a factor $N_f$, the number of
neutrino flavors. We will consider $N_f= 3$ massless neutrino
flavors.

The central quantity in \eqref{eq:emiss_pbf} is the polarization
tensor $\Pi_{\mu\lambda}(q)$, which describes the response of the
superfluid to weak vector and axial currents. A microscopic
approach for calculating the response function (or polarization
tensor) in superfluid matter was first developed by
\cite{Abrikosov:QFT} in the context of the electrodynamics of
superconductors. In this theory, the response of superconductors to
external probes is expressed in the language of GF at
non-zero temperature and density, with contact interactions that do
not distinguish among the particle-hole and particle-particle
channels. It is equivalent to the theories initially advanced for
metallic superconductors~\cite{Anderson1958,Bogolyubov:1958kj}, which
are based on equations of motion for second-quantized operators.  A
more general approach was subsequently developed within the
Fermi-liquid theory for superconductors and
superfluids~\cite{Leggett1966,LarkinMigdal1963}. The latter method
implements wave-function renormalization of the quasiparticle spectrum
and higher-order harmonics in the interaction channels, and allows for
particle-hole (ph) and particle-particle (pp) interactions having
different strengths and/or signs.

\begin{figure}[bt]
\begin{center}
\includegraphics[angle=0, width=0.99\columnwidth]{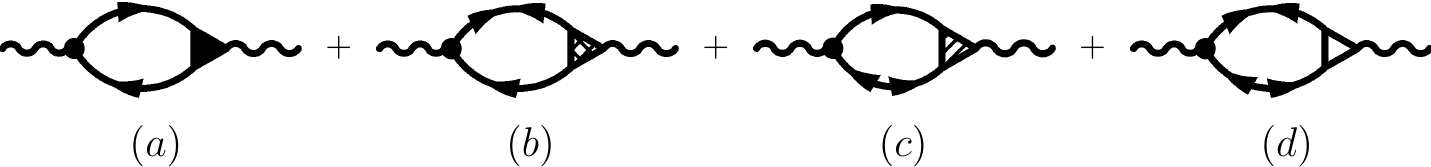}
\end{center}
\caption{ The sum of polarization tensors contributing to the
  vector-current neutrino emission rate.  Note that diagrams $b$, $c$,
  and $d$ are specific to superfluid systems and vanish in the
  unpaired state.  }
\label{fig:pol_PBF}
\end{figure}

Computation of the polarization tensor proceeds in three steps. In the
first step, one solves the coupled integral equation for the
three-point vertices (shown in Fig.~\ref{fig:Ladders_PBF}) in the
superfluid matter.  Next, the four polarization tensors shown in
Fig.~\ref{fig:pol_PBF} are resummed to obtain the full response
function. Finally, this function is expanded, to first non-vanishing
order, in the small parameter $v_F/c\ll 1$, where $v_F$ is the Fermi
velocity of nucleons and $c$ is the speed of light.  In the case of
vector-current response, a non-zero contribution is obtained at order
$v_F^4$, whereas in the case of axial vector coupling, one finds a
non-zero contribution at order $v_F^2$ (here and below we again set
$c=1$).
Next, the phase-space integrals in the emissivity \eqref{eq:emiss_pbf} 
are computed after contracting the polarization tensor with the trace over 
leptonic currents.  The final result for three neutrino flavors ($N_f =3$)
can be cast in the form \cite{LeinsonPerez2006,SteinerReddy2009,KolomeitsevVoskresensky2008,Sedrakian2012PRC}
\bea\label{eq:e_vector_Swave}
\epsilon_{\nu\bar\nu}^{V}(z) &=&\frac{16 G_F^2 c_V^2 \nu(p_F) v_F^4 }{1215\pi^3} I^{V}(z) T^7 ,
\eea
where $z = \Delta/T$, $c_V = 1$ for neutrons and $0.08$ for protons, 
$\nu(p_F)$ is the density of states at the
Fermi momentum $p_F$ and the integral is given by 
\bea I^{V}(z)=z^7 \int_{1}^{\infty}\!\!\!\!  
\frac{dy\  y^5}{\sqrt{y^2-1}} f \left(z y\right)^2 
\left[1+  \beta(y)v_F^2\right],
\label{eq:I}
\eea 
where $f$ is the Fermi distribution.  The explicit functional form
of $\beta(y)$, which specifies the next-to-leading term in the
expansion of $\epsilon_{\nu\bar\nu}(z)$, is given
in~\cite{Sedrakian2012PRC}.  The result to order $v_F^4$ has been
obtained by a number of authors within comparable theoretical
frameworks~\cite{LeinsonPerez2006,KolomeitsevVoskresensky2008,SteinerReddy2009,Sedrakian2012PRC}.
A computation of the vector-current emissivity to order $v_F^6$ shows
that the corrections to the leading non-zero term are below $10\%$ for
values of $v_F\le 0.4$ characteristic of baryons in compact
stars~\cite{Sedrakian2012PRC}. This result provides evidence of the
convergence of the series expansion of the vector-current polarization
tensor in the regime where the momentum transfer is small compared to
other relevant scales.

\begin{figure}[t]
\begin{center}
\includegraphics[angle=0, width=0.8\columnwidth]{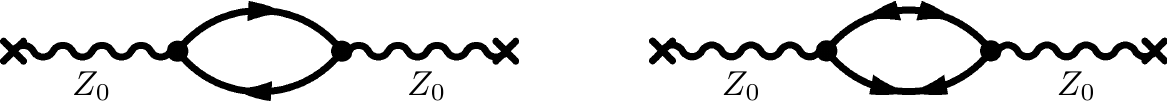}
\end{center}
\caption{The two diagrams contributing to the axial response of
  baryonic matter. Conventions are the same as in
  Fig.~\ref{fig:Ladders_PBF}.  }
\label{fig:pol_axial}
\end{figure}

Turning to the axial-vector contribution, one finds that the corresponding 
polarization tensor is unaffected by the vertex corrections; hence
it is comprised of the two diagrams shown in Fig.~\ref{fig:pol_axial}.
The emissivity of this processes is given by \cite{KolomeitsevVoskresensky2008}
\bea\label{eq:e_axial_Swave}
\epsilon^A_{\nu\bar\nu}&=&\frac{4G_F^2g_A^2}{15\pi^3} \zeta_A\nu(0)
{v_F}^2T^7  I^ A ,\\
\label{eq:i_axial_Swave} I^ A &=& z^7
\int_1^{\infty} dy \frac{y^5}{\sqrt{y^2-1}} f^2\left(z y\right)
\left[1 + O(v_F^4)\right] 
\eea
with $\zeta_A = 6/7$ and $g_A\simeq 1.26$, a result exhibiting
the $v_F^2$ scaling of the axial neutrino emissivity compared to 
the $v_F^4$ scaling found for $\epsilon^V_{\nu\bar\nu}$ in 
\eqref{eq:e_vector_Swave}.  Thus we see that the axial neutrino 
emissivity dominates the vector current emissivity, other factors 
having the same order of magnitude.

The temperature dependence of the pair-breaking processes can be
understood from dimensional analysis~\cite{ShapiroTeukolsky1983,Pehick1992RvMP}. 
First, we observe that the initial- and final-state (degenerate) neutrons, being 
confined to a narrow band $\sim T$ around the Fermi surface, each 
contribute a factor $T$, while the final-state neutrino and antineutrino 
each contribute a factor $T^3$. Energy and momentum conservation provide 
an additional factor $T^{-2}$ (the momentum exchange being thermal 
because the neutrinos are thermal).  Another factor $T$ arises from 
the fact that one is computing the energy production rate.

The rate as given by Eqs.~\eqref{eq:e_axial_Swave} and
\eqref{eq:i_axial_Swave} is applicable for the $^1S_0$ neutron
condensate in the neutron-star crust and the proton condensate in the
core of the star.  A calculation similar to that described above can
be carried out for the $^3P_2$--$^3F_2$
condensate~\cite{YakovlevKaminker1999,Leinson2013}.  The main
difference from the $S$-wave case is that the leading-order
contribution already appears at first order ($\propto 1)$ in the 
small-$v_F$ expansion, other factors and temperature dependence 
being the same.

The hyperonic $S$-wave condensates introduced above will also
contribute to the neutrino radiation through pair-breaking processes,
in full analogy to their nucleonic counterparts. These processes were
initially considered without vertex
corrections~\cite{YakovlevKaminker1999,JaikumarPrakash2001} and
revised later to account for them in~\cite{RadutaSedrakian2017}.
The inclusion of vertex corrections implies that the contribution from
vector-current coupling is negligible for $S$-wave paired hyperons,
compared to that from axial-vector coupling. By the same argument as
made for nucleons, the former contribution scales as $v_{{\rm Y}F}^4$, 
where $v_{{\rm Y}F}$ is the Fermi velocity of the Y hyperons, whereas 
the latter scales as $v_{{\rm Y}F}^2$. In analogy to the nucleon case, 
this last contribution is given by Eqs.~\eqref{eq:e_axial_Swave} and
\eqref{eq:i_axial_Swave}, but with nucleonic quantities replaced by
their hyperonic analogs \cite{RadutaSedrakian2017} including the
weak coupling constants given in \cite{Savage1997}.

\subsection{Collective modes and entrainment}
\label{sec:collective_modes}

The set of polarization tensors shown in Fig.~\ref{fig:pol_PBF} also
determines the collective modes of the fermionic superfluid. Indeed the
vector and axial-vector responses are associated in the
non-relativistic limit with the vertices 
\bea
\label{eq:vertices2}
\Gamma_0^{D\mu}=\left(1,\vecv_F\right), \quad
\Gamma_0^{S\mu}=\left(\vecsigma \cdot \vecv_F,\vecsigma\right), 
\eea 
which are the same as for the density and density-current (subscript
$D$) and the spin-current and spin-density perturbations (subscript
$S$); here $\mu$ is the Dirac index. By definition, the dispersion
relations of the collective modes are obtained from the poles of the
polarization tensor~\cite{Leggett1966,LarkinMigdal1963}. For
single-component neutral superfluids, one finds two branches of
density modes. First, there is the Anderson-Bogolyubov  mode, which 
is an acoustic mode having dispersion $\omega = c_s k$, with $\omega$ 
and $k$ the mode energy and momentum.  At zero temperature, the mode 
velocity is given by~\cite{LarkinMigdal1963,Leggett1966}
\be \label{eq:AB}
c_s = \frac{v_F}{\sqrt{3}} (1+F_0)^{1/2} \left(1+\frac{F_1}{3}\right)^{1/2},
\ee
where $v_F$ is the Fermi velocity and $F_{0,1}$ are the $l=0,1$ Landau parameters
defined in Eq.~\eqref{eq:expansion}.  The second mode is the so-called
Higgs mode and has a finite threshold as $k\to 0$ of the order of the
pair-breaking energy $2\Delta$. These modes were studied in nuclear matter 
long ago~\cite{LarkinMigdal1963}, but the case of neutron matter below 
saturation density has been considered only
recently~\cite{Keller2013,KolomeitsevVoskresensky2011,MartinUrban2014}.

An approximation based on pure neutron matter may not be accurate in a
number of problems associated with the physics of neutron star crusts. 
The lattice of nuclei in which the neutron fluid is embedded
affects its properties in a number of ways. The elementary excitations
of the lattice are phonons, and they will affect the neutron spectrum
via neutron-phonon coupling, which contributes to the neutron effective mass
as well as to the pairing interaction~\cite{Sedrakian1996}. Furthermore, 
the Anderson-Bogolyubov mode \eqref{eq:AB} of the neutron superfluid couples 
to the phonon modes of the lattice, in which case there is a mixing among the
modes~\cite{Cirigliano2011,ChamelPage2013,UrbanOertel2015}.

The problem of collective excitations in neutron-star crusts can
also be approached starting from theories of collective modes of
finite nuclei, by accounting for the possibility of the continuum
neutron states outside of the clusters. In such a treatment, the
oscillatory modes of the surface of the cluster play a new dynamical
role. In the absence of pairing, these models have been studied using 
the operator $(dU/dr)Y_{LM}$, where $U$ is the mean-field potential 
and the $Y_{LM}$ are the spherical harmonics~\cite{BaroniPhysRevC2010}. 
The quadrupole and octupole
  giant resonances were found to be similar to those of ordinary
  atomic nuclei. However, the strength functions (in other words, the
  imaginary parts of the relevant response functions) were found to be
  broadened over the energy range with corresponding reduction of the
  peak value.  The Hartree-Fock-Bogolyubov method was combined with
  the quasiparticle random-phase approximation to find the nuclear
  collective dipole excitations that correspond to density
  perturbations driven by an operator of the form $ r^LY_{LM}$in the
  same treatment, but including pairing correlations of neutrons in
  the Wigner-Seitz cell~\cite{Grasso2008NuPhA,KhanPhysRevC2005}.  A
  single low-lying mode was found in such systems that reaches its
  peak at an energy significantly lower than that of the dipole mode
  ($L=1$) in finite nuclei. This suggests that one is dealing here with
  the Anderson-Bogolyubov mode of neutron matter modified by the
  existence of a nuclear cluster in the center of the WS cell.  The
  presence of the dipole Anderson-Bogolyubov mode was confirmed in a
  later study~\cite{Inakura2017PhRvC}, which finds that this mode is
  present outside the cluster and is strongly suppressed inside
  it. That this mode shows up only for neutrons is consistent with the
  fact that these are continuum modes associated with the dripped
  neutron fluid. Its characteristics are also weakly dependent on the
  proton number $Z$ assumed for the WS cell. An additional mode was
  found at zero energy, which arises from the displacements of the
  cluster as whole, the corresponding quanta being the lattice
  phonons. The coupling of these two modes was found to be
  weak~\cite{Inakura2017PhRvC}.  

Of particular interest for the phenomenology of neutron stars is the
hydrodynamical limit, in which case a crustal layer can be considered
as a two-fluid system consisting of the unbound neutron superfluid and
plasma comprising crustal nuclei and
electrons~\cite{Carter2006,Chamel2012,Chamel2013,Chamel2017,Watanabe2017,MartinUrban2016,Pethick2010PThPS,Kobyakov2013}. 
The normal fluid is locked into the motion of the star by the magnetic
field. Its identification is not unambiguous: there are neutrons
inside and outside of nuclear clusters, and there is a ``transfusion''
from one to another under non-stationary conditions. The coupling
between the neutron superfluid (labeled $n$) and the crustal plasma
(labeled $p$) is reflected at the hydrodynamical level in the
{\it entrainment effect}, which states that the (mass) currents $\vecp$ of
the fluids are given by 
\begin{equation}
\label{eq:entrainment}
 \vecp_{ n} =\rho_{ nn} \vecv_{ n} +\rho_{ np}
 \vecv_{ p}, \hskip 0.5cm
 \vecp_{ p} =\rho_{ np}\vecv_{ n} +\rho_{ pp}\vecv_{ p}, 
\end{equation}
where $\rho_{ nn}$, $\rho_{ pp}$ are diagonal and
$\rho_{ np} = \rho_{ pn}$ are off-diagonal densities, which form a
$2\times 2$ entrainment matrix, while
$\vecv_{i} \equiv (1/2m_i) \vecnabla\phi_i-(e_i/m_i)\vecA$ with
$i \in n,p$, where $\phi_i$ is the phase of the pairing amplitude,
$\vecA$ is the vector pontential, and $e_i$ and $m_i$ are the
charge and mass of component $i$.  Note that the vector $\vecv$
transforms as a co-vector and should not be confused with the proper
velocity of the fluid, which is a contravariant
vector~\cite{Carter2006,Chamel2012}. The off-diagonal densities
$\rho_{np} = \rho_{pn}$ account for the fact that the mass current of
a given component is not aligned with the gradient of the amplitude phase.

Before proceeding, we point out that entrainment was originally introduced
in the context of mixtures of superfluid phases of
He~\cite{Andreev1976}.  It was then applied in the context of neutron
stars to describe the mixtures of neutron and proton superfluids in
the star's
core~\cite{Vardanyan1981,Alpar1984,SedrakyanShakhabasyan1980}.  See
the discussion in Sec.~\ref{sec:mutual_friction}.  The elements of the
entrainment matrix are related to each other by Galilean invariance,
so it is sufficient to determine, for example, only the ratio
$\rho_{ pn}/\rho_{pp}$, known as the {\it entrainment
coefficient}. Relations \eqref{eq:entrainment} demonstrate that static
computations of the density of neutrons outside nuclear clusters
cannot be used as a measure of the density of the superfluid neutron
component.

The classical hydrodynamical interaction between neutron superfluid and
crustal nuclei has been studied extensively~\cite{Epstein1988,Sedrakian1996,Bulgac2004,MartinUrban2016}.  Flow of a neutron superfluid past a nucleus
induces a backflow, thereby endowing the nucleus with a hydrodynamic
mass.  The amount of ``free'' neutron superfluid determined in this manner 
(\ie, that moving with velocity $\vecv_{n}$) can be expressed in terms
of the ratio~\cite{MartinUrban2016,Chamel2017}
\begin{equation}\label{eq:k_ent}
\frac{\rho_{nn}}{\rho_{ n}}=
1+3\frac{V_A}{V_{\rm cell}}\frac{\delta-\gamma}{\delta+2\gamma},
\end{equation}
where $\gamma$ is the density ratio of the neutrons outside and inside
of the nucleus in the static limit, $\delta$ is the fraction of
superfluid neutrons within a nuclear cluster of volume $V_A$, and
$V_{\rm cell}$ is the volume of the Wigner-Seitz cell. Consider for
illustration the case $\delta=0$, which corresponds to the limit of an
impenetrable cluster; then the ratio \eqref{eq:k_ent} is independent
of $\gamma$ and one finds a lower bound
$\rho_{nn}/\rho_{ n}= 1- (3/2)(V_A/V_{\rm cell})$ with
$V_A/V_{\rm cell}\ll 1$.  We see that $\rho_{nn}\simeq \rho_{ n}$,
\ie, the entrainment is weak, which is confirmed by detailed
computations~\cite{MartinUrban2016}.  The hydrodynamic models of
entrainment assume that the coherence length of the neutron superfluid
is much smaller than other scales in the problem, in particular, the
size of the nucleus. On the other hand, band-structure calculations
(analogous to those in the theory of solids) predict a strong
entrainment, with the density of superfluid neutrons reduced by an
order of magnitude~\cite{Chamel2012,Chamel2013,Chamel2017}. While
these results were obtained in the limit where the pairing can be
neglected compared to other scales, specifically the depth of the
lattice potential, a semi-analytical model \cite{Watanabe2017} that
includes pairing correlations suggests a rather weak entrainment. It
is in the range predicted by the hydrodynamical models. The depth of
the potential in this study is of the order of the pairing gap, so it
cannot be neglected. Further studies of this problem are needed in
order to resolve this discrepancy.  Phenomenologically, strong
entrainment would imply that there is not enough moment of inertia in
the superfluid component of the crust to account for pulsar glitch
dynamics~\cite{Chamel2013,Watanabe2017,Andersson2012}.

In the high-density region, corresponding to the quan\-tum-liquid core
of the star, the proton-electron component supports plasma modes,
which couple to the modes of the neutron superfluid. These modes were
studied microscopically on the basis of the linear response theory
in~\cite{BaldoDucoin2009,BaldoDucoin2017}.  An addition new degree of
freedom is associated with the spin of Cooper pairs in a $P$-wave
superfluid. As a consequence, such superfluid admits additional
modes~\cite{Bedaque2015,BedaqueNicholson2013,BedaqueReddy2014,Bedaque2003,Leinson2012,Leinson2011}.
Apart from the usual first and second sound modes, small-amplitude
hydrodynamical oscillatory modes of the nucleonic fluids give rise to
new modes due to the entrainment and coupling to plasma oscillations
of the electron-proton component~\cite{Epstein1988,Vardanyan1981}.

Phenomenologically, as new degrees of freedom, the collective modes
contribute to the thermodynamics of the superfluid. The specific heat
is of particular interest for the cooling of neutron stars
\cite{Keller2013,MartinUrban2014,Gallo2011,ChamelPage2013}. The
collective modes (phonons) can lose energy by neutrino
emission~\cite{Bedaque2003,Leinson2011} and can contribute to the
transport~\cite{Manuel2014,Mannarelli2013,Manuel2011,Aguilera2009}.

\subsection{Urca process in superfluid phases}
\label{sec:Urca_superfluid}

Urca processes involving nucleons \eqref{eq:urca} or  hyperons
\eqref{eq:UrcaLambda}-\eqref{eq:Xizero} are operative at high
densities for many models of the equation of state of dense matter.
Consequently, it is important to understand how the rates of these
processes are affected by 
superfluidity of the
baryons. At
asymptotically low temperatures $T\ll {\rm min}(\Delta_n, \Delta_p)$,
where $\Delta_{n/p}$ are the neutron/proton gaps, the neutrino
radiation is suppressed by a Boltzmann factor
${\rm exp}(-\Delta_{\rm max}/T)$, with
$\Delta_{\rm max} = {\rm max}(\Delta_n, \Delta_p)$  the larger of
the neutron and proton gaps~\cite{LattimerPrakashHaensel}.  However, a
substantial part of the neutrino cooling of a neutron star occurs in
the temperature range $0.2\le T/T_c\le 1$, where $T_c$ is
the relevant critical temperature for either neutron or proton
pairing. Hence
 a more accurate description of the suppression of Urca processes is
required. Because pairing mainly affects the phase space of nucleons,
an initial step is to introduce the BCS spectrum in the distribution 
functions of nucleons when computing 
the rate of the Urca process~\cite{LevenfishYakovlev1994}.  In principle,
the matrix element of the process is also modified because one is
dealing with a coherent state composed of a superposition of particles
and holes as expressed by the coherence factors $u_p/v_p$.  
\begin{figure}[t]
\begin{center}
\includegraphics[angle=0, width=0.5\columnwidth]{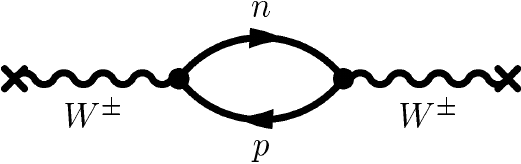}
\end{center}
\caption{The lowest order (one-loop) Urca process, which involves weak
  charged currents.  Due to the charge conservation in the weak
  vertex, the loop with anomalous GF (the analog of the
  second diagram in Fig.~\ref{fig:pol_axial}) does not contribute at
  the one-loop order. }
\label{fig:urca}
\end{figure}
It is convenient to carry out the computation using the GF
 for baryons~\cite{SedrakianUrca2005,Sedrakian2007PrPNP}. To
lowest order (\ie, neglecting the vertex corrections discussed above)
the one-loop contribution to the Urca process is given by the diagram
shown in Fig.~\ref{fig:urca}. A new feature in such computation that
keeps $u_p\neq 1$ and $v_p \neq 0$, is the emergence of the
pair-breaking process in the Urca channel. The polarization tensor
computed from the diagram in Fig.~\ref{fig:urca} contains
contributions not only from the scattering processes
$\propto [f_n(\vecp)-f_p(\vecp+\vecq)]$, where $f_{p/n}$ are the
proton and neutron distribution functions and $\vecq$ is the momentum
transfer, but also from processes
$\propto [1-f_n(\vecp)-f_p(\vecp+\vecq)]$ that are due to the breaking
of neutron and proton Cooper pairs. Close to the critical temperature
$0.5 \le T/T_c\le 1$ the scattering contribution is dominant, but at
lower temperatures, the pair-breaking contribution becomes comparable
to the scattering contribution, without changing cooling behavior
qualitatively.

Neutron stars are seismically active bodies.  Density oscillations
(more specifically first sound in a superfluid) can induce variations
in the chemical potentials of species which can modify the Urca
process rate in the superfluid phases. Large enough density
oscillations can displace the Fermi seas of nucleons and bridge the
gap~\cite{AlfordReddySchwenzer2012,AlfordPangeni2017}. This
super-thermal effect may strongly enhance the rate of the Urca process
in the superfluid up to levels comparable to that of the normal state.
In hadronic matter, the relative amplitudes of the density oscillations
required for this effect to be operative are of the order of
$\Delta n/n\sim 10^{-3}$. Consequently, an (unstable) growth of
oscillation amplitude in a superfluid can saturate due to the
dissipation of the energy of oscillations via neutrino
emission~\cite{AlfordMahmoodifar2012}.  Out-of-equilibrium Urca
processes in the superfluid phases are also important for
understanding the coupled rotational and chemical evolution of neutron
stars~\cite{PetrovichReisenegger2011,PetrovichReisenegger2015}.

\subsection{Axion radiation from superfluid phases}

Superfluid phases of neutron stars may radiate not only the three
neutrino flavors encountered in the Standard Model (SM), but
hypothetical particles that have been conjectured in various extension
of the SM. Confrontation of theoretical tracks of neutron star cooling
with measurements of X-ray flux from suitable neutron-star candidates
thus can constrain the properties of such particles and their
coupling to the SM sector. We discuss this possibility using the
specific example of QCD axions, which were originally introduced
in~\cite{1978PhRvL..40..279W} and \cite{1978PhRvL..40..223W}
to solve the strong-CP problem in
QCD~\cite{PecceiQuinn1977,1976PhRvL..37....8T}.

Stellar physics has indeed been widely used to put constraints on 
models of particle physics beyond SM.  As non-SM particles can be
produced in stellar environments, they can contribute to transport
and losses of energy.  This allows setting constraints on the strength
of coupling of these particles to SM matter, by requiring that their
existence does not introduce contradictions in estimates of stellar
lifetimes and energy-loss
rates~\cite{Raffelt1996,Raffelt2008,Giannotti2017}. This kind of
astrophysical limit has been obtained from the physics of the Sun, red
giants and horizontal-branch stars in globular clusters, white dwarfs,
and neutron stars, and from the duration of the neutrino burst of the
supernova SN1987A~\cite{Agashe:2014kda}.
In the case of neutron stars, we need to assume that axion
emission, which carries additional energy away from the stellar interior,
does not significantly alter the agreement between theoretical cooling
models and observations.

The computation of the pair-breaking process
\begin{eqnarray}
\label{eq:Na_PBF}
N+N\to {\cal C} + a 
\end{eqnarray}
involving emission of an axion $a$ is analogous to that of the axial-current
neutrino emission, since the axion couples to the nucleonic axial current. 
The required response function is represented by Fig.~\ref{fig:pol_axial}, where
now an axion is attached to the nucleonic loop instead of a $Z_0$ gauge boson.

To set the notation, we start with the interaction Lagrangian 
\be 
\mathscr{L }^{(B)}_{int} = \frac{1}{f_a} B^{\mu}  A_{\mu},
\ee
in which $f_a$ is the axion decay constant, and the baryon and axion current 
are given by
\be\label{eq:currents}
B^{\mu} =
\sum_{N} \frac{C_N}{2}  \bar\psi_N\gamma^{\mu}\gamma_5\psi_N,\quad \quad 
A_{\mu} = \partial_{\mu} a,
\ee
where $C_N$ is the Peccei-Quinn (PQ) charge of a baryonic
current and we denote nucleons collectively by
$N\in n,p$.  The dimensionless Yukawa coupling can be defined as
$g_{aNN} = C_Nm_N/f_a$, from which it follows that the axionic
``fine-structure constant'' is $\alpha _{aNN} = g^2_{aNN}/4\pi$.

\begin{figure*}
\begin{center}
\includegraphics[width=11cm]{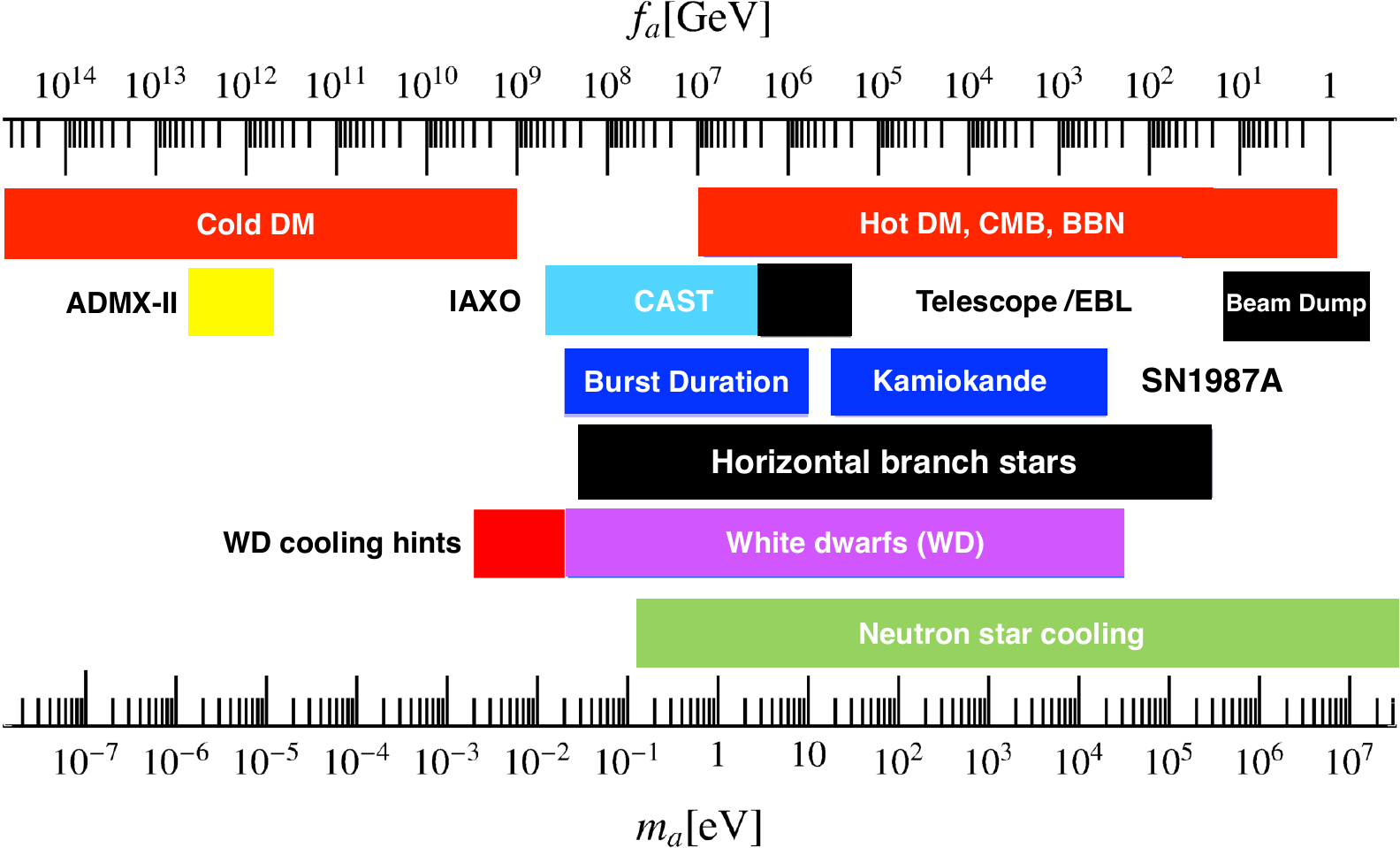}
\end{center}
\caption{Regions of exclusion for axion masses and coupling derived
  from combined experimental and theoretical studies. These are
  organized in rows with regions of exclusion derived from (top to
  bottom): (1) cosmology, (2) laboratory experiments, (3) supernova
  1987A event, (4) horizontal branch stars, (5) physics of white
  dwarfs, and (6) neutron-star cooling. The bounds (1)-(5) are taken
  from ~\cite{Agashe:2014kda}, while the last one is based on
  comparison between numerical simulations and X-ray data on surface
  photon luminosity of thermally emitting neutron
  stars.}
\label{fig:axion_limits}
\end{figure*}

The charges introduced above are given by generalized Goldberger-Treiman 
relations 
\bea 
C_p & =& (C_u-\eta)
\delta_u+(C_d-\eta z) \delta_d +(C_s-\eta w)\delta_s,\nonumber\\
\\
C_n & =& (C_u-\eta) \delta_d+(C_d-\eta z) \delta_u 
+(C_s-\eta w)\delta_s, \nonumber\\
\eea 
where $\eta = (1+z+w)^{-1}$, with $z = m_u/m_d$, $w = m_u/m_s$, and
$\delta_u =0.84\pm 0.02$, $\delta_d = -0.43\pm 0.02$, and
$\delta_s = -0.09\pm 0.02$. The main uncertainty is associated with
$z = m_u/m_d = 0.35$--$0.6$. While there are numerous models of
axions, a particularly useful model is the hadronic axion model 
\cite{KimPhysRevLett1979,Shifman1980} with $C_{u,d,s} = 0$; 
in this model, the nucleonic charges vary in the ranges
\bea \label{eq:axion_range} -0.51 \le C_p\le -0.36, \quad -0.05 \le
C_n\le 0.1.  
\eea 
The axion mass is related to $f_a$ by
\bea 
\label{eq:axion_mass}
 m_a = \frac{z^{1/2}}{1+z} \frac{f_\pi m_\pi}{f_a} = \frac{0.6 
  ~\textrm{eV}}{f_a/10^{7}~\textrm{GeV}} 
\eea 
in terms of the pion mass $m_\pi=135$ MeV and the decay constant
$f_\pi = 92$ MeV, having adopted the value $z=0.56$ from the range of
$z$ values quoted.  Equation~\eqref{eq:axion_mass} translates a lower
bound on $f_a$ to an upper bound on the axion mass.

Computations analogous to those for neutrinos lead to the result 
\bea\label{eq:axion1} \epsilon^S_{aN}  &=& \frac{ 2 C^2_N}{3\pi} \, f_a^{-2}
\,\nu_N(0)\, v_{FN}^2 \, T^5 \, I^S_{aN}, 
\eea 
for the axion emissivity from $S$-wave condensates~\cite{2013NuPhA.897...62K}, 
where
\bea \label{eq:Ia}
I^S_{aN} =
z_N^5\int_1^{\infty}\!\! dy ~ \frac{y^3}{\sqrt{y^2-1}} 
f_F^2\left(z_N y\right) 
\eea 
and $z_N= \Delta^S_N(T)/T$.  Here $\Delta^S_N$ refers to the $S$-wave
nucleonic gap. In a first approximation a bound on $m_a$ can
be obtained by requiring that the axion cooling does not
overshadow the neutrino cooling (which is assumed to be dominated by
the $S$-wave neutrino radiation), \ie,
\bea \label{eq:ratio2}
\frac{\epsilon^S_a}{\epsilon^S_{\nu\bar\nu}} &\simeq & 15 
\frac{C_N^2}{ f_a^{2} G_F^2 } \frac{r(z)}{\Delta^S_N(T)^2}\le 1,
\eea
where $r(z)$ is the ratio of the phase-space integral for axions
\eqref{eq:Ia} to its counterpart \eqref{eq:i_axial_Swave} 
for neutrinos and is numerically bounded from above by
$r(z)\le 1$. Hence this factor can be dropped from the bound 
on $f_a$.  Substituting into Eq.~\eqref{eq:ratio2} the value 
of the Fermi coupling constant $ G_F = 1.166\times 10^{-5}$ GeV$^{-2}$,
we may convert this bound to
\be\label{eq:fbound} 
\frac{f_a/10^{10}\textrm{GeV}}{C_N} >  0.038
\left[\frac {1~\textrm{MeV}}{\Delta^S(T)}\right].
\ee
Using Eq.~(\ref{eq:axion_mass}), this translates to an upper bound
on the axion mass of
\be \label{eq:mbound} 
m_a\,C_N\le 0.163~\textrm{eV}\,\left(\frac{\Delta^S_N(T)}{1~\textrm{MeV}}\right).
\ee 
The nucleon pairing gap on the right-hand side can, in fact, be replaced
by the critical temperature $T_c$, because in the temperature range
which is important for pair-breaking processes, \ie,
$0.5\le T/T_c< 1$, BCS theory predicts $\Delta (T) \simeq T_c$.

As explained previously, the neutron condensate in neutron-star cores
is paired in the $^3P_2$--$^3F_2$ channel, \ie,  in a state which features an
anisotropic gap~\cite{ZverevClarkKhodel2003}.  The corresponding axion
emissivity is found to be~\cite{Sedrakian2016PRD}
\bea\label{eq:axion2}
\epsilon^P_{an} &=& \frac{ 2 C^2_n}{3\pi} \, f_a^{-2} \,\nu_n(0)\,
 T^5 \, I^P_{an}, 
\eea
where 
\bea 
\label{eq:IaP} I^P_{an} =\int \frac{d\Omega}{4\pi}
z_N^5\int_1^{\infty}\!\! dy ~ \frac{y^3}{\sqrt{y^2-1}} f_F^2\left(z_N 
  y\right). 
\eea 
Here $\int d\Omega$ denotes integration over the solid angle, and $z_N
= \Delta^P(T, \theta)/T$ depends on the polar angle $\theta$, where
$\Delta^P(T,\theta)$ is the pairing gap in the $P$-wave channel.
Note that $C_n=0$ is not excluded; \ie, it is conceivable that 
axions are not emitted by the neutron $P$-wave condensate. 

The axion emissivities \eqref{eq:axion1} and \eqref{eq:IaP} scale with
temperature as $\propto T^5$. This scaling differs from its neutrino
counterpart \eqref{eq:e_axial_Swave}, which is $\propto T^7$.
Accordingly, axionic cooling processes would change the slope of the
cooling curves in the temperature-age diagram.  Detailed numerical
simulations of axionic cooling \cite{Sedrakian2016PRD} yield the
regions of exclusion of axion masses and couplings illustrated in
Fig.~\ref{fig:axion_limits}. As seen in this figure, the results from
axion cooling simulations of neutron stars and their comparison with
the X-ray data on thermally emitting neutron stars, which depend
crucially on the axion emission by superfluid phases,
are compatible with other constraints derived from stellar physics.

\section{Quantum vorticity}
\label{sec:vorticity}

\subsection{Motivation}

The motivation for the study of vorticity in nuclear systems derives
from the fact that neutron stars are rotating and that neutrons, which
form a neutral superfluid, must rotate by forming quantized rotational
vortices. Although it has been conjectured that vortex states exist in
finite nuclei, the coherence length of the nucleonic condensate, which
sets the size of the vortex core, is of the order or larger than the
nuclear radius.  Vorticity is generic to superfluids and superconductors 
and, apart from ordinary metallic superconductors, is also encountered 
at the atomic level in Bose-condensed liquid $^4$He and fermionic 
superfluid $^3$He~\cite{volovik2009universe}, as well as in ultracold atomic
gases of bosons and fermions~\cite{PitaevskiiBEC,pethick_smith_2008}.

In neutron stars, rotation at angular velocity
$\Omega$ induces a array of neutron vortices with number density per
unit area
\bea
\label{eq:vortex_nn}
n_n^{(V)} = \frac{2\Omega}{\kappa}, \qquad \kappa = \frac{\pi}{m_n},
\eea
where $\kappa$ is the quantum of circulation and  $m_n$ is the neutron 
mass. In the parameter range where the proton superconductor in neutron 
stars is of type II, electromagnetic vortices are formed with a 
density 
\bea\label{eq:AS:np}
n_p^{(V)} =\frac{B}{\phi_0}, \qquad \phi_0 = \frac{\pi}{e} ,
\eea
where $\phi_0$ is the flux quantum and $B$ is the mean magnetic-field
induction.  The  vortex lattices of neutron ($n$) and proton 
($p$) superfluids are triangular with basis-vector lengths given by
\bea
   d_n= \left(\frac{\kappa}{\sqrt{3}\,  \Omega}   \right)^{1/2},
\quad \quad  d_p=\left(\frac{2\, \phi_0}{\sqrt{3} \, B}   \right)^{1/2},
\eea
which are of order $10^{-4}$ cm and $10^{-9}$ cm, respectively, for 
rotation periods of the order of a fraction of second and
fields $B\sim 10^{12}$ G. The latter scale $d_p$ is larger than the
penetration depth of the magnetic field, $\lambda \simeq 10^{-11}$ cm,
set by the Meissner mass of a photon inside the proton superconductor.

These length scales define a new {\it mesoscopic scale} for the description
of neutron-star superfluids and superconductors, since an averaging
over a large number of vortices is required to obtain the
hydrodynamical fluid velocity and the macroscopic value of the
magnetic field. The microscopic scale is set by the size of the vortex
core, which for charged and neutral fermionic superfluids alike is
given by the coherence length $\xi$. Within the region $r\le \xi$,
where $r$ is the radial cylindrical coordinate, the order parameter of
the superfluid is suppressed linearly for $r\to 0$, vanishing at its
center. From the microscopic point of view, the core of a vortex
contains a new type of excitation -- a quasiparticle bound state that
emerges from solution of the microscopic Bogolyubov-De Gennes
(BdG) theory~\cite{Gennes1999superconductivity,Caroli1964,Yu2003}. 
This section is devoted to the physics of these excitations and their 
interactions with matter, which give rise to mutual friction. The primary
motivation for studies of mutual friction in neutron stars is a deeper
understanding of the non-stationary dynamics of neutron-star rotation,
in particular, the phenomena of glitches and post-glitch relaxation in
pulsars; for a recent review and further references
see~\cite{Haskell2017}.

\subsection{Vortex core quasiparticles}

The microscopic theory of bound states of a fermionic vortex was
initially developed in \cite{Caroli1964} for a vortex in a type-II
superconductor. Their approach is based on the solution of the BdG
equations for the pairing amplitudes $u(\vecr)$ and $v(\vecr)$ given by 
Eq.~\eqref{eq:u_and_v}, but expressed in configuration space.  These 
early results were soon adapted to neutron vortices, so as to obtain the
coefficients of mutual friction in the core of a neutron star in terms
of interactions of the neutron quasiparticles bound in the vortex core
with ambient electrons~\cite{Feibelman1971}.

An isolated neutron vortex was studied
in~\cite{deBlasio1999,deBlasio2001,Yu2003} by solving the BdG
equations in neutron matter.  Substantial depletion in the region of
the vortex core was found in \cite{Yu2003}, a feature uncharacteristic
of condensed-matter vortices. Density depletion in vortex cores is
important since it allows the vortices to be detected experimentally
in ultracold atomic gases~\cite{Zwierlein2006}.  Theoretically, the
vortex profile in an ultracold atomic gas was investigated in a
population-imbalanced
gas~\cite{Takahashi2007,Iskin2008,Warringa2011,Warringa2012} and
across the BCS-BEC
crossover~\cite{MachidaKoyama2005,Chien2006,MachidaOhashi2006}.

The BdG theory can be derived using the Green functions formalism introduced 
in Sec.~\ref{sec:GF}, with specialization to configuration space. In this 
case, the Dyson-Schwinger equation for the Nambu-Gor'kov GF 
takes the form
\begin{equation}
G^{-1}(X, X')
=
-  \left(\frac{\partial}{\partial \tau} + 
H \right) \delta(X-X'),
\label{eq:dsB}
\end{equation}
where $X = (\vecr,\tau)$ is the four-coordinate including the imaginary
time $\tau$, while 
\begin{multline}
\label{eq:hbdg}
H = \left(
\begin{array}{cc}
 h(\Omega) \!-\! \mu_\uparrow \! +\! g n_{\downarrow}(\vecr)
&
\Delta(\vecr)
\\
\Delta^*(\vecr)
&
- h(\Omega)^* \!+\! \mu_\downarrow \!-\! g n_{\uparrow}(\vecr) 
\end{array}
\right).
\end{multline}
In this expression, $h(\Omega)$ denotes the single-particle Hamiltonian 
in the frame rotating with frequency $\Omega$, the symbols 
$\downarrow,\uparrow$ refer to spin-down and spin-up particles 
with chemical potentials $\mu_{\downarrow,\uparrow}$ and densities 
$n_{\downarrow,\uparrow}$, and $g$ is the strength of the assumed 
four-fermion contact interaction.  

The solutions of the Dyson-Schwinger 
equation are obtained by inversion of Eq.~(\ref{eq:dsB}).   This is 
achieved by solving the BdG equation
\begin{equation}
H\left(
\begin{array}{c}
u_i(\vecr)
\\
v_i(\vecr)
\end{array}
\right)
= 
E_i
\left(
\begin{array}{c}
u_i(\vecr)
\\
v_i(\vecr)
\end{array}
\right)
\label{eq:BdGB}
\end{equation}
for the amplitudes $u_i(\vecr)$ and $ v_i(\vecr)$, where the index $i$
refers to the particle's spin state and the energies $E_i$ are the
eigenvalues of the BdG equation~\footnote{At this point, it is
    worthwhile to draw reader's attention to the analogy between the
    BdG equations and the HFB equations used to describe finite
    nuclei. The BdG equations are written in the presence of external
    vector field $\vec\Omega$ in cylindrical geometry and thus
    describe bound states in the plane orthogonal to $\vec\Omega$. The
    HFB equations for nuclei, in contrast, describe bound states in a
    finite three-dimensional volume and in the absence of external
    electromagnetic fields or rotation are invariant against rotation
    in space. (Note that some nuclei may be spontaneously deformed, in
    which case the rotational O(3) symmetry will be broken down to
    some subgroup).  This analogy implies that the same numerical
    methods can be effectively applied for the solution of BdG and HFB
    equations.}.
The functions $u_i(\vecr)$ and $v_i(\vecr)$ are normalized by
$\int d^3 r\, \left[ \vert u_i(\vecr) \vert^2 + \vert v_i(\vecr)
  \vert^2 \right] = 1$.
The densities of up-spin and down-spin fermions, written as
\begin{eqnarray}
 n_{\uparrow}(\vecr) &=& \sum_i f(E_i) \vert u_i(\vecr) \vert^2,
\label{eq:densupA}
 \\ 
 n_{\downarrow}(\vecr) &=& \sum_i f(-E_i) \vert v_i(\vecr) \vert^2
\label{eq:densdownB}
\end{eqnarray} 
in terms of the Fermi-Dirac distribution function $f(E) $, are to be
determined simultaneously with the solution of the BdG equation.  The
gap function $\Delta(\vecr)$ is obtained from the anomalous component
of the GF, which is given by
\begin{equation}
F_{\uparrow \downarrow}(\vecr, \tau; \vecr', \tau) 
= \sum_i f(E_i) u_i(\vecr) v_i^*(\vecr')
\label{eq:gupdown}
\end{equation}
in the limit $\vecr' \rightarrow \vecr$.  One finds a relation
between the gap and the GF in Eq.~\eqref{eq:gupdown} of the following 
form
\begin{equation}
  F_{\uparrow \downarrow}(\vecr, \tau; \vecr', \tau) = 
  -\frac{ m\Delta(\vecr)}{4 \pi } \frac{1}{\vert \vecr- \vecr' \vert}
 + F_{\uparrow \downarrow}^{\mathrm{reg}}(\vecr, \tau; \vecr, \tau),
\end{equation}
where $m$ is the (effective) mass of fermions and the regular part 
$F_{\uparrow \downarrow}^{\mathrm{reg}}(\vecr, \tau; \vecr, \tau)$ 
of the GF  can be found elsewhere~\cite{Yu2003}.
The Helmholtz free energy ${\cal F}$ can be now evaluated using the 
 solutions of the BdG equation, according to
\begin{eqnarray}
\label{eq:thermopotC1}
{\cal F}&= &- 
\sum_{i} \left[ \frac{ \vert E_i \vert}{2}   
+ \frac{1}{\beta} \log \left(1+\exp^{-\beta \vert E_i \vert} \right)
\right]+ \sum_{i} \epsilon_i\nonumber\\
& - &
\int  d^3 r \,
F_{\uparrow \downarrow}(\vecr, \tau; \vecr, \tau)^*
\Delta(\vecr)\nonumber\\ 
& - &
g \int  d^3 r \, 
n_{\uparrow}(\vecr) n_{\downarrow}(\vecr)
+ \mu_\uparrow N_\uparrow + \mu_\downarrow N_\downarrow ,
\end{eqnarray}
where $\epsilon_i$ are the eigenvalues of the Hartree-Fock Hamiltonian
$H_{\mathrm{HF}} = H(\Omega = 0) - \mu + g n(\vecr)$.
It should be mentioned that some of the individual terms in 
Eq.~\eqref{eq:thermopotC1} are ultraviolet-divergent, but their sum, 
and hence the Helmholtz free energy, is ultraviolet-finite.  This equation
allows one to determine the parameter space spanned in the phase diagram 
by the coupling $g$, the population imbalance, the rotation frequency, 
and relevant thermodynamic quantities.

We now present approximate solutions of BdG equations that
provide insight into recent numerical work. The states of the vortex
core can be approximated as~\cite{1965PKM.....3..380C}
\be\label{13}
\left(\begin{array}{c}
u_{q_{\pp},\mu}(\vecr_{\perp})\\ v_{p_{\pp}, \mu}(\vecr_{\perp})\end{array}\right) 
= e^{ip_{\pp}z} \left(e^{i\theta(\mu-\frac{1}{2})}
~e^{i\theta(\mu+\frac{1}{2})}\right) \left(\begin{array}{c}
u'_{\mu}(r)\\ v'_{\mu}(r)\end{array}\right), 
\ee
where the vector $\vecr = (r,\theta, z)$ has been decomposed into cylindrical
coordinates with the $z$-axis along the vortex circulation, $\pp$ and
$\perp$ being its components parallel and perpendicular to the vortex
circulation. Here $\mu$ labels the azimuthal quantum number, which
assumes half-odd-integer positive values. It is seen that the vortex-core 
states are plane waves along the vortex circulation, but are quantized in the
orthogonal direction. The radial part of the wave function is given by
\be\label{14}
\left(\begin{array}{c}
u'_{\mu}(r)\\ v'_{\mu}(r)\end{array}\right)=2\left(\frac{2}{\pi p_{\perp}r}\right)^{1/2}
e^{-K(r)} \left(\begin{array}{c}
{\rm cos}\left(p_{\perp}r - \frac{\pi\mu}{2} \right) \\ 
{\rm sin}\left( p_{\perp}r - \frac{\pi\mu}{2} \right)\end{array}\right),
\ee
where $p_{\perp} = \sqrt{p^2-p_F^2}$, $p_F$ being the neutron Fermi momentum. 
The function in the exponent is 
\be \label{15}
K(r) = \frac{p_F }{\pi p_{\perp} \Delta_{\infty}} \int_0^r\Delta(r') dr'\simeq
\frac{p_F r}{\pi p_{\perp}\xi}\, \left(1 + \frac{\xi e^{-r/\xi}}{r}\right),
\ee
where $\Delta_{\infty}$ is the asymptotic value of the gap far from
the vortex core, while $\xi$ is the coherence length. For small momenta,
the vortex core quasiparticles have energies given by
\be \label{eq:vortex_eigenstate}
\ep_{\mu}(p) 
 \simeq \frac{\pi\mu\Delta^2_{\infty}}{2\ep_{F}}\left( 1+\frac{p^2}{2p_F^2}\right),
\ee
where $\ep_{F}$ is the Fermi energy. 

\subsection{Vortex dynamics and pinning}
\label{sec:dynamics}

Following the suggestion in \cite{1975Natur.256...25A} that the
neutron superfluid dynamics is driven by the interaction of vortices
with the nuclear lattice in the inner crust of a neutron star, many
calculations have been performed in efforts to understand the
pinning-type interactions between vortices and
nuclei.  This is, in general, a time-dependent problem,
but the static interactions are of great interest as well.  Indeed,
the stationary minimum energy state of a neutron vortex could require
its pinning to a nucleus (with geometrical overlap), or,
alternatively, its pinning in the space between nuclei if the
vortex-nucleus interaction is repulsive.

Stationary studies of pinning in neutron stars compare the energy
difference between a configuration where nucleus and vortex are well
separated with a configuration in which they intersect. A naive
picture suggests that the energy required to create the vortex core
quasiparticles out of the condensate is gained if the vortex passes
through the nucleus~\cite{Alpar1984,Pines1980}. A more flexible and
quantitative basis is offered by Ginzburg-Landau
theory~\cite{EpsteinBaym1988}, recognizing that other contributions to the
Ginzburg-Landau functional besides the condensation energy can play
key roles.  In this approach, whether the vortices pin on nuclei or in
between them depends on the density; typically high densities favor
pinning to nuclei. Similar conclusions have also been reached in
semi-classical models that assume a realistic Argonne
interaction~\cite{DonatiPizzochero2004,DonatiPizzochero2006}; however,
the magnitude of the pinning energy or force is smaller by an order of
magnitude compared to what is found in the Ginzburg-Landau
models. Microscopic solutions of the BdG equations for the pinning
problem exist~\cite{Avogadro2008}, but the results for pinning
energies are not conclusive.

A number of time-dependent formulations of the vortex-nucleus
interaction go beyond static considerations that simply compare
the energy differences between stationary pinned and unpinned
configurations. Dynamical studies have included (i) purely
hydrodynamical modeling~\cite{Sedrakian1995,Link2009}, (ii) modeling
based on Gross-Pitaevskii-like equations~\cite{BulgacForbes2013} and,
most recently, (iii) application of time-dependent superfluid density
functional theory~\cite{Sekizawa2016}.  The last study captures most
of the microphysics, and it concludes that nuclei repel vortices in
the neutron-star crust, \ie, if pinned, vortices reside in between the
nuclear clusters.

\subsection{Mutual friction}
\label{sec:mutual_friction}

Mutual friction arises through the interaction of vortices with the
ambient non-superfluid components of neutron star matter. Analogous
phenomena have been investigated extensively in the context of liquid
He-II hydrodynamics~\cite{Sonin2016}, but the context of neutron stars
is unique because both the ambient fluid and the vorticity are of
fermionic nature. We next review the microphysics and kinetics of
particle interactions with the bound states in the vortex cores of
quantized vortices.  Electrons will couple to the core quasiparticles
of the neutron vortex via the interaction of the electron charge $-e$
with the neutron magnetic moment $\mu_n = -1.913\mu_N$, where
 $\mu_N = e/2m_p$ is the nuclear
magneton~\cite{Feibelman1971}. The relaxation time scale for the
electron momentum due to scattering by neutron vortex-core
quasiparticles is given by~\cite{1989ApJ...342..951B}
\bea \label{eq:S_wave_relax}
\tau_{eV}[^1S_0] &=& \frac{1.6\times
  10^3}{\Omega} \frac{\Delta}{T}
\left(\frac{\ep_{Fe}}{\ep_{Fn}}\right)^2\nonumber\\
&\times&\left(\frac{\ep_{Fn}}{2m_n}\right)^{1/2}
\exp\left(\frac{\ep^0_{1/2}}{T}\right),
\eea 
where $\ep_{Fe}/\ep_{Fn}$ are the electron/neutron Fermi energies,
$\Delta$ is the $S$-wave neutron pairing gap, $\ep^0_{1/2}$ is given
by Eq.~\eqref{eq:vortex_eigenstate} with $\mu=1/2$, and $\Omega$ is
the angular velocity of the superfluid, which enters through the
number of scattering centers per cm$^2$ according to
formula~\eqref{eq:vortex_nn}.  We see that the relaxation time is
inversely proportional to the Boltzmann factor that measures the
probability of finding core quasiparticle states at a given
temperature.

The electron dynamics in the stellar core is strongly affected by the proton
component, but we assume for the time being that electrons interact
exclusively with neutron vortices in a $P$-wave superfluid.  The
order parameter in the $P$-wave case has a tensor character and 
can be written as a traceless and symmetric function $A_{\mu\nu}$,
$\mu, \nu = 1,2,3$. This function can be decomposed in cylindrical 
coordinates ($r,\phi,z$) as \cite{1982PhRvD..25..967S} 
\bea\label{Pparameter}
A_{\mu\nu}  &=& \frac{\Delta}{\sqrt{2}} e^{i\phi} 
\Big\{
[f_1 \hat r_{\mu}\hat r_{\nu}
+f_2 \hat\phi_{\mu}\hat\phi_{\nu}\nonumber\\
&-&(f_1+f_2) \hat z_{\mu}\hat z_{\nu}
+if_3 ( r_{\mu}\hat\phi_{\nu}+ r_{\nu}\hat\phi_{\mu}
)]\Big\},
\eea
where $f_{1,2,3}(r)$ are the radial functions describing the vortex
profile and $\Delta$ is the average value of the gap in the $^3P_2$
channel.  The $P$-wave vortices that are described by the order
parameter \eqref{Pparameter} possess intrinsic magnetization because
the relevant Cooper pairs are spin-1 objects.  Thus, the interaction
of electrons with $P$-wave superfluid vortices is driven by the
electromagnetic interaction associated with coupling of the electron
charge to the magnetization of the vortex. The relaxation time for the
electron-vortex scattering is obtained as~\cite{1982PhRvD..25..967S}
\bea\label{eq:P_wave_relax}
\tau_{eV}[^3P_2] \simeq \frac{7.91 \times 10^{8}}{\Omega}
\left(\frac{k_{Fn}}{\textrm{fm}}\right)
\left(\frac{\textrm{MeV}}{\Delta_n}\right)
\left(\frac{n_e}{n_n}\right)^{2/3}. \nonumber\\
\eea
In contrast to the case of scattering off the quasiparticles, the
relaxation time \eqref{eq:P_wave_relax} is nearly independent of 
temperature, the only temperature-dependent quantity being the gap.
The result \eqref{eq:P_wave_relax} sets a lower limit on the
scattering rate at low temperatures ($T\ll \Delta$), where the
relaxation time-scale $\tau_{eV}[^1S_0]$ of Eq.~\eqref{eq:S_wave_relax} is
very large.

Allowing now for a proton component, we identify additional interaction 
channels, which actually turn out to be dominant in most cases.
Let us first consider the case of non-superfluid protons, since 
at sufficiently high densities the proton $^1S_0$ gap closes.
The neutron quasiparticles in the cores of vortices will then couple
to proton excitations, in much the same way as they coupled to the 
electron component~\cite{Feibelman1971}. However, a crucial
distinction is that the protons couple to neutrons by the strong 
nuclear force, instead of the much weaker electromagnetic interaction. 
The corresponding relaxation time becomes~\cite{1998PhRvD..58b1301S}
\bea\label{eq:np_relax}
 \tau_{pV}[^1S_0] &=& \frac{0.71}{\Omega_s} 
\frac
{m_n^*m_p^*}{m_n\mu_{pn}^*}\left(\frac{\ep_{Fp}}{\ep_{Fn}}\right)^2
\frac{\ep_{1/2}^0}{T} \nonumber\\
&\times&\exp\left(\frac{\ep^0_{1/2}}{T}\right)
\frac{\xi_n^2}{\langle \sigma_{np}\rangle},
\eea
where $\mu_{pn}^* = m_p^*m_n^*/(m_n^*+m_p^*)$ is the reduced mass of the 
neutron-proton system (entering the relation between the cross-section and 
the scattering amplitude squared), $\ep^0_{1/2}$ is the lowest energy
of vortex-core excitations according to Eq.~\eqref{eq:vortex_eigenstate}, 
and $\langle \sigma_{np}\rangle$ can be viewed as an average neutron-proton
cross-section. Eq.~\eqref{eq:np_relax} suggests a much stronger coupling
between the electron-proton plasma and the neutron vortices than implied 
by any of the previously quoted time scales. 

Consider next the case of superconducting protons, in which no
quasiparticle excitations are available for coupling to vortex-core
quasiparticles. Nevertheless, in this case, there is an entrainment
effect that induces a new type of magnetization of the neutron
vortex~\cite{Vardanyan1981,Alpar1984,SedrakyanShakhabasyan1980}.  In
effect, neutron vortices carry a non-integral multiple of the flux
quantum,
\bea\label{eq:effective_flux}
\phi^* = k_{\rm ent}\phi_0, \quad k_{\rm ent} = \frac{m_p^*}{m_p},
\eea
which leads to a magnetic field larger by four orders of magnitude
than that due to the spontaneous magnetization of neutrons in the
vortex core~\cite{1982PhRvD..25..967S}. The relaxation time scales are
correspondingly shorter. It is now convenient to define the relaxation
time in terms of a zero-range counterpart given by
\bea
\tau_{0}^{-1} &=& \frac{ 2 n_v  }{k_{eF}}\left(\frac{\pi^2\phi_*^2}{4\phi_0^2}\right).
\eea
The term in parentheses is an approximation to the exact Aharonov-Bohm
scattering result, in which $\sin^2 (\pi/2)(\phi_*/\phi_0)$ appears
instead~\cite{AlfordWilczek1989}, see
also the discussion in~\cite{AlfordSedrakian2010}. The finite-range result can then be
written as~\cite{Alpar1984}
\bea
\tau^{-1}_{e\phi} = \frac{3\pi}{32} \left(\frac{\ep_{Fe}}{m_p}\right)
\frac{\tau_{0}^{-1} }{k_{eF}\lambda},
\eea
where $\lambda$ is the penetration depth.  We 
call attention to the weak dependence
of the scattering relaxation time on the temperature, reflecting
the fact that the coupling is to the magnetic field and not to the thermally 
excited quasiparticles.

A more complete discussion of mutual friction requires consideration of the
interaction between neutron and proton vortices and their intertwined
dynamics, which however is beyond the scope of our focus on
microphysics.  We refer the reader to a recent
review~\cite{Haskell2017} for such a discussion.

\section{Conclusions} 
\label{sec:conclusions}

This review has covered a range of topics on nucleonic superfluidity
with an emphasis on extended systems such as neutron stars and matter
created in nuclear collisions.  The pairing problem at the level of
mean-field BCS theory, in which the pairing interaction is extracted
directly from free-space nuclear interactions, is essentially solved
within the density range corresponding to energies where the
scattering phase shifts are known. There still exist discrepancies
between various methods for microscopic many-body calculation of
pairing properties, notably in relation to the issue of suppression of
$S$-wave pairing in neutron matter by long-range collective
fluctuations in the nuclear medium.  Theories that incorporate such
fluctuation corrections, as well as the effects of short-range
correlations due to the repulsion of the two-nucleon potential at
short distances, have been emerging in recent years.  The goal of
achieving convergent results for pairing in low-density nuclear matter
appears to be within sight.  Other important objectives that arise at
higher densities are harder to achieve.  These include especially the
challenge of accurate evaluation of pairing gaps in the
$^3P_2$--$^3F_2$ channel, which is complicated by their
characteristically small magnitude, high sensitivity to the two-body
pairing interaction, which is not well constrained theoretically, and
the increasingly important role of the three-nucleon forces.
Additionally, the off-shell behavior of the pairing gap and its impact
on the phenomenology of nucleonic superfluids remain largely
unexplored.

Superfluid phases with broken space-time symmetries have received much
attention from theorists during the past two decades.  Recent
experimental realization of imbalanced superfluids in ultracold
fermionic atomic gases has created the possibility of laboratory tests of
the predictions of the many-body theory under highly controlled
conditions.  There are excellent prospects for future
cross-fertilization of nuclear theory and experimental activity in
cold atomic gases, especially in identifying the phases of imbalanced
superfluids and in exploring the physics of the BCS-BEC crossover.  The
phase diagram of imbalanced superfluids, as outlined in this review,
offers a broad arena for mutual interaction and enrichment of quantum
many-body theories and experimental studies of trapped atomic gases.

As discussed in detail in this review, the physics of the thermal
evolution of neutron stars is a sensitive probe of their interior
physics, particularly their composition. Accurate weak-interaction
rates in the superfluid phases of neutron stars are of great
importance for reliable modeling of neutron-star cooling.  The quantum
many-body methods involved in computations of these rates, some of
which existed already in the 1960s, have been recently applied to
compute  the weak response of nucleonic superfluids, thereby
providing accurate rates of neutrino emission from nucleonic and
hyperonic superfluids.  Future observational progress in measuring and
modeling the surface radiation of neutron stars, in conjunction with
improved theoretical input and simulations of neutron stars, can yield
further clues on their interior composition and on the couplings of
non-standard-model particles (e.g. axions) to matter.

Quantum vortex states, reviewed in the last section, are fundamental to
an understanding of the rich spectrum of observed rotational anomalies in
pulsars.  This is  an area in which models and theories 
developed for nuclear systems can be tested in laboratory experiments 
on ultracold atomic gases.  Further theoretical studies of vortex 
dynamics, combined with pulsar timing observations, can be expected 
to shed new light on the internal structure of the superfluid phases 
of neutron stars, especially on the microphysics of mutual friction 
as surveyed in this review.

\section*{Acknowledgments}

We are grateful to our colleagues who have helped us to shape our
views on the topics covered in this review.  A.S.\ acknowledges the
support by the DFG (Grants No.\ SE 1836/4-1 and No.\ SE 1836/3-2), by
the Helmholtz International Center for FAIR, and by the the European
COST Actions ``PHAROS'' (CA16214) and ``NewCompStar'' (MP1304).
J.W.C.\ acknowledges support from the McDonnell Center for the Space
Sciences and is grateful for the hospitality of the Centro de
Investiga\c{c}\~{a}o em Matem\'{a}tica e Aplica\c{c}\~{o}es,
University of Madeira, Funchal, Portugal.

%
%
%
\providecommand{\href}[2]{#2}\begingroup\raggedright\endgroup

\end{document}